\documentclass[twocolumn]{aastex63}
\usepackage[normalem]{ulem}

\usepackage{amsmath}
\usepackage{xcolor}
\usepackage{footmisc}
\usepackage{rotating,longtable}

\received{----, 2021}
\revised{----, 2021}
\accepted{----, 2021}

\shorttitle{Sh 2-301}
\shortauthors{Pandey et. al.}

\begin{document}

\title{Sh 2-301: a blistered H\,{\sc ii} region undergoing star formation}

\correspondingauthor{Rakesh Pandey}
\email{pandey.rakesh405@gmail.com}

\author{Rakesh Pandey}
\affil{Aryabhatta Research Institute of Observational Sciences (ARIES),
Manora Peak, Nainital 263 002, India}
\affil{School of Studies in Physics and Astrophysics,
Pt. Ravishankar Shukla University, Raipur, (C.G.), 492010, India}
\author[0000-0001-5731-3057]{Saurabh Sharma}
\affil{Aryabhatta Research Institute of Observational Sciences (ARIES),
Manora Peak, Nainital 263 002, India}
\author{Lokesh K. Dewangan}
\affil{Physical Research Laboratory, Navrangpura, Ahmedabad - 380 009, India}
\author{Devendra K. Ojha}
\affil{Tata Institute of Fundamental Research (TIFR),
Homi Bhabha Road, Colaba, Mumbai - 400 005, India}
\author{Neelam Panwar}
\affil{Aryabhatta Research Institute of Observational Sciences (ARIES),
Manora Peak, Nainital 263 002, India}
\author{Swagat Das}
\affil{Indian Institute of Science Education and Research(IISER), 
Tirupati, 517507, India}
\author{D. P. Bisen}
\affil{School of Studies in Physics and Astrophysics,
Pt. Ravishankar Shukla University, Raipur, (C.G.), 492010, India}
\author{Arpan Ghosh}
\affil{Aryabhatta Research Institute of Observational Sciences (ARIES),
Manora Peak, Nainital 263 002, India}
\affil{School of Studies in Physics and Astrophysics,
Pt. Ravishankar Shukla University, Raipur, (C.G.), 492010, India}
\author{Tirthendu Sinha}
\affil{Aryabhatta Research Institute of Observational Sciences (ARIES),
Manora Peak, Nainital 263 002, India}



\begin{abstract}
{
We present a multiwavelength study of the H\,{\sc ii} region Sh~2-301 (S301) using deep optical data, near-infrared data, radio continuum data and other archival data at longer wavelengths. A cluster of young stellar objects (YSOs) is identified in the north-east (NE) direction of S301. The H$\alpha$ and radio continuum images trace the distribution of the ionized gas surrounding a massive star ALS~207, and the S301 H\,{\sc ii} region is bounded by an arc-like structure of gas and dust emission in the south-eastern direction. The north-western part of S301 seems to be devoid of gas and dust emission, while the presence of molecular material between the NE cluster and the central massive star ALS 207 is found. The distribution of warm dust emission, ionized gas, and neutral hydrogen together suggests a blistered morphology of the S301 H\,{\sc ii} region powered by ALS 207, which appears to be located near the edge of the cloud.
The location of the NE cluster embedded in the cold molecular cloud is found opposite to the blistered morphology. There is a noticeable age difference investigated between the massive star and the NE cluster. This age difference, pressure calculation, photodissociation regions (PDRs), and the distribution of YSOs favour the positive feedback of the massive star ALS 207 in S301. On a wider scale of S301, the H\,{\sc ii} region and the young stellar cluster 
are depicted toward the central region of a hub-filamentary system, which is evident in the infrared images.
}
\end{abstract}




\section{Introduction}


Previous studies of bubbles associated with the H\,{\sc ii}
regions suggest that their expansion probably triggers 14\% to 30\% of
the star formation in our Galaxy \citep[e.g.,][]{2010A&A...523A...6D,2012MNRAS.421..408T,2012ApJ...755...71K}.
Feedback from the massive stars ionizes the surrounding molecular cloud
through their immense UV radiation and powerful winds, resulting in an H\,{\sc ii} region, a glowing nebula of ionized gas.
\citet{2014ApJ...795..121L} have discussed the various feedback processes in detail.
The feedback from the massive star may inhibit or terminate further star formation in the immediate
vicinity (termed as `negative feedback') or it can also promote and accelerate the star formation (known as `positive feedback'). Out of these two outcomes which one  will dominate depends not only on the process itself but also on the property of clouds \citep{2017MNRAS.467..512S}.
Stars born out of these processes are generally assembled in a group or cluster,
an entity  having a collection of physically-related stars. The physical features such
as shape, size, age, mass distribution of these clusters vary with their hosting environment and seems to
show the imprints of star formation processes itself  \citep{2003ARAA..41...57L, 2007prpl.conf..361A, 2017ApJ...842...25G, 2018MNRAS.481.1016G}.
Therefore, stellar clusters constitute the nearest laboratories for direct astronomical investigation of the
physical processes of star formation and early evolution.
The first generation of  massive stars in these regions also play a very important role as they
can give very crucial clues on the star formation through entirely different physical processes. Thus, we can have a mix of stars in the same star-forming region, giving hints of their origin. \citet{2015MNRAS.450.1199D} have discussed other star formation processes like cloud-cloud collision,
filamentary interactions etc. In recent years many authors have pointed out the active role of filamentary
structures and their subsequent interaction in star formation
\citep{2012A&A...540L..11S,2017ApJ...851..140D,2020ApJ...903...13D}.
Filamentary structures are often seen harboring young stellar clusters and massive star-forming clumps,
yet their precise role in star formation is not very well understood.

\begin{figure}
\includegraphics[width=0.51\textwidth]{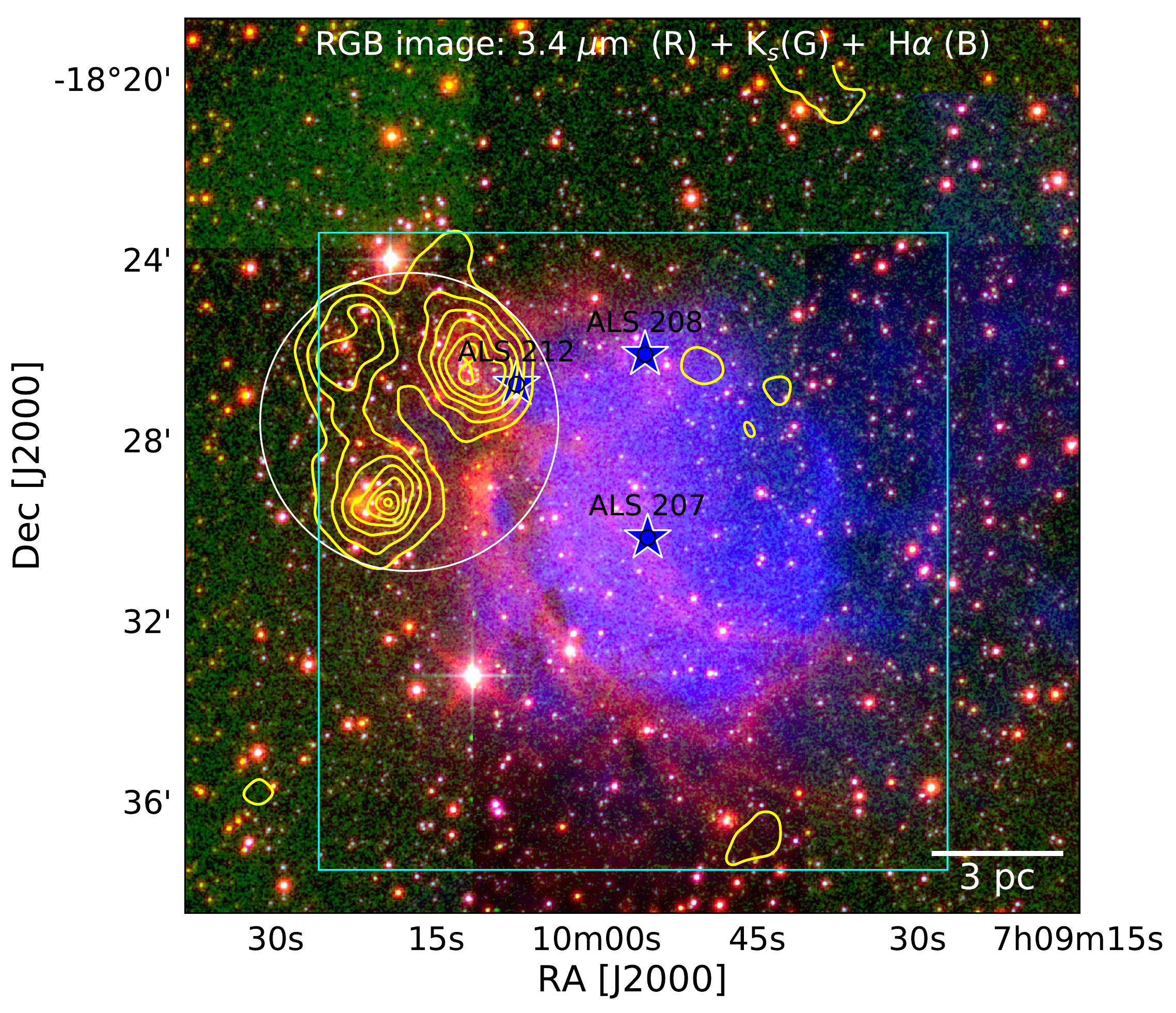}
\caption{\label{image1} Color-composite image of the $\sim18^\prime.5\times 18^\prime.5$ FOV around S301 H\,{\sc ii} region obtained using the  
WISE 3.4 $\mu$m (red), $K$-band (green) and UK Schmidt telescope (UKST) H$\alpha$ (blue).The locations of three massive stars ALS 207, ALS 208 and  ALS 212 are also shown in the figure by star symbols \citep{1979A&AS...38..197M}. The white circle represents the extent of the clustering identified in the present analysis (cf. Section 3.2). The stellar surface density of the NIR sources generated using nearest neighbor method (cf. Section 3.2) is shown with yellow contours.
The lowest contour is 1 $\sigma$ above the mean value of stellar density (i.e., 2.6 stars/arcmin$^2$) where the step size is 1$\sigma$=2  stars/arcmin$^2$. 
The cyan box represents the FOV covered in NIR (TIRSPEC) observation.
}
\end{figure}

With an aim to investigate the stellar clustering and their origin, star formation,
shape of the mass function (MF), and effects of the feedback from massive stars on these processes,
we have performed a multiwavelength study of the H\,{\sc ii} region `Sh 2-301' (hereafter, S301).
This is a southern sky ($\alpha$$_{2000}$ =07$^{h}$09$^{m}$55$^{s}$, $\delta$$_{2000}$ = -18$\degr$29$\arcmin$36$\arcsec$)
Galactic H\,{\sc ii} region located in a very large molecular cloud complex ($\sim6^\circ \times 3^\circ$) \citep{1995A&AS..114..557R}.
\citet{1989BAICz..40...42A} placed this region in a star-forming region `SFR 231.44-4.41' along with reflection nebulae Bran 6 and Bran 7.
This region is thought to be ionized by a massive O type star ALS 207 and also harbors two B type stars
ALS 208 and ALS 212 \citep{1979A&AS...38..197M,2015AJ....150...41G,2016ApJS..224....4M}.
Despite being showing very interesting features at different wavelengths,
this region is one of the poorly studied H\,{\sc ii} regions in our Galaxy.
In Figure \ref{image1}, we show the color-composite image made by using the 3.4 $\mu$m (red), $K$-band (green), H$\alpha$ (blue) images.
Clearly, the image shows a heated environment in the central region near massive star  ALS 207 as indicated by the H$\alpha$ emission.
The heated region is surrounded by gas and dust indicated by the infrared (IR) emissions.
All these morphologies suggest a prominent feedback from the massive star which is influencing its surroundings.
Thus, S301 is an ideal site for our investigation of feedback of massive star in the region.

We organize this work as follows.
In Section 2, we describe the optical/IR observations and data reduction
along with the archival data sets used in our analysis.
In Section 3 we describe the schemes to study the stellar densities,
membership probability, distance, reddening, age, MF, identification of YSOs etc.
The main results of the present study are summarized and discussed in Section 4 and we conclude in Section 5.
       
\section{Observation and data reduction}

\subsection{Imaging data}

We have observed S301 region in broad-band optical ($UBV{(RI)}_c$) filters using the 1.3 m Devasthal fast optical telescope (DFOT) of
Aryabhatta Research Institute of Observational Sciences (ARIES), Nainital, India.
The imaging camera on the telescope is a 2K$\times$2K CCD  covering $\sim18^\prime.5\times18^\prime.5$ field-of-view (FOV) of the sky.
The readout noise and gain of the CCD are 8.29 $e^-$ and 2.2  $e^-$/ADU, respectively.
The images of target field (S301) and standard field (SA98) \citep[$\alpha_{J2000}$: 06$^{h}$52$^{m}$14$^{s}$, $\delta_{J2000}$: -00$\degr$18$\arcmin$59$\arcsec$,][]{1992AJ....104..340L},
along with flat and bias frames were taken during the observations.
We have also observed S301 in broad-band Near-IR (NIR, $JHK$) filters using the
TIFR Near Infrared Spectrometer and Imager (TIRSPEC)\footnote{http://www.tifr.res.in/$\sim$daa/tirspec/}
mounted on the 2~m Himalayan $Chandra$ Telescope (HCT), Hanle, Ladakh, India. The specifications of the instrument and the detector array are mentioned in \citet{2014JAI.....350006N}.
The FOV of the instrument is $307^{\prime\prime}\times 307^{\prime\prime}$ in the imaging mode.
We took 9 pointing covering $15^{\prime}\times 15^{\prime}$  FOV around the
central massive star (ALS 207). 
Five dithered positions with 7 frames of 20 secs in each position have been used to create average combined images in the $J,H$ and $K$ filters.
The complete log of observation is provided in the Table \ref{log}.
For image cleaning, photometry and astrometry, we have used standard data reduction procedures
as mentioned in \citet[][]{2020ApJ...891...81P}.

We have followed the procedures outlined by \citet{1992ASPC...25..297S} for the calibration of the
optical instrumental magnitudes to the standard magnitudes (Vega) by using following
calibrations equations:

\begin{equation}
\begin{split}
u&= U + (5.328\pm0.021) -(0.046\pm0.01)(U-B)\\
&+ (0.490\pm0.061)X_U,
\end{split}
\end{equation}
\begin{equation}
\begin{split}
b&= B + (3.292\pm0.007) -(0.127\pm0.006)(B-V) \\
&+ (0.208\pm0.006)X_B,
\end{split}
\end{equation}
\begin{equation}
\begin{split}
v&= V + (2.488\pm0.007) +(0.072\pm0.005)(V-I_c) \\
&+ (0.131\pm0.009)X_V,
\end{split}
\end{equation}
\begin{equation}
\begin{split}
r_c&= R_c + (1.797\pm0.006) +(0.102\pm0.010)(V-R_c) \\
&+ (0.074\pm0.007)X_R,
\end{split}
\end{equation}
\begin{equation}
\begin{split}
i_c&= I_c + (2.529\pm0.009) -(0.013\pm0.006)(V-I_c) \\
&+ (0.011\pm0.007)X_I
\end{split}
\end{equation}

where,  $U,B,V,R_c,I_c$ and $u,b,v,r_c,i_c$ are the
standard and instrumental magnitudes of the standard stars in the SA98 field, respectively.  
Instrumental magnitudes were normalized for the exposure time and
X is the air mass at the time of observation.

The instrumental magnitudes in the NIR bands were transferred to the standard magnitudes (Vega)
by using the following transformation equations:

\begin{equation}
(J-K)= (0.99\pm0.02)\times (j-k)+ (0.61\pm0.01)
\end{equation}
\begin{equation}
(H-K)=(0.94\pm0.04)\times(h-k) + (0.66\pm0.01)
\end{equation}
\begin{equation}
(K-k)= (-0.12\pm0.05)\times(H-K) +  (-4.95\pm 0.01)
\end{equation}

where, $JHK$ and $jhk$ are the standard and instrumental magnitudes of the common stars
from the 2MASS catalog and HCT observations, respectively. 
The coefficients for the above equations were generated separately for each pointing.

We have compared our derived standard magnitudes with the available $V$ and $B$ band standard magnitudes
in `APASS'\footnote{The AAVSO Photometric All-Sky Survey, https://www.aavso.org/apass} archive and the photometric agreement is found to be reasonable.

Finally, we have made a photometric catalog by taking only those stars which are having photometric errors $<$0.1 mag.
Photometry of the stars that were saturated in our deep observations, have been taken from the short exposures and 2MASS catalog for the optical and NIR bands, respectively.
The number of sources detected in different bands and their detection limits are given in Table \ref{cftt}.
 
\subsubsection{Completeness of the photometric data}

\begin{figure}
\centering
\includegraphics[width=0.40\textwidth]{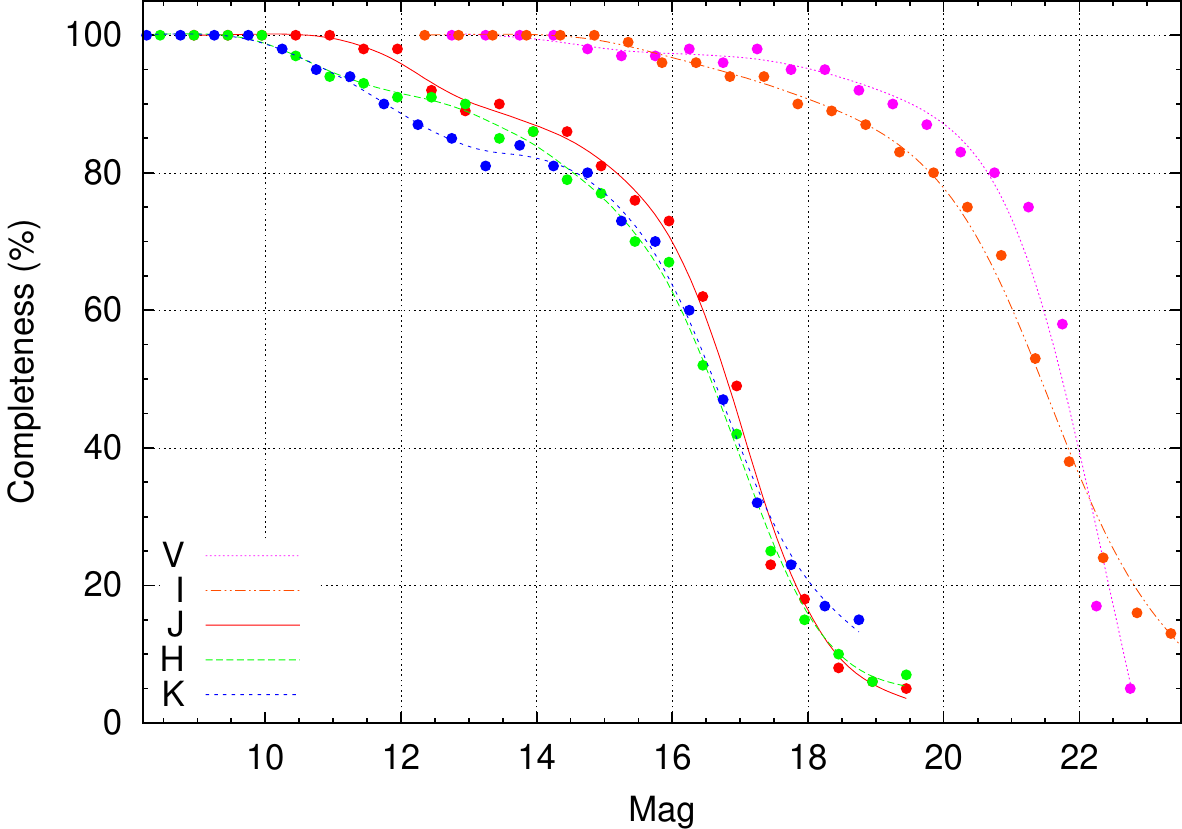}
\caption{\label{comp} Completeness levels as a function of magnitude in $V,I,J,H,K$ filters
derived from the artificial star experiments ({\it ADDSTAR}, see Section 2.1.1 for details).}
\end{figure}

Due to nebulosity, crowding of the stars, detection limit, etc the photometric data may be incomplete. It becomes very important to know the completeness limit in terms of mass for  MF analyses. We have used the procedure outlined in \citet{2008AJ....135.1934S} to determine the completeness factor. In this method we artificially added stars of known magnitudes and position in the original images using the $IRAF$ routine $ADDSTAR$. After getting images having artificially added stars, we performed photometry of those images as we did for the original images (keeping all the parameters same). Finally, the completeness factor can be easily determined by taking the ratio of the number of stars recovered to the added stars, in different magnitude bins as a function of magnitude. In Figure \ref{comp}, we have shown the completeness factor in different bands
as a function of magnitude. As expected, the completeness of the  data decreases with fainter magnitudes.
The completeness limits in terms of magnitudes and mass in different bands are given in Table \ref{cftt}.


\subsection{Spectroscopic data}

We have performed the spectroscopic observations of three bright sources ALS 207, ALS 208 and ALS 212 (see Figure \ref{image1}) in the S301 region,
using Hanle Faint Object Spectrograph Camera (HFOSC) instrument mounted on the HCT, Hanle, India.
The observations were performed with GRISM 7 (3800-6840 \AA~) with a resolution of 1200. The spectroscopic standard star was also observed in the same night for flux calibrations. The log of the spectroscopic observation is provided in Table \ref{log}.
 
\begin{figure*}
\centering
\includegraphics[width=0.95\textwidth]{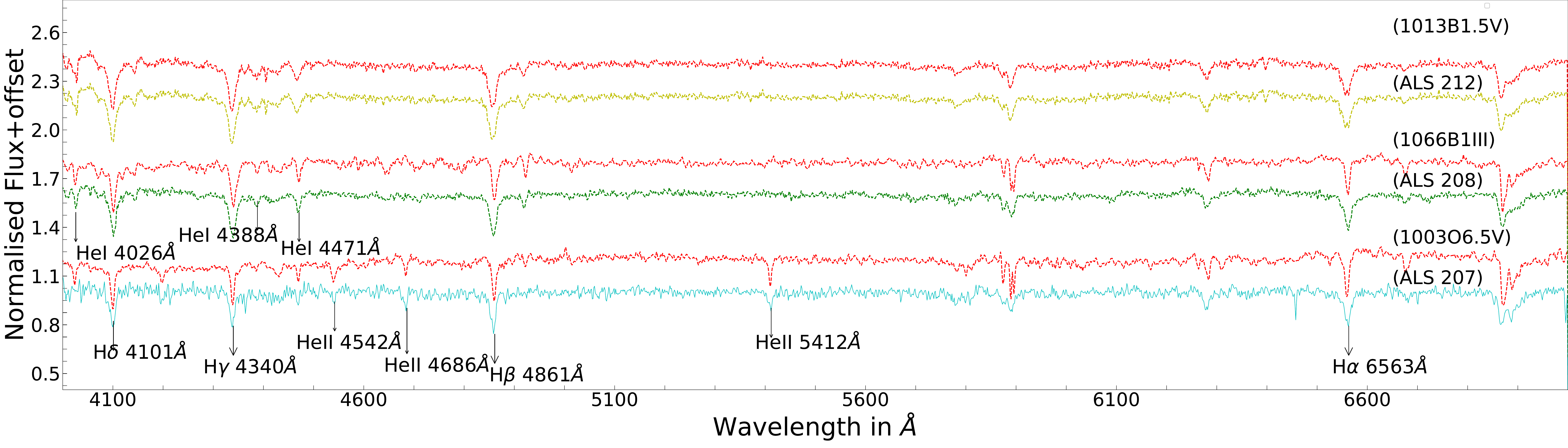}
\caption{\label{spec} The flux-calibrated and normalized spectra of the sources ALS 207, ALS 208 and ALS 212, shown with cyan, green and yellow color  curves, respectively.
For comparison, we have also plotted the standard spectra from the spectral library of \citet{1984lss..book.....J} in red color.
The important spectroscopic lines are also marked in the figure.}
\end{figure*}

Spectroscopic data reduction is done with IRAF packages using standard procedures \citep{2012MNRAS.424.2486J}.  
For wavelength calibration,  FeAr and FeNe arc lamps were observed during each night. For flux calibration,
standard stars such as Feige 34, Feige 110 and HZ 44 were observed.
Aperture extraction, identification of lines using lamps and
dispersion correction are done by tasks {\sc apall}, {\sc identify} and {\sc dispcor}, respectively.
{\sc standard}, {\sc sensfunc} and {\sc calibrate} tasks are used for flux calibration.
Finally, the spectrum is normalized using the task {\sc continuum} in IRAF.
The flux-calibrated normalized spectra of the three
sources are shown in the Figure \ref{spec}.

\subsection{Radio continuum data}

Radio continuum observation of the S301 region at 1280 MHz was carried out using GMRT facility on 2018 December 22 (Proposal Code: $35\_ 107$; PI: Rakesh Pandey). 
We have observed the flux calibrators for $\sim$ 15 mins in the beginning and end of the observation sequence. The phase calibrator is observed periodically along with the target source.
In each period, we observed the target for $\sim$ 30 minutes and the phase calibrator for $\sim$ 5 minutes.
The data has been reduced using the Astronomical Image Processing System (AIPS). We used VLA calibrators 3C147 and 3C286 as flux calibrators and 0744-064 as a phase calibrator.
Tasks {\sc tvflg} and {\sc Calib} were used to remove the bad data points and calibrations, respectively. After getting an image by doing several rounds of self calibration, primary beam correction was done using the task {\sc pbcor}.
In this way, we got our final reduced image at 1280 MHz with a synthesized beam size of $\sim$ $7^{\prime\prime}.5\times3^{\prime\prime}.4$.


\subsection{Archival data}

We have used the  point source catalog from the 2MASS in NIR \citep{2003yCat.2246....0C} and {\it Wide-field Infrared Survey Explorer} (WISE)
\citep{2010AJ....140.1868W} in mid-IR (MIR) from the  IRSA, the NASA/IPAC Infrared Science Archive\footnote{https://irsa.ipac.caltech.edu/applications/Gator/}.
We have used the recently released Gaia DR3\footnote{https://gea.esac.esa.int/archive/} data for the membership determination  \citep{2016A&A...595A...1G,2018A&A...616A...1G}.
Apart from that we have also used the multiband images from the surveys like AAO/UKST\footnote{https://www.roe.ac.uk/ifa/wfau/halpha/}, 2MASS\footnote{https://skyview.gsfc.nasa.gov/current/cgi/query.pl},
WISE\footnote{https://skyview.gsfc.nasa.gov/current/cgi/query.pl}, AKARI \footnote{https://irsa.ipac.caltech.edu/data/AKARI/} and Planck \footnote{https://www.ipac.caltech.edu/project/planck} \citep[cf.][]{2020ApJ...891...81P}.

To trace the distribution of neutral hydrogen gas towards the S301 region, we have also used the H\,{\sc i} (21 cm) line data from
the `HI Parkes All Sky Survey' (HIPASS)\footnote{https://www.atnf.csiro.au/research/multibeam/release/}. The angular resolution of the   HIPASS data is 15$^\prime$.5.
The channel spacing is 13.2 km s$^{-1}$ and the velocity resolution is 18.0 km s$^{-1}$.

\section{Results and Analysis}

\subsection{Spectral analysis of the bright stars }

The wavelength calibrated spectra of the bright stars in the S301 region are shown in the Figure \ref{spec}.
To classify the spectral type of these sources, we have used different
spectral libraries and criteria available in the literature \citep[e.g,,][]{1984lss..book.....J,1990PASP..102..379W}.
In the case of star ALS 207 (shown with the cyan color curve in Figure \ref{spec}), we can see  prominent hydrogen lines (3970, 4101, 4340, 4861, 6563 \AA)
along with ionized Helium lines, i.e.,  He II (4686, 4542 and 5411 \AA) and He I (4144, 4387, 4471, 4713 \AA). The line strength of He II usually gets weaker for
late O type stars and is last seen in B0.5 type stars \citep{1990PASP..102..379W}.
As this line can be seen in the spectrum of ALS 207 star, we can constrain its spectral type to be earlier than B0.5.  
The presence of He II lines (4686, 5411 \AA) and He II+I (4026 \AA) limits the spectral type of ALS 207 to O type.
In the case of ALS 208 and ALS 212 (shown with green and yellow color curves in Figure \ref{spec}) we can see stronger He I lines at 4026, 4388 and 4417 \AA,
while the He II 4686 \AA  ~line is absent. This implies that the ALS 208 and ALS 212 stars have spectral type later than B0.5.
The spectral types of these stars are further constrained by visually comparing their spectra with the spectral library of \citet{1984lss..book.....J}
as shown in the Figure \ref{spec} by red color curves. Thus, ALS 207, ALS 208 and ALS 212 stars are classified as O6.5 V, B1 III and B1 V spectral types, respectively. 
As we have assessed the spectral types of the massive stars based on the low-resolution spectra, we may have an uncertainty of $\pm$ 1 in the subclass identification.
The massive star ALS 207 was previously identified as O6.5 V ((f)) by \citet{2016ApJS..224....4M} which matches with our spectral classification 
with an additional ((f)) feature in the spectra. The ((f)) feature denotes the weak N III (4634-40-42 \AA) emission feature along with the 
strong He II (4686 \AA) absorption feature. These two features are also quite evident in the observed spectra of ALS 207 (Figure \ref{spec}). 

\subsection{Search for stellar clustering/grouping in the S301 region}
\subsubsection{Isodensity contours : The North-East cluster}

As most of the young star clusters are associated  with the active star-forming regions,
we have determined stellar surface density distribution to identify any kind of clustering/groupings present in the S301 region. We performed nearest neighbor (NN) method on the 2MASS NIR
catalog to generate surface density maps in the  $\sim18^\prime.5\times 18^\prime.5$ FOV around S301 H\,{\sc ii} region.
In a grid of 20$^\prime$$^\prime$$\times$20$^\prime$$^\prime$, the local surface density is determined by varying the radial distance as it encompasses
20$^{th}$ nearest star \citep[for details, refer][]{2020ApJ...891...81P}. In Figure \ref{image1}, the isodensity contours are shown with the yellow color contours.
The lowest contour is at 1$\sigma$ above the mean stellar density (2.6 stars/arcmin$^{2}$), while the step size is 1$\sigma$ (2 stars/arcmin$^{2}$).
The isodensity contours clearly reveal a clustering with two different peaks in the north-east (NE)
direction of S301. We refer to this stellar clustering as a NE-cluster. This clustering also includes one of the massive stars (ALS 212)
whereas the other two massive stars (ALS 207 and ALS 212) are located out of its boundary.
The approximate boundary of the NE-cluster is shown with a white circle in the Figure \ref{image1}.
The radius of the NE-cluster is found to be 3$^{\prime}$.3 centered at $\alpha_{2000}$: 07$^{h}$10$^{m}$17$^{s}$.51, $\delta_{J2000}$: -18$\degr$27$\arcmin$35$\arcsec$.

\begin{figure*}
\centering
\includegraphics[width=7.5cm,height=9cm]{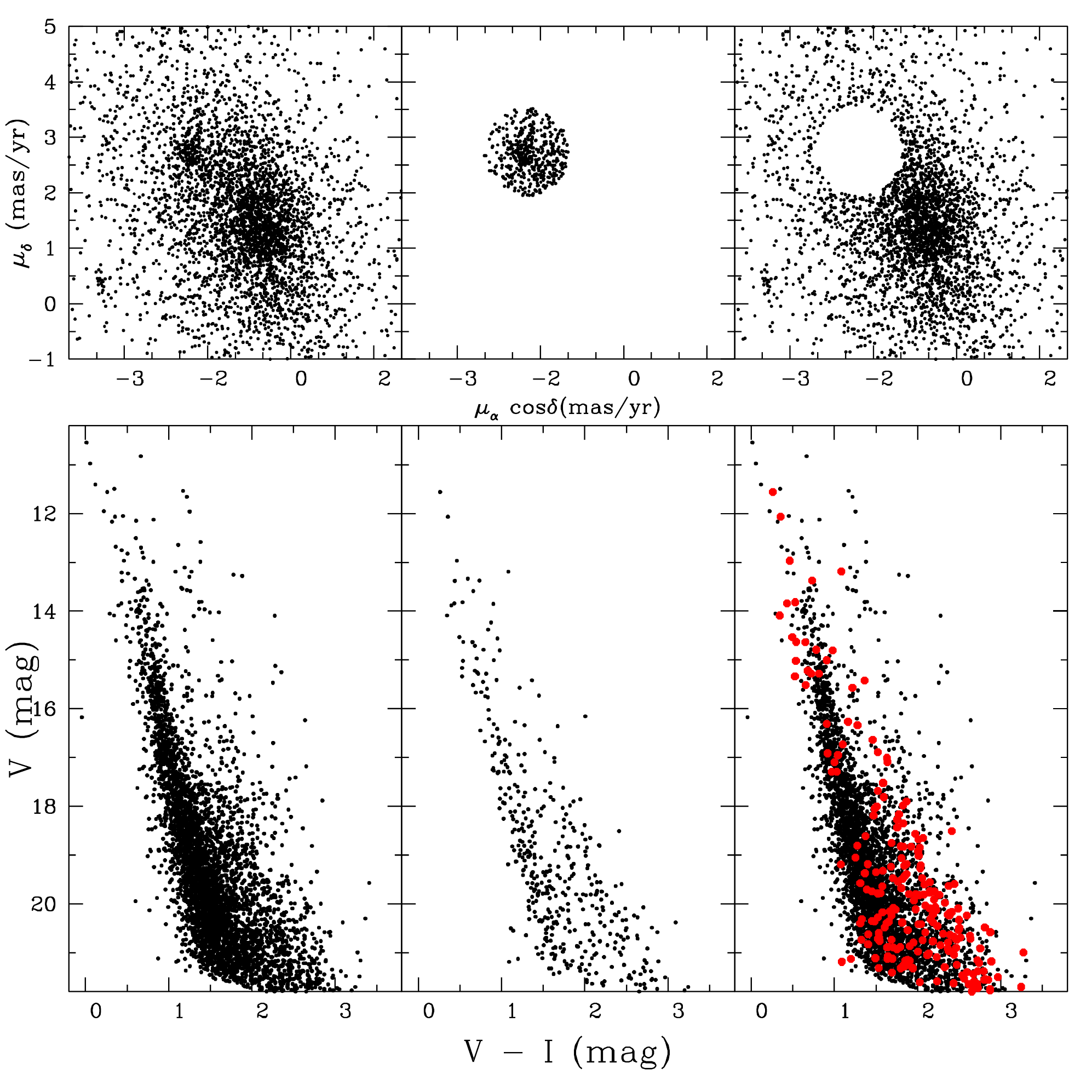}
\caption{\label{pm0} PM vector-point diagrams (VPDs; top sub-panels) and
$V$ vs. $(V-I)$ CMDs (bottom sub-panels) for the stars located inside the S301 region ($\sim18^\prime.5\times 18^\prime.5$ FOV). 
The left sub-panels show all stars, while the middle and right sub-panels show the probable cluster members and field stars, respectively. Red circles in the bottom right panel are probable member stars of the S301 region. 
}
\end{figure*}

\begin{figure*}
\centering
\includegraphics[width=8cm,height=9cm]{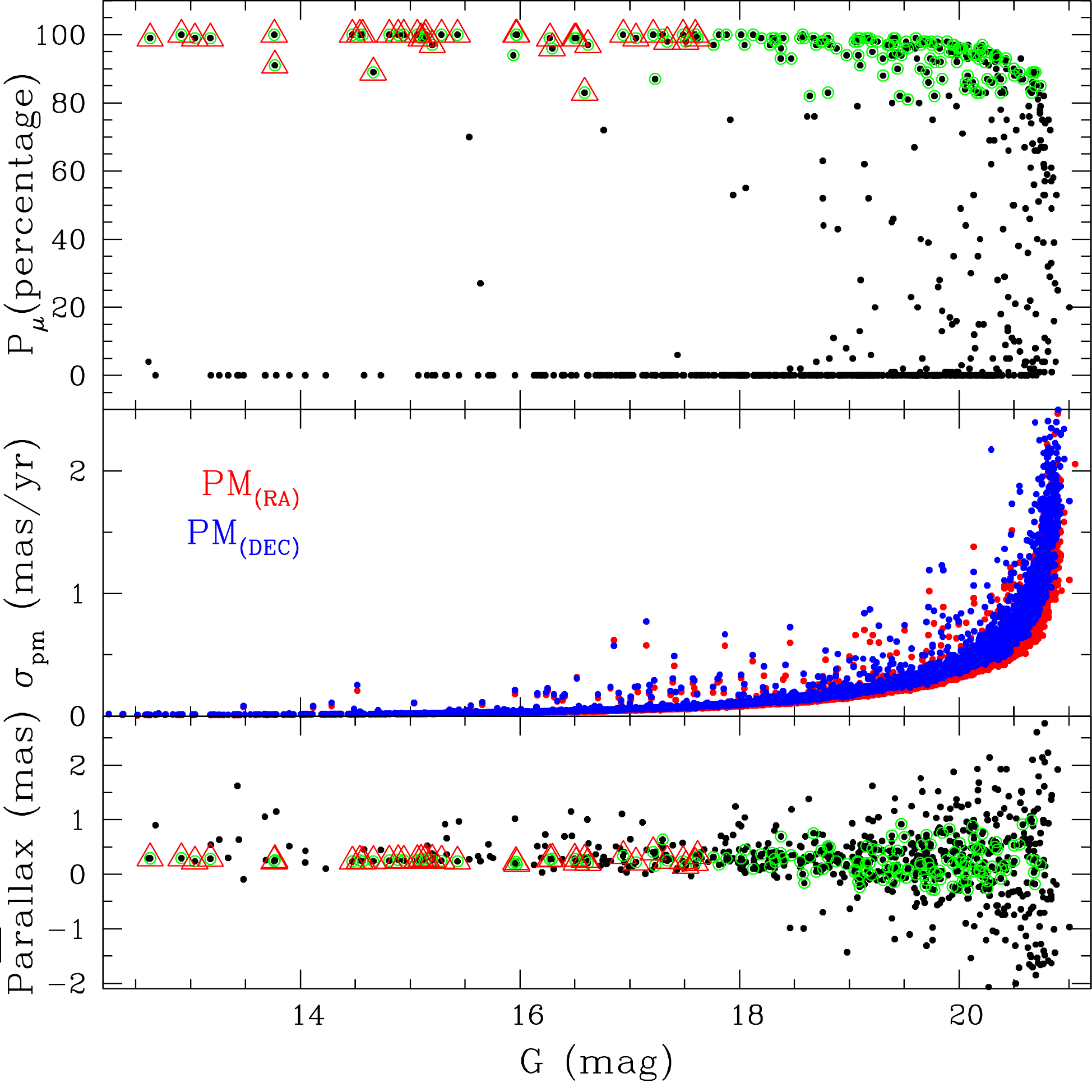}
\caption{\label{pm1} Membership probability P$_\mu$, PM errors $\sigma_{PM}$ and parallax of stars as
a function of $G$ magnitude for stars in the NE-cluster region. The probable member stars (P$_\mu>$80 \%) are shown by green circles while the 37 members of the S301
having parallax values with good accuracy (i.e., error$<$ 0.1 mas) are shown by red triangles.
}
\end{figure*}


\subsubsection{Membership probability of stars in the NE cluster}

The advent of Gaia DR3 data with its precise measurements of proper motion and parallax, has opened a new horizon in the study of star clusters.
We have used the $Gaia$ proper motion (PM) data  with $\sigma_{PM}<$3 mas/yr
to determine the membership probability of the stars belonging to the NE-cluster found in the S301 region using the method described in \citet{1998A&AS..133..387B}.
This method has been extensively used recently \citep[for example,][]{2020MNRAS.493..267S,2020MNRAS.498.2309S, 2020ApJ...896...29K, 2020ApJ...891...81P}.
        PMs in $RA$ and $Dec$  are plotted as vector-point diagrams (VPDs) in the top panels of
        Figure \ref{pm0}. The bottom panels show the corresponding $V$ versus $V -I$
        color-magnitude diagrams (CMDs).
        The dots in the top-left panel represent PM distribution of all the stars in the S301 region ($\sim18^\prime.5\times 18^\prime.5$ FOV)
        where a prominent  clump within a radius of $\sim$0.8 mas yr$^{-1}$ centered at
        -1.84 mas yr$^{-1}$ ($\mu_\alpha$cos($\delta$) and 2.74 ($\mu_\delta$) mas yr$^{-1}$ can be seen.
        This population of stars have almost similar PMs and have high probability for cluster membership.
        Remaining stars with scattered PM values are most probably a field population.
        This is more clearer in the VPDs and CMDs of probable cluster and field populations as shown in the
        middle and right panels, respectively.
        The probable cluster members are showing a well defined MS in the CMD which is usually seen for the similar population of stars.
        On the other hand, the probable field stars are quite obvious by their broad distribution in the CMD.
We first construct the frequency distributions of cluster stars ($\phi^\nu_c$) and field stars
($\phi^\nu_f$) using the equations 3 and 4 of \citet{1998A&AS..133..387B}. The input parameters such as PM center and its dispersion for the cluster stars ($\mu_\alpha$cos($\delta$) = -1.84 mas yr$^{-1}$, $\mu_\delta$ = 2.74 mas yr$^{-1}$, $\sigma_c$, $\sim$0.06 mas yr$^{-1}$)
and field stars ($\mu_{xf}$ = -0.81 mas yr$^{-1}$, $\mu_{yf}$ = 1.19 mas yr$^{-1}$, $\sigma_{xf}$ = 3.52 mas yr$^{-1}$ and $\sigma_{yf}$ = 3.62 mas yr$^{-1}$) are estimated in the same manner as have been discussed in our earlier work \citep{2020ApJ...891...81P}.

The membership probability (ratio of distribution of cluster stars with all the stars) for the $i^{th}$ star
is then estimated as:

\begin{equation}
P_\mu(i) = {{n_c\times\phi^\nu_c(i)}\over{n_c\times\phi^\nu_c(i)+n_f\times\phi^\nu_f(i)}}
\end{equation}

where $n_c$ (=0.26) and $n_f$ (=0.74) are the normalized numbers of stars for the cluster
and field regions ($n_c$+$n_f$ = 1).

We have plotted the estimated membership probability, errors in the PM  and parallax values as a function of $G$ magnitude in the Figure \ref{pm1}.
It is clear from the figure that the members are well separated from the field stars in the brighter $G$ magnitude while in the fainter end
uncertainty is larger. We have only considered only those stars as the members of the NE-cluster which have membership probability P$_\mu$ $>$80 \%.
Using this criterion, we have identified 194 stars as the members of the NE-cluster. We further cross-matched 136 of these sources to our optical catalog using a matching radius of 1 arc-second. The member stars are tabulated in Table \ref{PMT}.

\begin{figure*}
\centering
\includegraphics[width=0.45\textwidth]{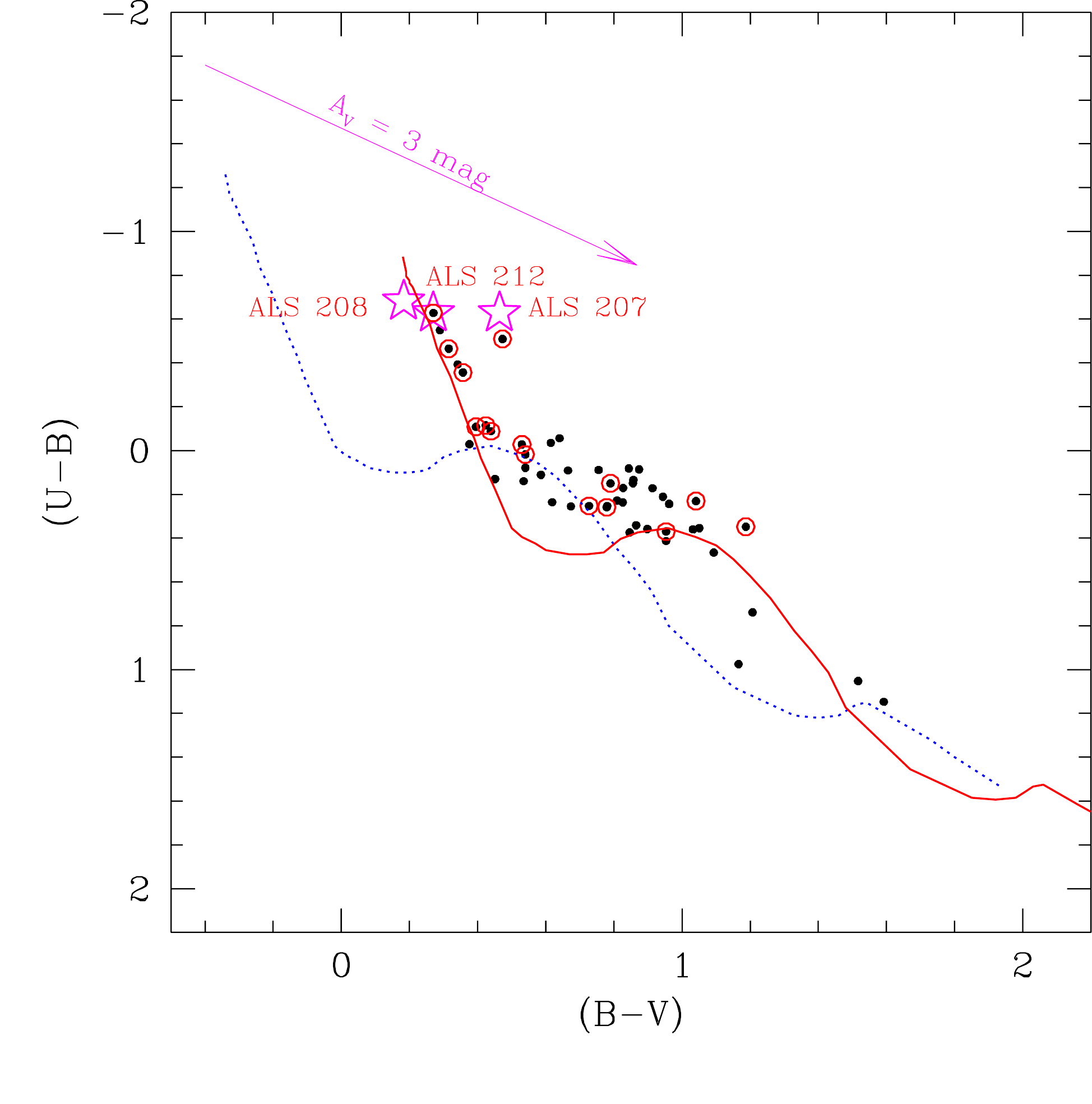}
\includegraphics[width=0.45\textwidth]{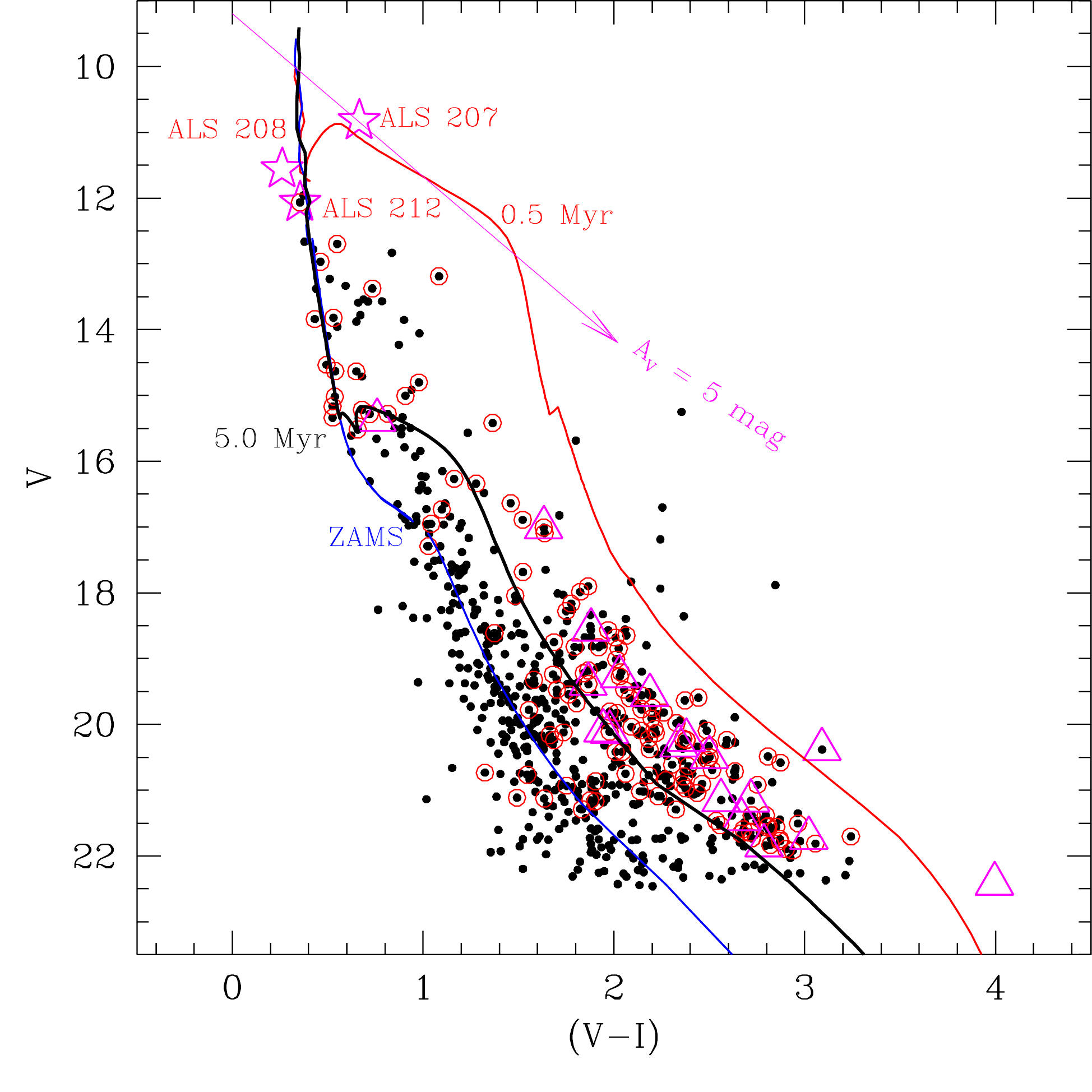}
\caption{\label{ccd}
Left panel: $(U-B)$ vs. $(B-V)$ TCD for the sources in the NE-cluster region (radius $<$ 3$^{\prime}$.3, black dots).
The identified member stars using PM analysis are also plotted with red circles.
The dotted blue curve represents the intrinsic ZAMS for $Z=0.02$ by \citet{Schmidt-Kaler1982}. The red continuous curve represents the ZAMS shifted along the reddening vector (see text for details) by $E(B-V)_{cluster}$ = 0.50 mag for the stars associated with the cluster. Right panel: $V$ vs. $(V-I)$ CMD for the same sources. The ZAMS (blue continuous curve) and PMS isochrone for 0.5 Myr and 5 Myr (red and black continuous curves) by \citet{2019MNRAS.485.5666P}, corrected for the distance of 3.54 kpc and reddening $E(B-V)=0.50$ mag, are also shown.
We have also plotted the location of massive stars and YSOs (see section 3.3) by star and triangle symbols in both the figures.}
\end{figure*}

\begin{figure}
\includegraphics[width=0.44\textwidth]{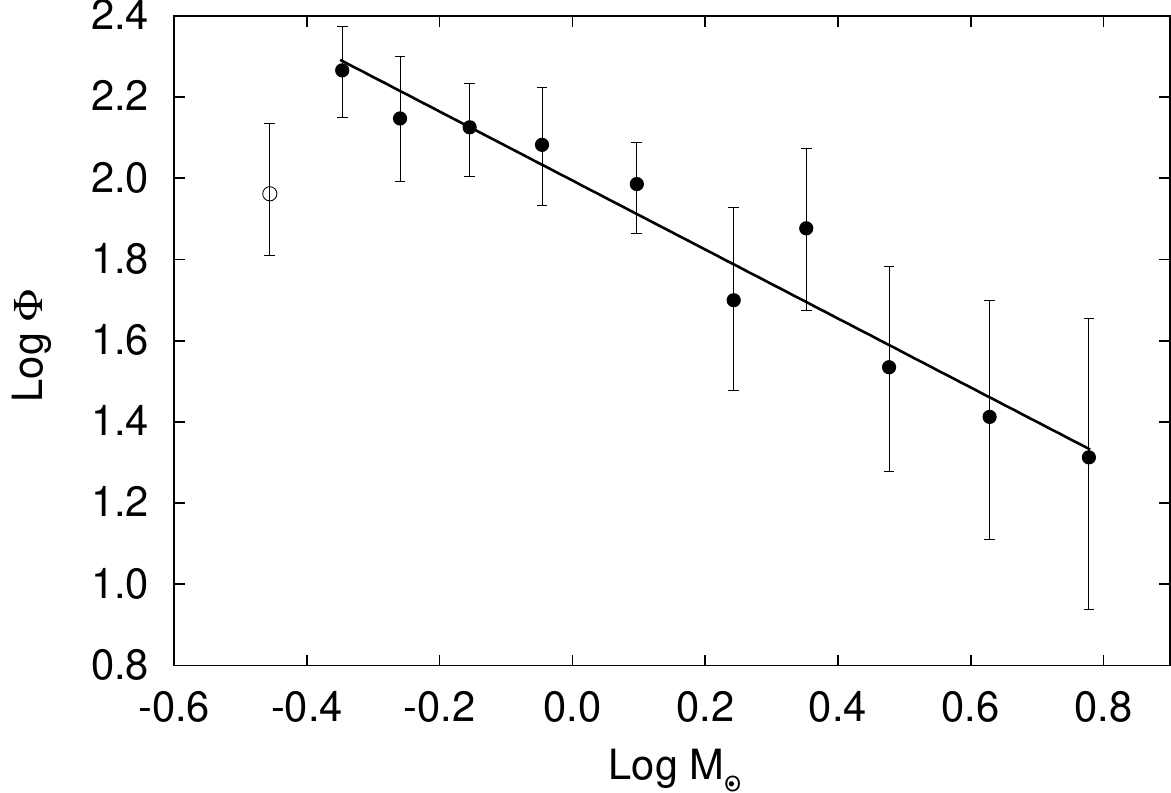}
\caption{\label{masf}  A plot of the MF for member stars in the NE-cluster.
        Log $\phi$ represents log($N$/dlog $m$). The error bars represent $\pm\sqrt N$ errors.
        The solid line shows the least squares fit to the MF distribution (black filled circles).
Open circle represents the data below the 80\% completeness limit.}
\end{figure}

\begin{figure}
\includegraphics[width=0.23\textwidth]{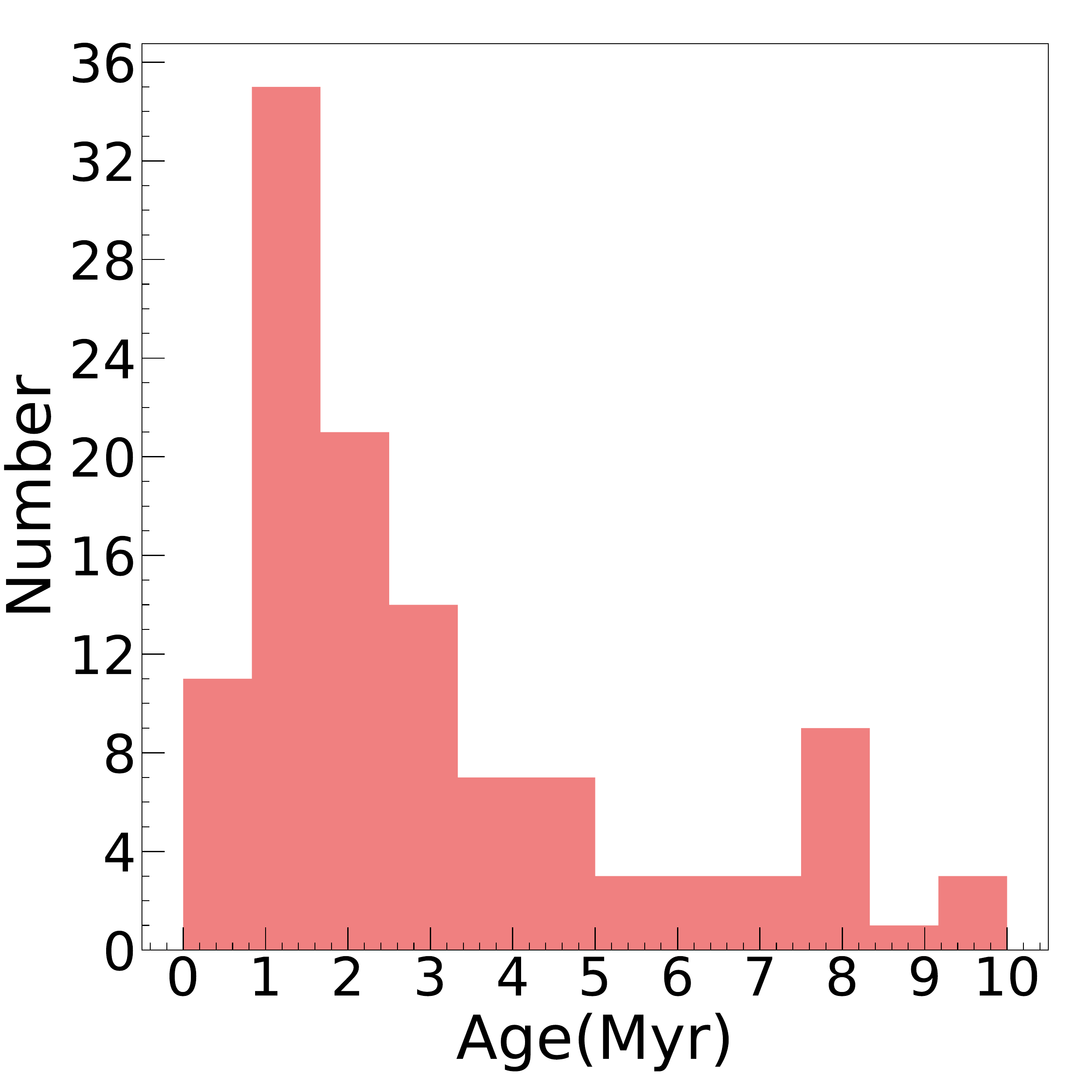}
\includegraphics[width=0.23\textwidth]{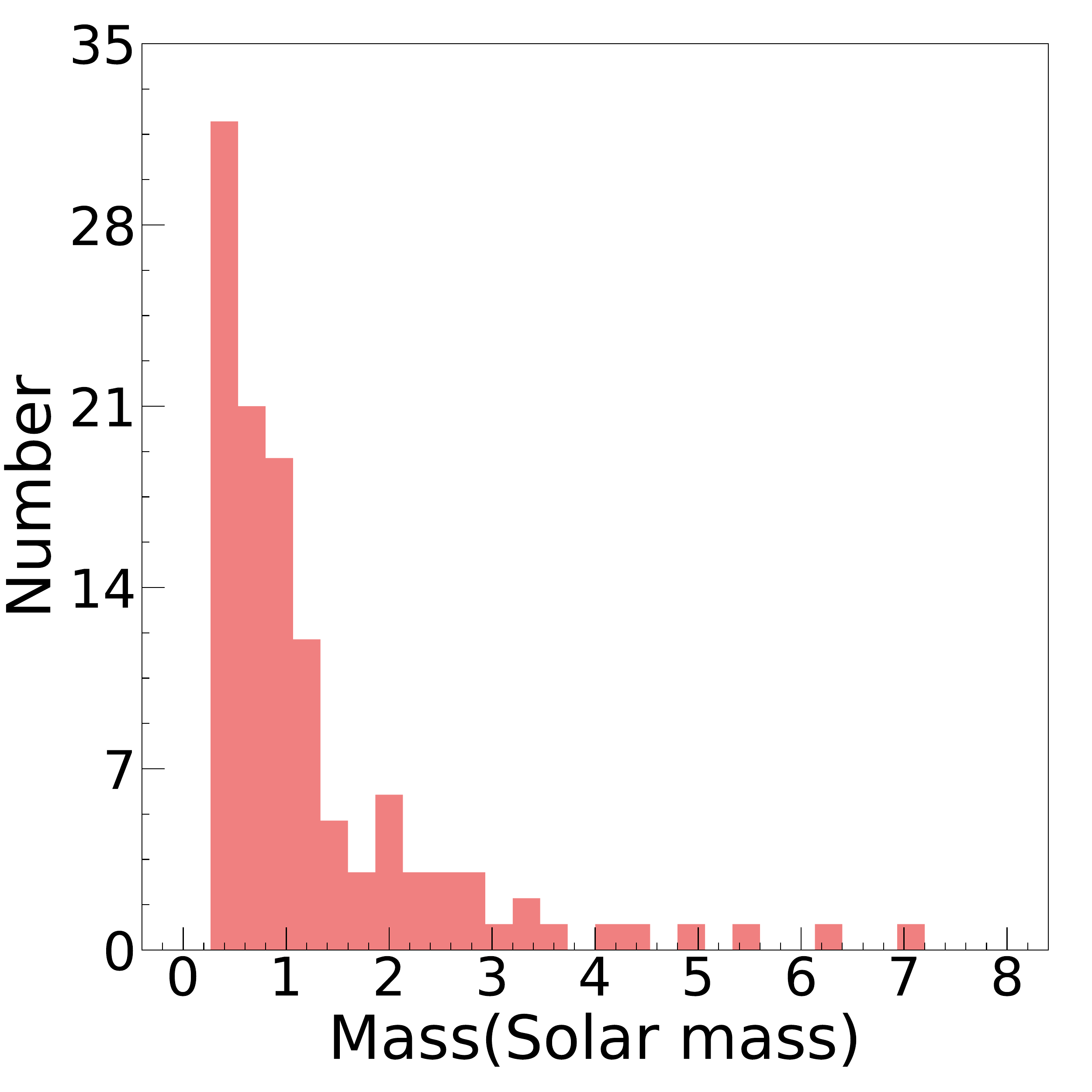}
\caption{\label{histogram-mem} Histograms showing the distribution of the ages (left panel) and mass of the member stars (right panel) of the NE-cluster
derived using the CMD (Figure \ref{ccd}).}
\end{figure}

\subsubsection{Reddening, distance and age of the NE cluster}

Two-color diagrams (TCDs) and CMDs of the member stars have been proven to be very good tools to constrain
reddening, distance and age of a cluster \citep[for example,][]{1994ApJS...90...31P,2006AJ....132.1669S, 2020MNRAS.498.2309S}.
In the left panel of  Figure \ref{ccd}, we have shown $(U-B)$ versus $(B-V)$ TCD of the stars located within the boundary of the NE-cluster (radius $<$ 3$^{\prime}$.3).
The intrinsic zero-age-main-sequence (ZAMS) from \citet[][blue dotted curve]{Schmidt-Kaler1982},
the member stars (red circles) identified with the Gaia data and most massive stars  are also shown in the figure.
We have shifted the ZAMS along the reddening vector i.e., $E(U - B)/E(B - V)$ = 0.72 (corresponding to $R_V$ $\sim$ 3.1, cf. Appendix A)
so that it matches with the distribution of member stars in the cluster region. The shifted ZAMS is shown with the red continuous curve.
By this method, we obtained a foreground reddening value $E(B-V)= 0.50$ mag for the NE-cluster region.

Gaia data releases have opened up the possibility of an entirely new perspective on the problem of distance estimation 
in cluster studies by providing the new and precise parallax measurements upto very faint limits\footnote{ https://www.cosmos.esa.int/web/gaia/}.
But, the uncertainty of the parallax is such that any parallax that comes with an uncertainty higher that 10\% can not be simply 
converted into distance in parsec. \citet{2018AJ....156...58B, 2021AJ....161..147B} have highlighted that for the vast majority of stars in the Gaia data releases, 
reliable distances cannot be obtained by inverting the parallaxes. 
Studies comparing the distances obtained using Gaia DR2 parallaxes with those from other methods show the 
existence of a systematic offset. The value of the offset ranges from $-0.082$ to $-0.029$ mas, depending on the objects 
and the method used \citep[for details, refer][]{2017A&A...599A..50A,2021MNRAS.504..356D}. 
Recently, \citet{2021MNRAS.504..356D} have found a systematic offset of $-0.05\pm0.04$ mas by comparing the open 
cluster distances obtained from isochrone fitting with those obtained from a maximum likelihood estimate of the 
individual member parallaxes. 
Using the Gaia data and maximum likelihood procedure, 
\citet{2018A&A...618A..93C} have estimated the distance of 1229 open clusters. However, they also highlighted that the 
distances to clusters with mean parallaxes smaller than $\sim$0.2 mas would be better constrained by a Bayesian approach using 
priors based on an assumed density distribution of the Milky Way \citep{2018AJ....156...58B} or photometric 
considerations \citep[e.g.][]{2018AJ....156..145A}, or simply with more classical isochrone fitting 
methods \citep[e.g.,][]{1994ApJS...90...31P,2006AJ....132.1669S,2017MNRAS.467.2943S,2020MNRAS.492.2446P,2020ApJ...891...81P}.
Therefore, for the present study, we relied on the distance estimates of \citet{2021AJ....161..147B}.  
We have calculated the mean of the distances \citep{2021AJ....161..147B} of  37 members of the NE-cluster
having parallax values with good accuracy (i.e., error$<$ 0.1 mas, as shown in Figure \ref{pm1} with red triangles) as 3.54$\pm$0.54 kpc.
\citet{1979A&AS...38..197M} have mentioned the distance of S301 as 5.8 kpc based on the spectroscopy of massive stars ALS 207, ALS 208 and ALS 212.
Later on \citet{1989BAICz..40...42A} have estimated the distance of S301 kinematically as 5.1 kpc. 
With the availability of the Gaia DR3 data, we can now constrain the distances of these stars with better kinematical measurements.
The distances of the massive stars ALS 207, ALS 208 and ALS 212 are found to be $3.21\pm0.14$ kpc, $2.99\pm0.20$ kpc and $2.93\pm0.19$ kpc, respectively \citep{2021AJ....161..147B}.
Similar distances of both the NE-cluster and massive stars implies that the NE-cluster is in fact associated with the S301  H\,{\sc ii} region.

The difference between spectro-photometric and the kinematical distances can be explained by the abnormal reddening law in this region. 
We have estimated the distances of massive stars in the region by assuming a normal reddening law, i.e,  $R_V=3.1$ \citep[as assumed by][]{1979A&AS...38..197M, 1989BAICz..40...42A} 
and an abnormal reddening law, i.e., $R_V=3.7$ (a typical reddening law observed in many Galactic star forming regions, 
for example, $R_V$ = 3.75 \citep[][NGC 6910]{2020ApJ...896...29K},  3.85 \cite[][NGC 7538]{2017MNRAS.467.2943S}, 3.7 \citep[][the Carina region]{2014A&A...567A.109K}, 
3.3 \citep[][NGC 1931]{2013ApJ...764..172P}, 3.5 \citep[][NGC 281]{2012PASJ...64..107S}, and 3.7 \citep[][Be 59]{2008MNRAS.383.1241P}. 
The individual $E(B-V)$ color excesses of the massive stars ALS 207, ALS 208, and ALS 212 are estimated 
as 0.80, 0.47, 0.57 mag (for $R_V =3.1$) and 0.84, 0.50, 0.60 mag (for $R_V=3.7$), respectively, 
based on the spectral types using the intrinsic color-spectral type relation given in \citet{Schmidt-Kaler1982}. 
Then, the individual spectro-photometric distances come out to be 5.9, 7.9, 5.9 kpc (for $R_V=3.1$)  and 4.3, 6.5, 4.1 kpc (for $R_V=3.7$) 
for ALS 207, ALS 208, ALS 212, respectively, using again the corresponding $M_V$-spectral type relation from \citet{Schmidt-Kaler1982}. 
Thus, assuming a normal reddening law can overestimate the spectro-photomeric distance, e.g., in the above case, it 
was overestimated by $\sim$1.6 kpc, and hence, it can explain the difference between the kinematical 
\citep{2021AJ....161..147B} and spectro-photometric \citep{1979A&AS...38..197M, 1989BAICz..40...42A} distance estimates. 
Also, from the above example, it is clear that the wrong $A_V$ correction in case of 
abnormal reddening law in the y-axis of the CMD is mostly compensated by wrong distance estimate and the change in 
the individual color excess value for the x-axis of the CMD is minimal, i.e., $E(B-V)\sim0.035$ mag. Thus, a change in the 
reddening law might have marginal effect on the derived physical parameters \citep[for more details, see][]{2017MNRAS.467.2943S}.


We have plotted the CMD
for the stars located within the boundary of the NE-cluster (radius $<$ 3$^{\prime}$.3) and the member stars  in the right-panel of Figure \ref{ccd}.
The most massive stars in the S301 region are also shown in the figure.
The blue continuous curve represents ZAMS from \citet{2019MNRAS.485.5666P} corrected for the extinction ($E(B-V)=0.50$ mag)
and distance (3.54 kpc).
The pre-main sequence (PMS) isochrones of 0.5 and 5 Myr by \citet{2019MNRAS.485.5666P} are also shown in the  right-panel of Figure \ref{ccd} with continuous red and black curves, respectively.
The shifted ZAMS and PMS isochrones match well with the distribution of member stars which further confirms the reddening and distance of this cluster.
An upper limit to the age of S301 region can be established
from the most massive member star. The location of the most massive star ALS 207 (O6.5V)
in the  $V$ versus $(V - I)$ CMD is traced back along the reddening
vector to the  turn-off point in the main-sequence (MS), which is equivalent to a 5 Myr old isochrone
(cf. black curve in Figure \ref{ccd} right panel).
The ages and masses of the member stars of the NE-cluster have also been derived
by applying the procedure described in our earlier work \citep{2009MNRAS.396..964C,2017MNRAS.467.2943S} using the $V$ versus $(V - I)$ CMD.
Briefly, we created an error box around each observed data point of a star in the CMD using the errors associated with photometry
as well as errors associated with the estimation of reddening and distance.
Five hundred random data points were generated by using Monte-Carlo simulations in this box.
The age and mass of each generated point were then estimated from the nearest passing
isochrone in this CMD.
For accuracy, the isochrones and  evolutionary tracks  were used in a bin size of 0.1 Myr
and were interpolated by 2000 points.
At the end we have taken the mean and standard deviation of age/mass of
the above simulated 500 data points, as the final derived values and errors of each star, and are provided in the Table \ref{PMT}.
The distributions of estimated age and mass of the member stars are shown in the Figure \ref{histogram-mem}
which peak around 1.5 Myr and 0.8 M$_\odot$, respectively.
The mean values of age and mass of the member stars are found to be 3.9$\pm$3.2 Myr and 1.3$\pm$1.2 M$_\odot$, respectively.

\subsubsection{Mass function}

The MF is usually expressed by the relation $N (\log m) \propto m^{\Gamma}$, where  the slope of the MF is given by  $\Gamma = d \log N (\log m)/d \log m  $ and $N (\log m)$ is the number of stars per unit logarithmic mass interval.
The MF is generally used as a statistical tool to understand the formation process of stars. It is basically the distribution of mass of stars formed in a star-forming event.
To calculate the MF in the NE-cluster of S301, we have used the distribution of the masses of the member stars estimated in the previous section.
We have used only those stars for which the age was less than  5 Myr, i.e., the upper age limit of this region is based on the age of the most massive star.
After applying the correction factor for the incompleteness of the photometric data (Section 2.1.1) on the number of stars in different mass bins, we have plotted the MF in Figure \ref{masf}.
The distribution seems to have no turn-off point. 
The slope of the MF ($\Gamma$) in the mass range $\sim$0.4$<$M/M$_\odot$$<$7.0
comes out to be $-0.85\pm0.07$ for the stars in the NE-cluster region.

%

\subsection{YSOs in the S301 region and their physical properties}

The spatial distribution of YSOs and their physical properties can be used
to infer the star formation scenario and evolution of a star-forming region.  
In this  study, we have used NIR and MIR observations of the S301 region
($\sim$ $18^\prime.5\times18^\prime.5$ FOV) to identify 37 candidate YSOs based on their excess IR emission.
The identification and classification schemes are described in Appendix B.
In Table \ref{data1_yso}, we provide the list of YSOs along with their magnitudes in different bands and their classification.

\begin{figure}
\centering\includegraphics[width=0.45\textwidth]{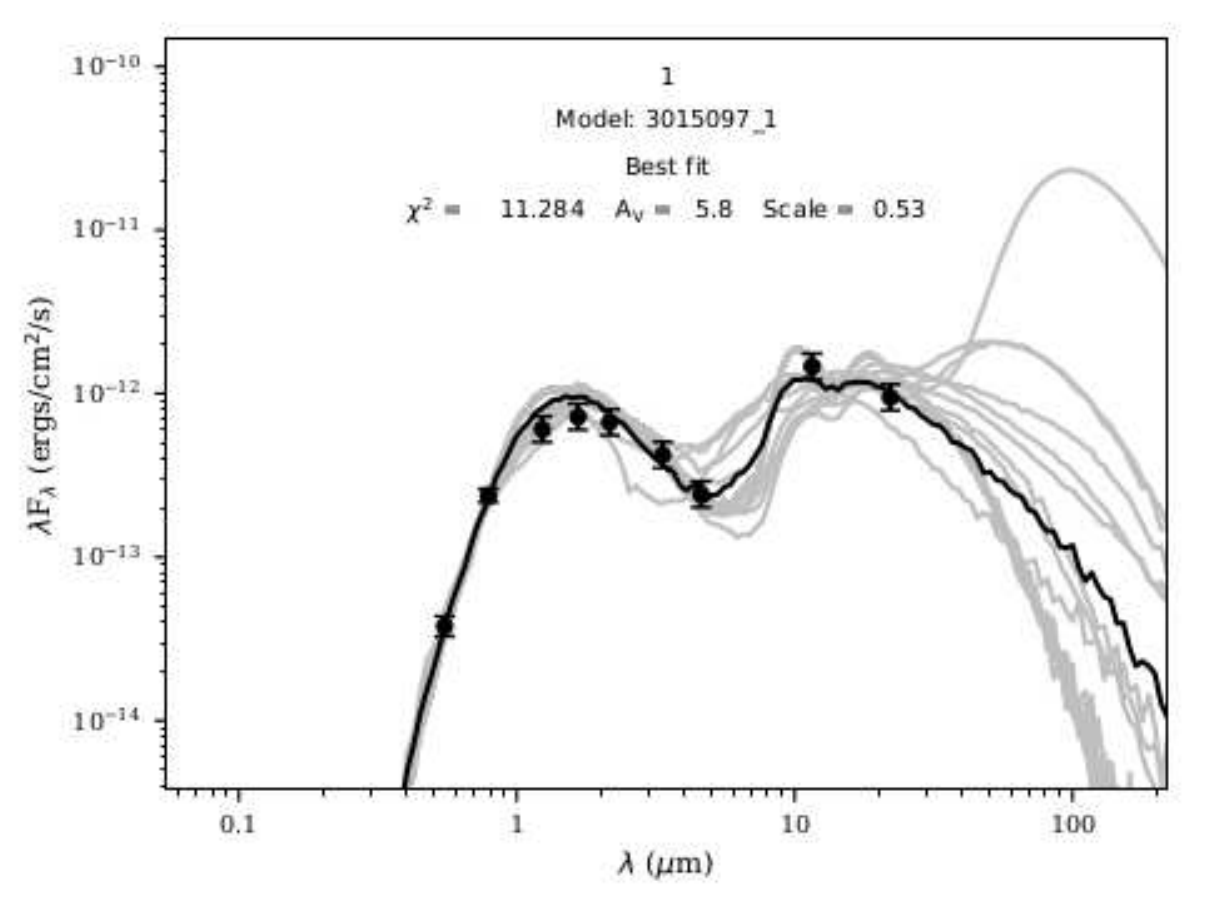}
\caption{\label{sed} Sample SED for a Class\,{\sc ii} YSO
created by the SED fitting tools of \citet{2007ApJS..169..328R}.
The black curve shows the best fit and the gray curves show the subsequent well fits.
The filled circles with error bars denote the input flux values.}
\end{figure}

\begin{figure}
\includegraphics[width=0.23\textwidth]{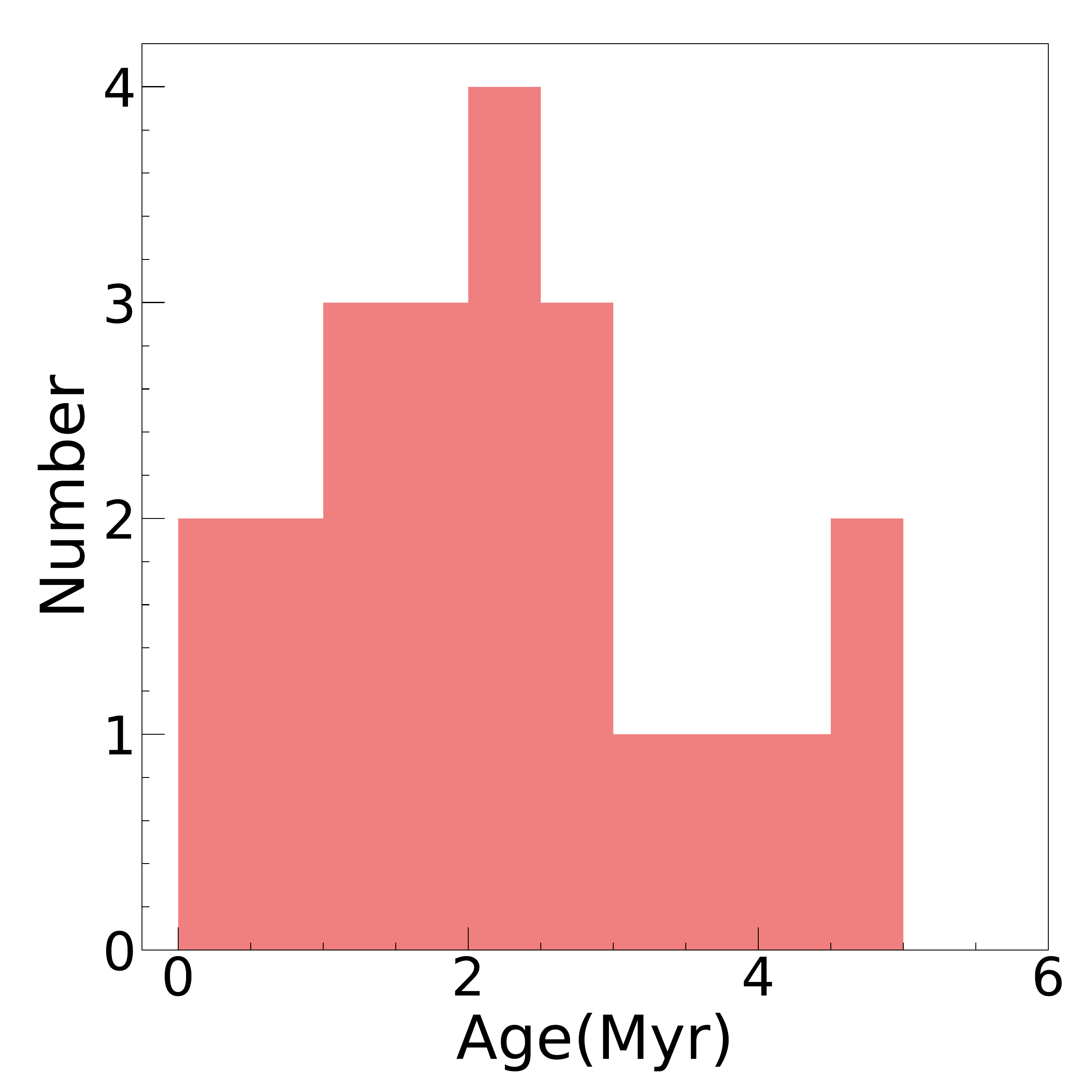}
\includegraphics[width=0.23\textwidth]{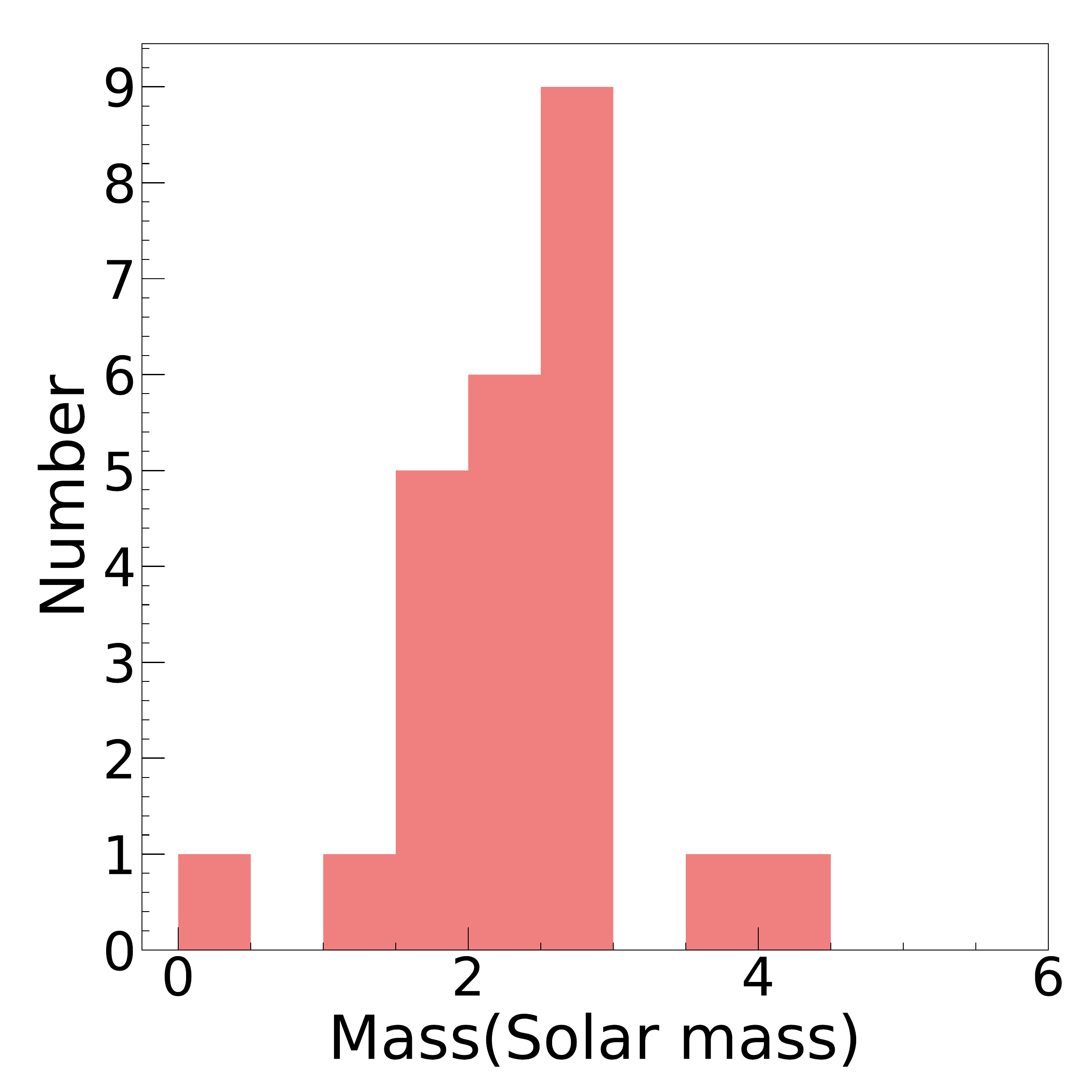}
\caption{\label{histogram-yso} Histograms showing the distribution of the ages (left panel) and
masses (right panel) of the YSOs (24) in the S301
as derived from the SED fitting analysis (cf. Section 3.5).
}
\end{figure}

We have derived physical properties like age and mass of the identified YSOs using the SED fitting analysis.
The grid models and fitting tools of  \citet{2003ApJ...598.1079W,2003ApJ...591.1049W, 2004ApJ...617.1177W} and \citet{2006ApJS..167..256R,2007ApJS..169..328R} have been used to construct SEDs of the YSOs.
This method has been used and described in our previous publication \citep{2020ApJ...891...81P}.
We have provided fluxes in optical to MIR wavelengths (0.37, 0.44, 0.55, 0.65, 0.80, 1.2, 1.6, 2.2, 3.4, 4.6, 12 and 22 $\mu$m) as input data points.
Taking extinction and distance as free parameters, the SED fitting tool fits the models to the input data points with a condition that minimum five data points should be there for fitting.
We gave input distance range from 3.0- 4.0 kpc keeping in mind the error associated with the distance estimation (cf. Section 3.2.3).
The $A_V$ range is given as 1.6 mag (foreground reddening value) to 30 mag (to accommodate deeply embedded YSOs).
In Figure \ref{sed}, a sample SED of a Class\,{\sc ii}  YSO is shown. Black curve represents the best fit model while the Grey curve represents the subsequent well fit models.
The well-fit models for each source are defined by
$\chi^2 - \chi^2_{min} \leq 2 N_{data}$, where $\chi^2_{min}$ is the goodness-of-fit parameter for the
best-fit model and $N_{data}$ is the number of input data points. From the well-fit models for each source derived
from the SED fitting tool, we calculated the $\chi^2$ weighted
model parameters such as the stellar mass and stellar age of each YSO
and they are given in Table \ref {data1_yso}. The histograms for age and mass of the YSOs are shown in Figure \ref{histogram-yso}.
They peak around 2.5 Myr and 1 M$_\odot$. The mean values of age and mass of the YSOs are found to be 2.5$\pm$1.6 Myr and 2.3$\pm$0.8 M$_\odot$, respectively.
These values are similar to those for the member stars of the NE-cluster.

\subsection{Distribution of molecular cloud around S301 region}

In the absence of CO data, we have used the extinction map as a proxy to trace the molecular gas
in the S301 region \citep{2009ApJS..184...18G, 2011ApJ...739...84G,2017MNRAS.467.2943S}.
We derived extinction map using the $(H - K)$ colors of the MS stars after removing the sources showing excess IR emission \citep[cf.][]{2011ApJ...739...84G}. 
The method is described in \citet{2005ApJ...632..397G,2009ApJS..184...18G} and extensively used recently \citep{2020ApJ...891...81P,2020ApJ...905...61P}.  
We have used the NN method to derive the extinction value in a grid of 20$^\prime$$^\prime$$\times$20$^\prime$$^\prime$, by taking the mean extinction value of the 20 nearest stars. 
To obtain the extinction value, we used the relation 
E =$(H - K)_{obs}$ - $(H - K)_{int}$, where $(H - K)_{obs}$ and  $(H - K)_{int}$ are the observed and intrinsic colors of the MS stars
and then calculated the $A_V = 15.87\times E(H-K)$ \citep[cf.][]{2011ApJ...739...84G}.
The sources deviating above 3$\sigma$ were excluded to calculate the final extinction value of each point.
The sources showing excess emission in IR can lead to overestimation of extinction
values in the derived maps. Therefore, the candidate YSOs and probable contaminating sources are excluded for the calculation of extinction.
Here, it is also worthwhile to note that the derived $A_V$ values are the lower limits of their 
values as the sources with higher extinction may not be detected in our study and the spatial resolution of the map depends on the local stellar density.

\begin{figure*}
\centering
\includegraphics[width=0.47\textwidth]{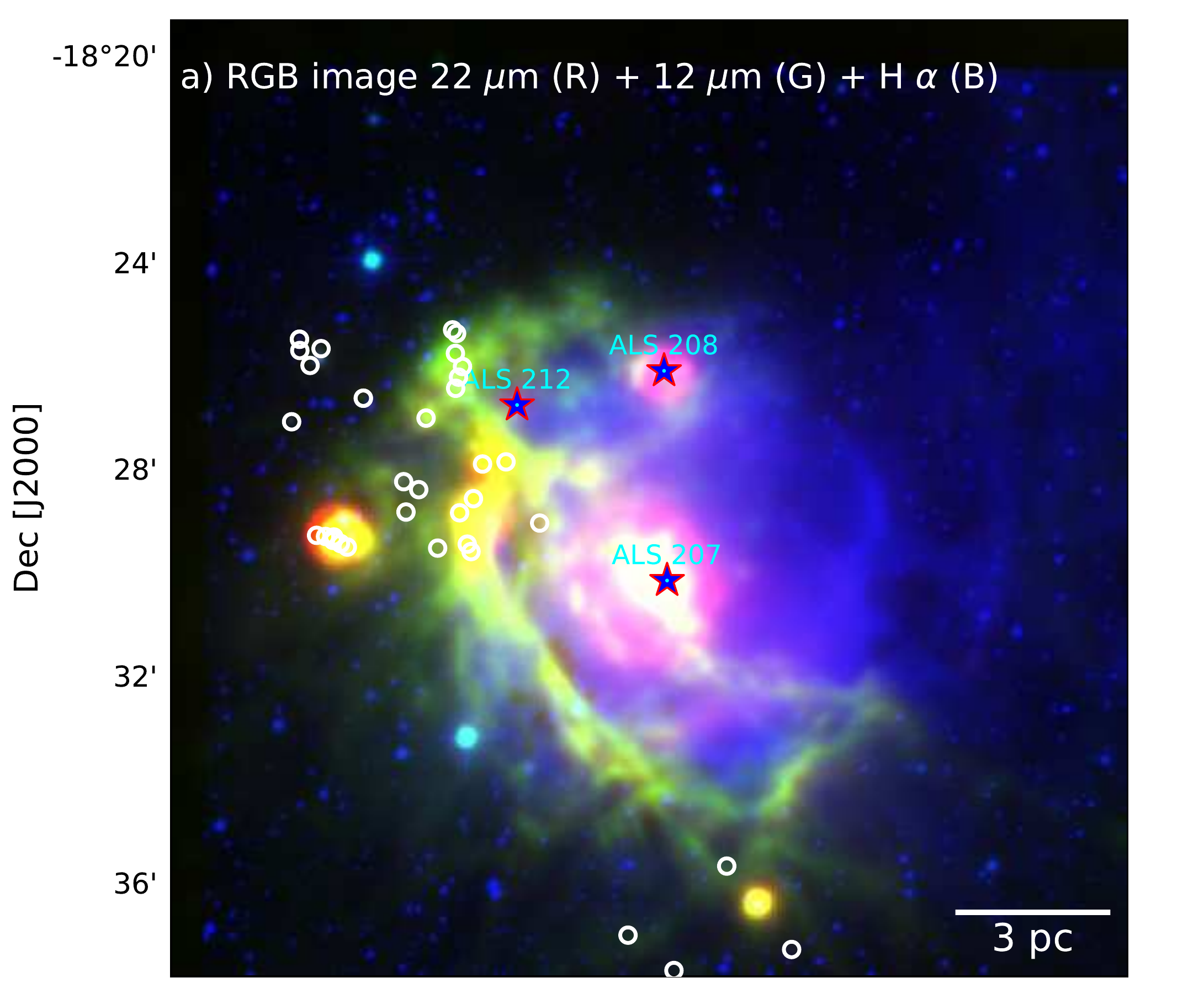}
\hspace{-0.49cm}\includegraphics[width=0.44\textwidth]{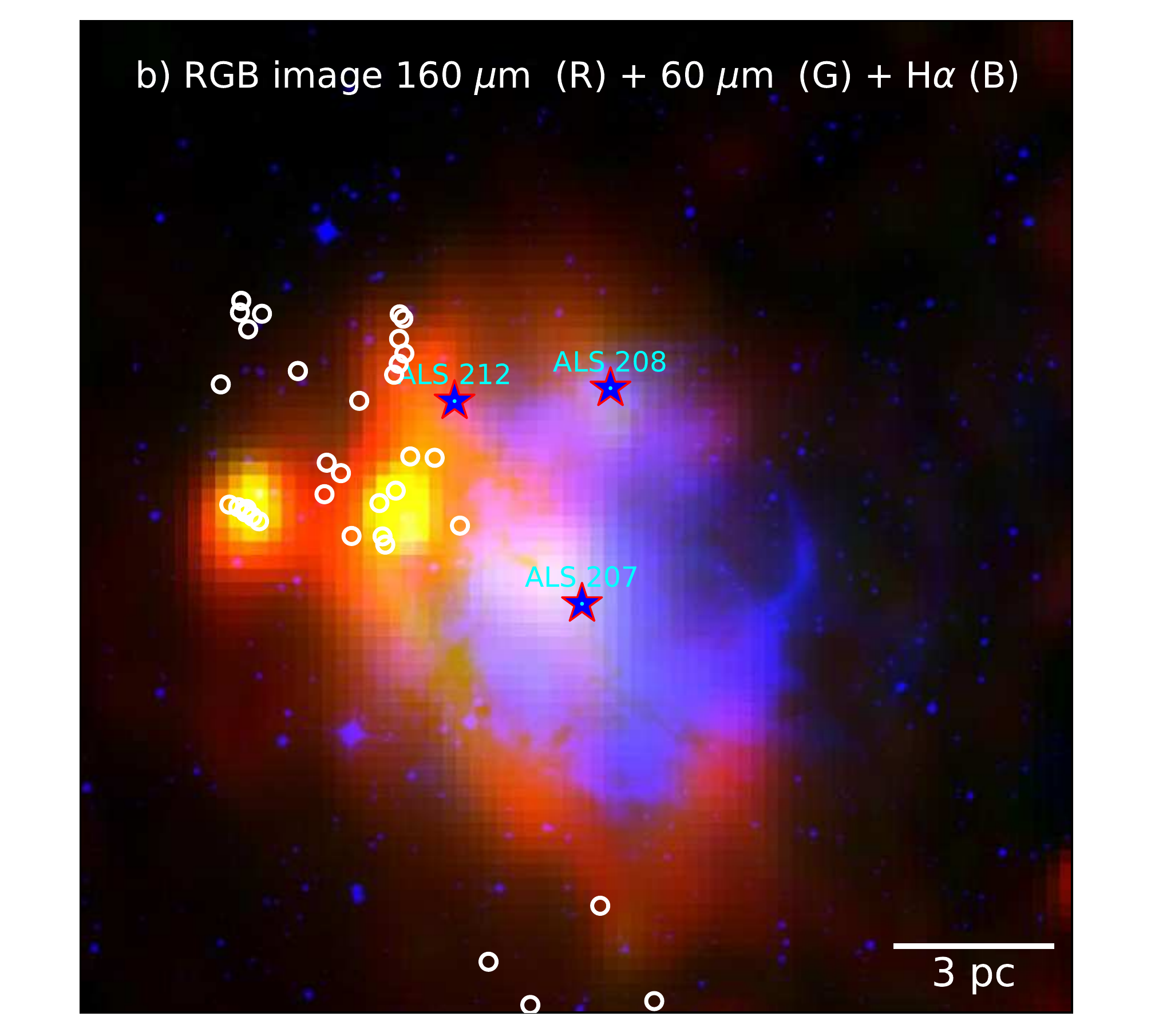}
\includegraphics[width=0.45\textwidth]{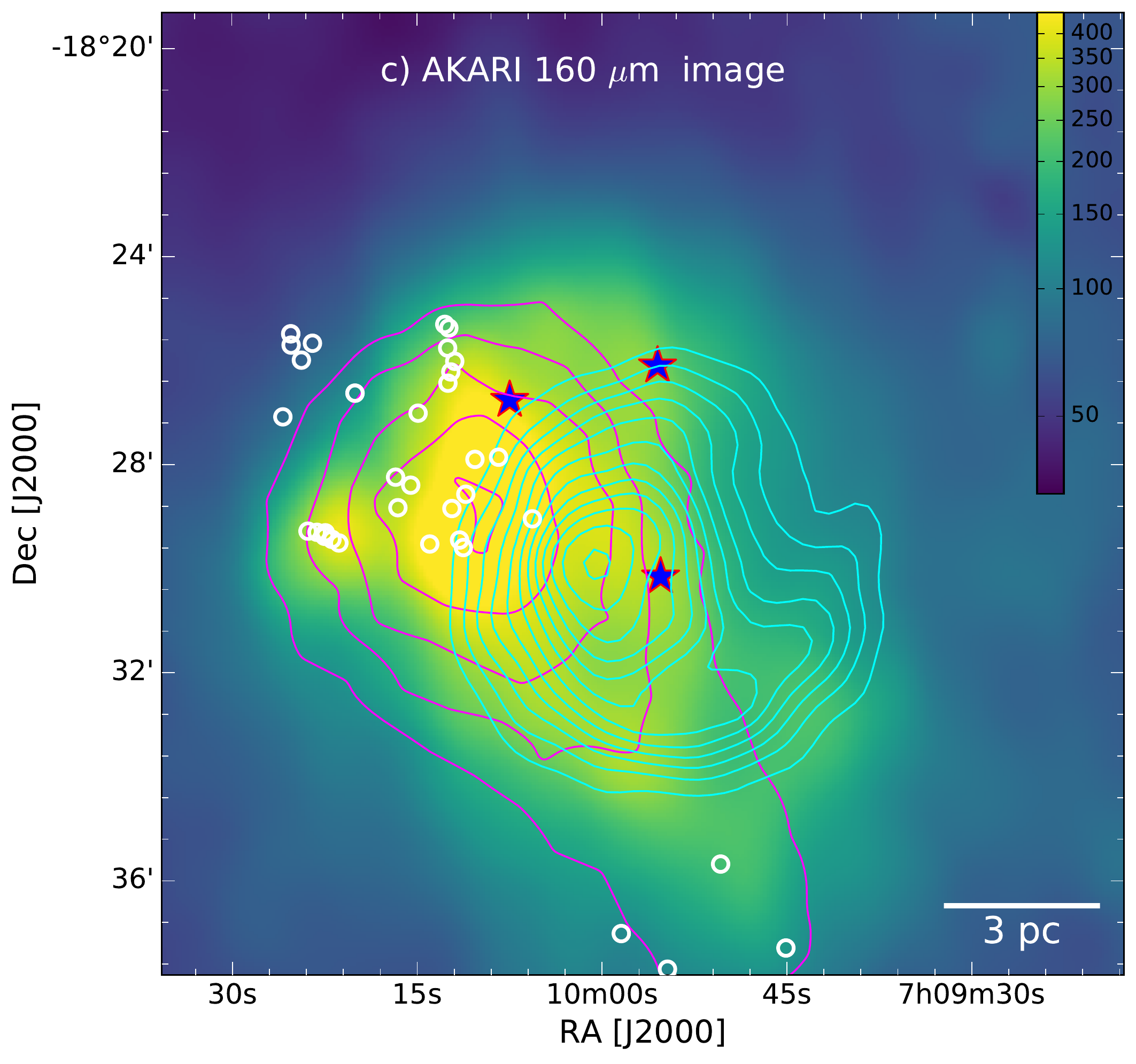}
\includegraphics[width=0.44\textwidth]{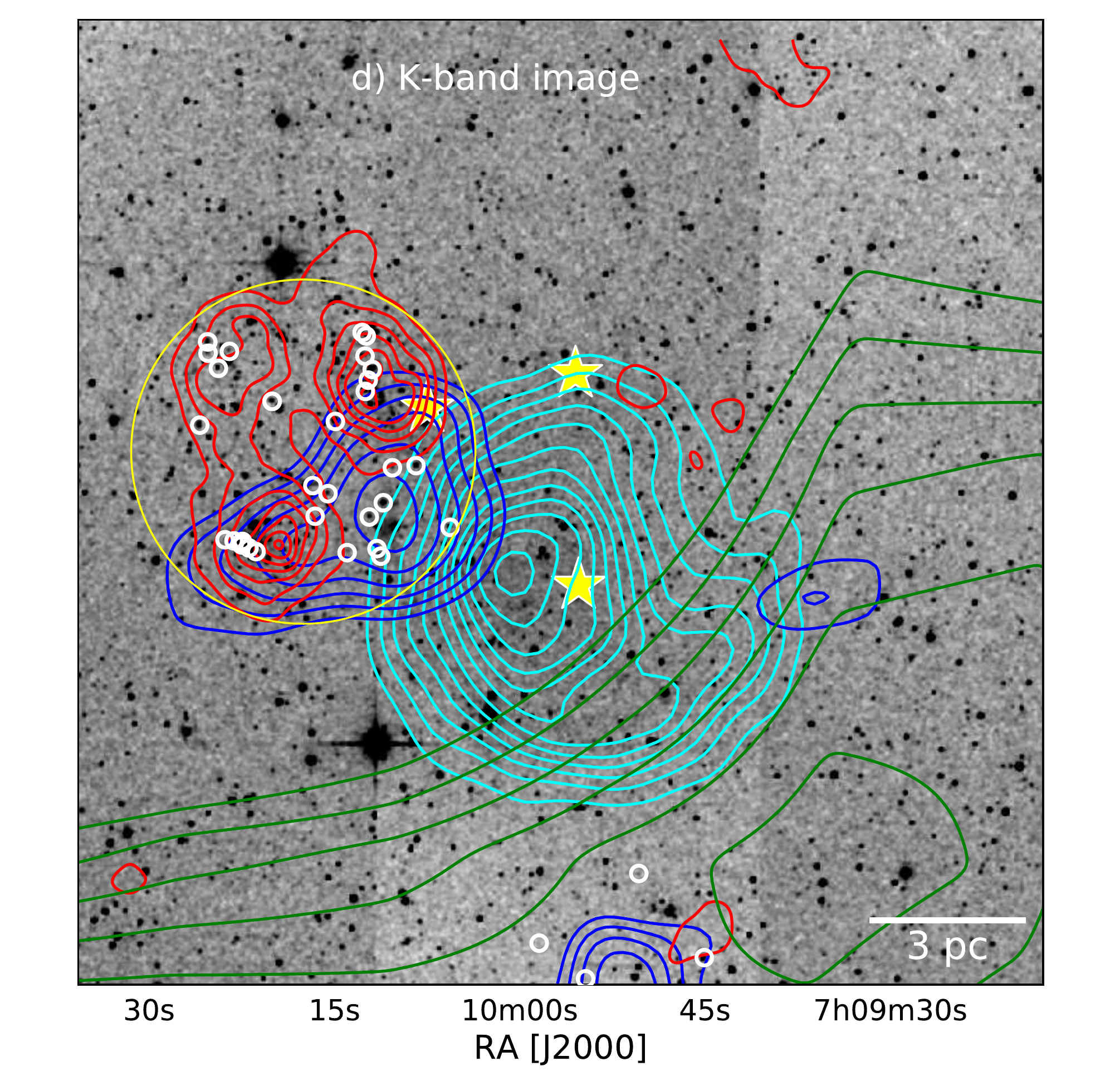}
	\caption{\label{multi} (a) Small scale view ($\sim$18$^\prime$.5$\times$18$^\prime$.5)  of the S301 shown with a color-composite image made using WISE 22 $\mu$m (red), WISE 12 $\mu$m (green) and H$\alpha$ (blue). (b) color-composite image of the S301 created  using AKARI 160 $\mu$m (red), AKARI 60 $\mu$m (green) and H$\alpha$ (blue) images. 
(c) The NVSS (cyan) and Planck 870 $\mu$m (magenta) contours superimposed on the AKARI 160 $\mu$m image.
The lowest NVSS contour is at 0.006 Jy/beam and the step size is 0.006 Jy/beam. The lowest Planck contour is at 0.021506 K and the step size is 0.004 K.
(d) Cluster surface density contours (red), extinction map contours (blue), NVSS radio contours (cyan) and H\,{\sc i} map contours (green) superimposed on the 2MASS $K$ band image. 
The lowest extinction contour is 1 $\sigma$ above the mean value of extinction ($A_V$=4.6 mag) where step size is 1$\sigma$=0.2 mag. 
The lowest  H\,{\sc i} contour is at 290 K and the step size is 1 K. The cluster surface density contours are at same level as in Figure \ref{image1}.
The circle and star symbol represent the location of YSOs and massive stars, respectively, in the S301 region.
}  
\end{figure*}

\subsection{Environmental conditions of S301}

In Figure \ref{multi}a, we have shown the color-composite image of  the S301 region made by 
using WISE $22$ $\mu$m (red), WISE $12$ $\mu$m (green) and H$\alpha$ (blue), images. 
The figure shows a stunning view of the gas and dust content distributed in the southern and eastern part of this region. 
$WISE$ 12 $\mu$m image covers the prominent polycyclic aromatic hydrocarbon (PAH) features
at 11.3 $\mu$m, indicative of photo-dissociation regions (or photon-dominated regions, or PDRs).
The southern and eastern parts of the complex seem to contain PDRs produced under the influence of massive stars.
These features are distributed as the arc-like structures which are probably swept-up gas and dust 
under the influence of the nearby massive stars.
In addition to above, YSOs are also seen in and around these arc-like structures.
The distribution of warm dust at $22$ $\mu$m near the massive stars show that 
the most massive star ALS 207 has created a warm dust envelope which seems to be bigger in size as 
compared to that from the less massive star ALS 208. The other massive star ALS 212 probably lacks a warm dust envelope.
The H$\alpha$ emission seems to trace the  full extension of the  S301 H\,{\sc ii} region.
Similar structures have been found in several H\,{\sc ii} regions excited by the massive stars \citep[e.g.,][]{2010A&A...523A...6D, 2012ApJ...760..149P}.
Though the  H$\alpha$ emission is in general spherically distributed,
the south-eastern peripheries of the S301 region are brighter with few dark lanes whereas the north-western part has low-intensity extended emission.
We notice that the most massive star, ALS 207, is at the center of H$\alpha$ emission whereas the other two massive stars are at its boundary.

We have also examined the distribution of colder dust (FIR emission) in the S301 region using AKARI 160 $\mu$m (red), AKARI 60 $\mu$m (green) and H$\alpha$ (blue) images shown in Figure \ref{multi}(b). 
The cold dust emission at FIR (traced by 160 and 60 $\mu$m) clearly surrounds the hot gas and dust in the south and east direction and is slightly opened-up in the north-west direction.
Several YSOs are located at the peak of the distribution of cold dust emission in the north-eastern part of this region.
This distribution almost mimics the arc-like structure seen at 12 $\mu$m, confining the heated region in the south and east directions.
The north-western part of S301 region is found to be devoid of gas and dust.

In Figure \ref{multi}c, we have also shown the NVSS 1.4 GHz radio contours that trace the ionized emission (cyan contours) on the  AKARI 160 $\mu$m image.
Planck 353 MHz emission (magenta contours) which traces cold cloud-clumps is also shown along with the distribution of YSOs and massive stars.
Almost all the YSOs are associated with cold gas and dust traced by the AKARI 160 $\mu$m and Planck 353 MHz emissions whereas the massive star ALS 207 is surrounded by ionized gas.
The peaks of 160 $\mu$m and 353 MHz emission are matching well with many YSOs located around these peaks.
Some of the YSOs are also located at the boundary of the ionized region where the ionized emission seems to be interacting with the cold or warm dust and gas. By looking at the morphology of this region, it appears that the most massive star is at the edge or near the outer surface of the molecular cloud where one can see the embedded YSOs in it.

\begin{figure}
\includegraphics[width=0.45\textwidth]{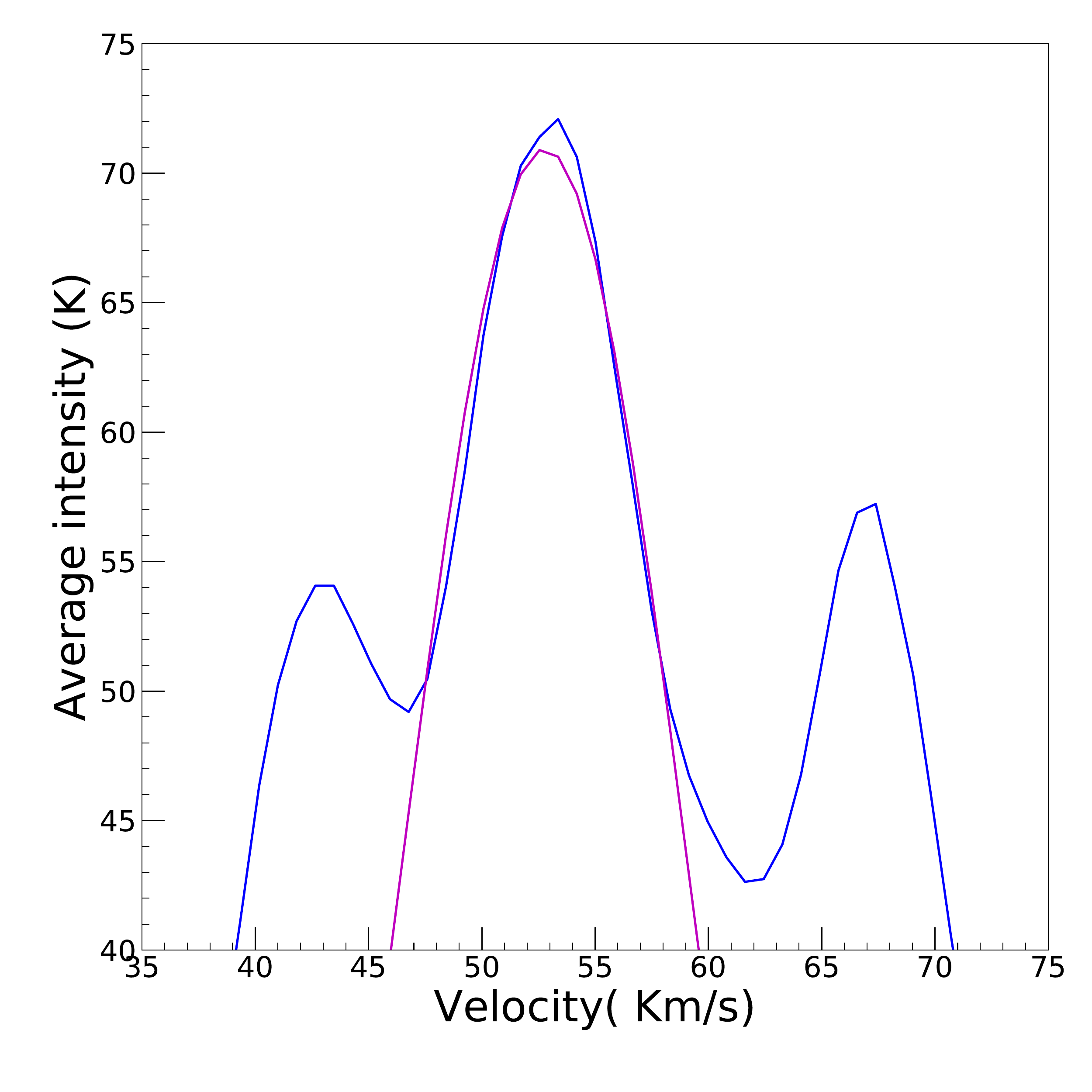}
\caption{\label{gauss} H\,{\sc i} (21cm) spectrum towards the S301 region. A Gaussian (shown with the magenta color) is fitted to determine the peak of the velocity.} 
\end{figure}

In Figure \ref{multi}d,  YSOs along with massive stars are superimposed 
on the $K$-band image. We also show the radio emission at NVSS 1.4 GHz (cyan contours), extinction contours (blue contours) and stellar surface density contours (red contours) in the figure.
The boundary of the newly identified NE-cluster region (see Section 3.2.1) is shown by a yellow circle.
Almost all the YSOs are distributed in the north-eastern direction within the boundary of the cluster region with few exceptions in the southern direction. 
The presence of YSOs inside the cluster region implies that the cluster is 
very young and thus might have formed via a recent star formation activity.
The extinction map shows that most of the molecular material is in front of the cluster at the eastern boundary of the radio emission.
The molecular cloud traced by the extinction map is bow-shaped pointing towards the massive star ALS 207. The peak of radio emission
is near to the tip of the bow-shaped molecular cloud which hints that the radio emission is density bounded in this direction.
The distribution of YSOs starts where the ionized emission interacts with the surrounding gas and dust distribution, which could 
lead us to assume that the formation of YSOs may be due to the interaction of the ionization front with the surrounding material. 
We have also examined the H\,{\sc i} (21cm) emission
in the direction of S301 and found a prominent velocity component ranging from 45 to 60 km$^{-1}$ (see Figure \ref{gauss}). 
We fitted a Gaussian in the spectrum and found the peak of the velocity component as 52.28 km$^{-1}$. 
\citet{1990AJ.....99..622F} and \citet{1988A&A...191..323W} have also reported  a radial velocity of $\sim$ 56.1 kms$^{-1}$ for  both the ionized and the molecular gas in the direction of S301.
\citet{2006ApJS..165..338Q} have done a radio recombination lines survey in this region and reported the line of site radial velocity as $\sim$55.0 kms$^{-1}$ and $\sim$53.7 km$s^{-1}$ 
for H and He lines, respectively. 
These velocities suggest the association of ionised/molecular/neutral gases in the region.
In Figure \ref{multi}d, we have also shown the distribution of H\,{\sc i} gas integrated over a velocity range of 45 to 60 km$^{-1}$.
The distribution of neutral gas represents the low density matter in the south-western part of this region.

\begin{figure*}
\centering
\includegraphics[width=0.44\textwidth]{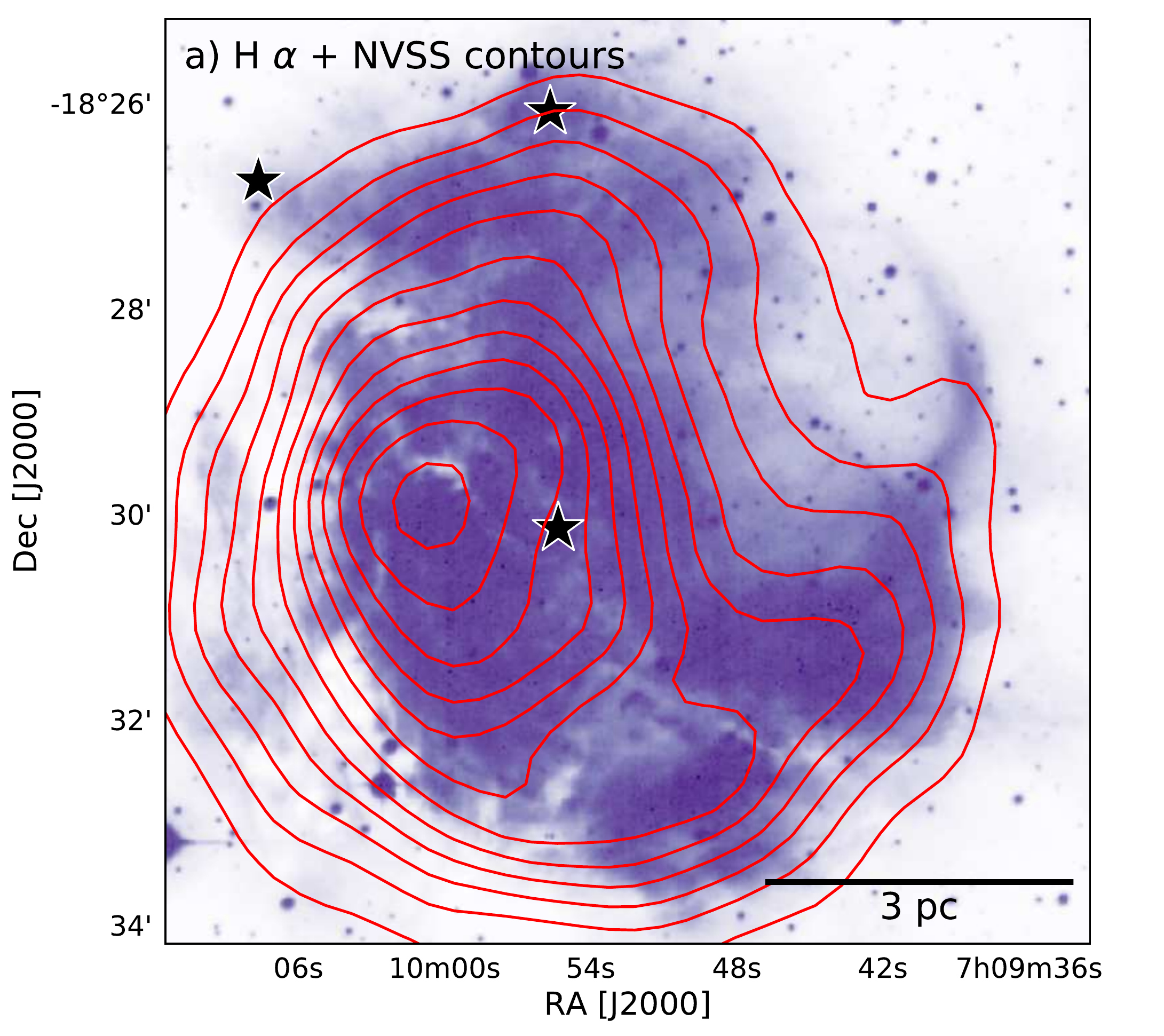}
\includegraphics[width=0.405\textwidth]{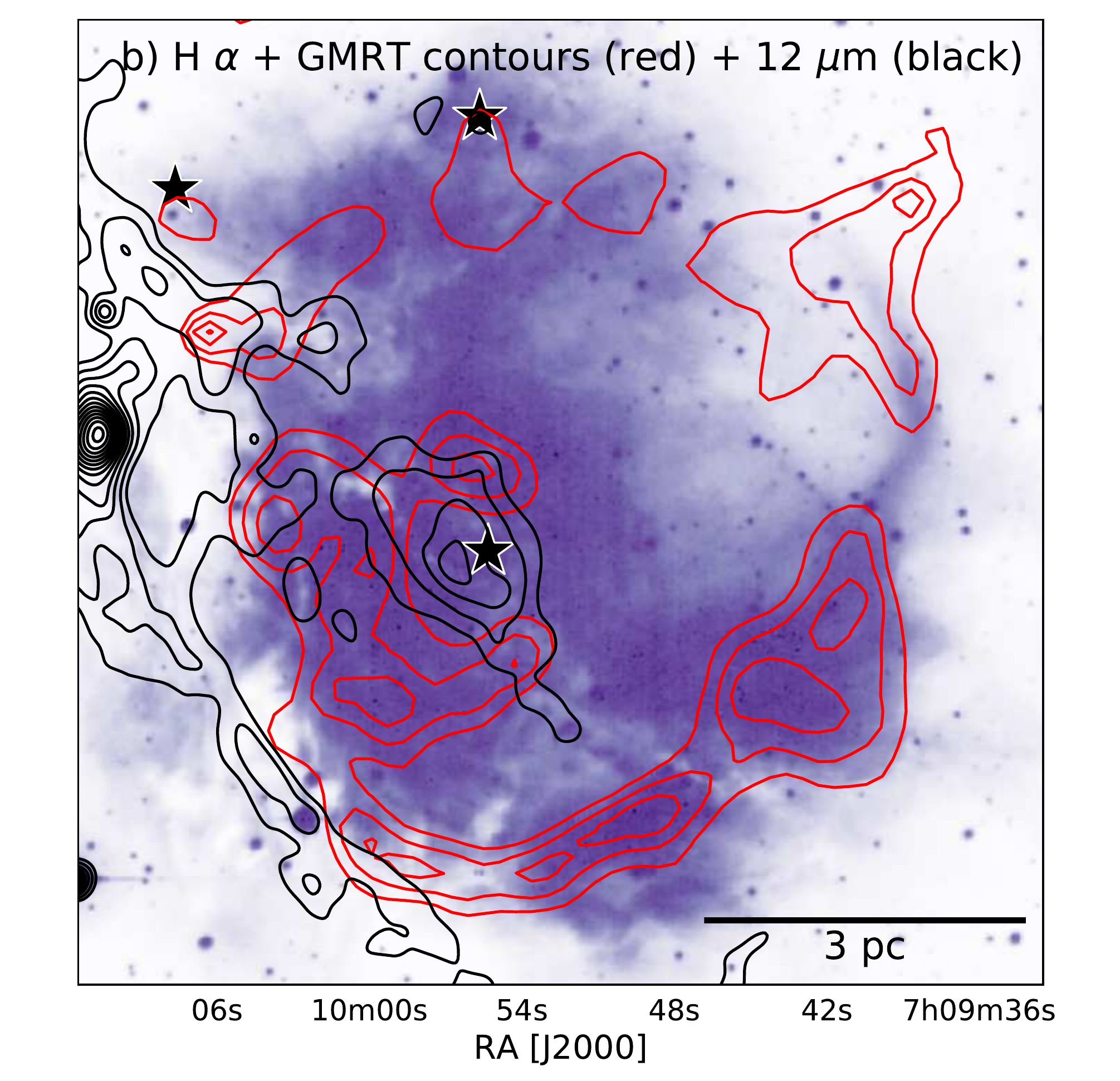}
\caption{\label{ionis} (a) H$\alpha$ image of the S301 region (FOV$\sim 9^\prime\times 9^\prime$) superimposed with NVSS (1.4 GHz) contours.
The lowest NVSS contour is at 0.006 Jy/beam and the step size is 0.006 Jy/beam.
 (b) H$\alpha$ image of the S301 region superimposed with GMRT (1280 MHz) contours.
The lowest GMRT contour is at 0.001 Jy/beam and the step size is 0.001 Jy/beam.
The star symbols are the location of massive stars in the region.
The beam size of the radio map is shown by a black circle in the lower right corners of the figures. The 12 $\mu$m intensity contours are shown with the black color curves with lowest contour at 640 counts and the step size  is 115 counts.  
	}
\end{figure*}

\subsection{Distribution of the ionized gas}

In Figure \ref{ionis}a, we have shown the distribution of ionized gas using the NVSS 1.4 GHz map, 
overplotted on the H$\alpha$ image of the S301 region. The ionized gas is more or less circularly distributed 
with extended emission in the west direction of the region.
The most massive star ALS 208 is located near the peak of the radio emission whereas  the other 
two massive stars are located at the boundary of the radio emission.
The center of the ionized gas is slightly toward the east direction of the region.
The observed distributions of the ionized gas, PDR and cold gas/dust indicate that most of the ionized emission is due to a massive star ALS 207
and the radio emission is density bounded towards the south and east peripheries.
In the other direction, the radio emission is more extended due to the absence of gas and dust material.

Due to the coarse beam size of NVSS ($40^{\prime\prime}\times 40^{\prime\prime}$), small scale structures are absent in  the image and we did not get finer details of the ionized gas distribution. To investigate them, we have used the GMRT 1280 MHz emission map (beam size $\sim 7^{\prime\prime} \times 7^{\prime\prime}$), over-plotted on the H$\alpha$ image of S301 and as shown in Figure \ref{ionis}b. 
The GMRT 1280 MHz emission shows a fragmented dense shell of the ionized gas
which seems to be slightly opened up towards the north-west direction. 
The ionized gas is density bounded by the PDRs in the south-eastern direction as seen by the 12 $\mu$m emission (black contours) in Figure \ref{ionis}b.
The H$\alpha$ emission also shows similar signatures as discussed in the previous section.
It looks like the ionized gas might have streamed out in the lower-density interstellar medium (ISM) 
through the north-west direction which is having very less or no gas and dust distribution.
We have also estimated the value of  Lyman continuum  flux `log(N$_{UV}$)' as 48.41 (see next Section) which corresponds to a O8.5V spectral type star \citep[see][]{1973AJ.....78..929P}. 
Since the main ionizing source `ALS 207' of this H\,{\sc ii} region has a spectral type of O6.5V (see Section 3.1), 
the estimated value of Lyman continuum  flux is far less than  the expected Lyman continuum flux of an O6.5V star (i.e., log(N$_{UV}$)=49.02). 
Even, if we consider the absorption by dust grains prior to the ionization \citep{2001AJ....122.1788I, 2018ApJ...864..136B}, we could not compensate 
for the remaining value of ionization flux. This also supports our argument that the
ionized gas has streamed out from the north-western direction of this complex.
This kind of H\,{\sc ii} region morphology is already seen in the
literature \citep{2020ApJ...905...61P,2009ApJ...703.1352K,2005ApJ...627..813H,1979A&A....71...59T}, 
where the H\,{\sc ii} regions open up in the direction away from the cloud edge and is known as ``blister-type" H\,{\sc ii} region. 
This mechanism of flowing ionized gas in the surrounding ISM is known as Champagne-flow \citep{2012A&A...537A.149D, 2020ApJ...905...61P}.

\subsection{Lyman continuum emission and Dynamical age of the S301 H\,{\sc ii} region}

We have estimated the  Lyman continuum flux associated with the ionized gas in the S301 region by using the following equation by \citet{2016A&A...588A.143S}:

\begin{equation}
\begin{split}
N_\mathrm{UV} (s^{-1})& = 7.5\, \times\, 10^{46}\, \left(\frac{F_\mathrm{th}(\nu)}{\mathrm{Jy}}\right)\left(\frac{D}{\mathrm{kpc}}\right)^{2} \\
&\left(\frac{T_{e}}{10^{4}\mathrm{K}}\right)^{-0.45} \times\,\left(\frac{\nu}{\mathrm{GHz}}\right)^{0.1}
\end{split}
\end{equation} 

where, N$_{UV}$ is the Lyman continuum photons per second, T$_e$ is the electron temperature, $\nu$ is the frequency, S$_\nu$ is the integrated flux, D is the distance of the region, and $\theta$ denotes the angular size of the region.
We have adopted the values of $T_{e}$ as 10\,000~K assuming all ionizing flux was generated by a single massive O type star. 
We have calculated the integrated flux density S$\nu$=2.5 mJy by integrating flux down to the lowest contour at 3 $\sigma$ level in the GMRT 1280 MHz map.
The $\sigma$ value which is the rms noise of the GMRT radio map was estimated to be 0.093 mJy/beam. The size of the H\,{\sc ii} region is found to be $\simeq$ 4$^\prime$.7 (4.8 pc at a distance of 3.54 kpc). 
Using the distance `D' of 3.54 kpc, we have estimated the log(N$_{UV}$) as 48.41. 

The dynamical age of the S301 H {\sc ii} region is estimated by using the following equation \citep{1980pim..book.....D}:

\begin{equation}
t_{dyn} = \Big(\frac{4R_s}{7c_s}\Big) [\Big(\frac{R_{HII}}{R_s}\Big)^{7/4}-1]
\end{equation}
 
where, c$_s$ is the isothermal sound velocity in the ionized gas (c$_s$ = 11 km s$^{-1}$) \citep{2005fost.book.....S}, 
R$_{HII}$ is the radius of the H\,{\sc ii} region, and R$_s$ is the  Str\"{o}mgren radius of the H\,{\sc ii} region which is  given by:

\begin{equation}
R_s = \Big(\frac{3S_{\nu}}{4\pi{{n_0}^2}{\beta_2}}\Big)^{1/3}
\end{equation}

where, n$_0$ is the initial ambient density (in cm$^{-3}$) and $\beta$$_2$ 
is the total recombination coefficient to the first excited state of hydrogen $\beta$$_2$ = $2.6\times10^{-13}$ \citep{2005fost.book.....S}.
The dynamical age of the S301 H\,{\sc ii} region is estimated as 4.8 Myr corresponding to n$_0$= 10$^4$ cm$^{-3}$  \citep{2006A&A...447..221B}.

\subsection{Feedback pressure from massive star}
\label{sec:feedb}

To quantitatively examine the effect of a massive star on its surrounding, 
we have determined the feedback pressure from the massive star ALS 207. 
Total feedback pressure  from a massive star consists of three components, pressure of a H\,{\sc ii} region (P$_{H\,{\sc II}}$), radiation pressure (P$_{rad}$), 
and stellar wind ram pressure (P$_{wind}$) \citep{2012ApJ...758L..28B,2017ApJ...851..140D}.
These pressure components can be estimated by using the following equations \citep[see for details,][]{2012ApJ...758L..28B}:

\begin{equation}
P_{HII} = \mu m_{H} c_{s}^2\, \left(\sqrt{3N_{UV}\over 4\pi\,\alpha_{B}\, D_{s}^3}\right);   
\end{equation}

\begin{equation}
P_{rad} = L_{bol}/ 4\pi c D_{s}^2; 
\end{equation}

\begin{equation}
P_{wind} = \dot{M}_{w} V_{w} / 4 \pi D_{s}^2; 
\end{equation}

Where N$_{UV}$ is the Lyman continuum photons, c$_{s}$ is the sound speed in the photoionized region \citep[=11 km s$^{-1}$;][]{2005fost.book.....S}, ``$\alpha_{B}$'' is the radiative recombination  coefficient \citep[=  2.6 $\times$ 10$^{-13}$ $\times$ (10$^{4}$ K/T$_{e}$)$^{0.7}$ cm$^{3}$ s$^{-1}$; see][]{1997ApJ...489..284K}, $\mu$ is the mean molecular weight in 
the ionized gas \citep[= 0.678;][]{2009A&A...497..649B}, m$_{H}$ is the hydrogen atom mass, $\dot{M}_{w}$ is the mass-loss rate, V$_{w}$ is the wind velocity of the ionizing source, and L$_{bol}$ is the bolometric luminosity of the ionizing source.
We adopted $L_{bol}$ = 234422 L$_{\odot}$ \citep[][]{1973AJ.....78..929P},
$\dot{M}_{w}$ $\approx$ 1.21 $\times$ 10$^{-7}$ M$_{\odot}$ yr$^{-1}$ \citep[][]{2009A&A...498..837M}, 
V$_{w}$ $\approx$ 5244 km s$^{-1}$ \citep[][]{2017A&A...598A..56M}, N$_{UV}$ = 1.047 $\times$ 10$^{49}$   \citep[][]{1973AJ.....78..929P} for ALS 207 (O6.5V). 
We took $D_{s}$ =4.24 pc as the projected distance of the cluster center from the massive star.
We got the value of $P_{HII}$= 8.98 $\times$ 10$^{-11}$,  $P_{rad}$ = 1.39 $\times$ 10$^{-11}$ and $P_{wind}$ = 1.86 $\times$ 10$^{-12}$ dynes\, cm$^{-2}$. 
The total pressure ($P$= $P_{HII}$+$P_{rad}$+ $P_{wind}$) comes out to be 1.06 $\times$ 10$^{-10}$ dynes\, cm$^{-2}$. 

\section{Discussion}

Broadly, the S301 cloud complex possesses gas and dust, massive stars, an H\,{\sc ii} region, a young embedded NE-cluster, 
and  several YSOs. All of these features hint towards 
a recent star forming activities in the region. 
As there are many active star-forming regions in the literature where massive stars have triggered the formation 
of YSOs \citep[][and references therein]{2006A&A...446..171Z, 2009A&A...494..987P, 2016MNRAS.461.2502Y,2017MNRAS.467.2943S,2020ApJ...891...81P,2020ApJ...896...29K}, 
in the present work we also explored the impact of massive stars on the star formation activities in the S301 region.      

\subsection{Feedback from massive star}

The morphological features discussed in Section 3.5 and 3.6 point S301 towards a blistered H\,{\sc ii} region which is created when a massive star 
forms at the edge of the molecular cloud. The ionization front soon reaches at the edge of the cloud and H\,{\sc ii} region opens away 
from the cloud edge devoid of gas and dust. In the other directions, where the molecular material is distributed, 
feedback from massive stars can trigger the formation of a new generation of stars.
Observational evidences of the triggered star formation have been discussed for 
several H\,{\sc ii} regions such as Sh 2-104, RCW 79, Sh 2-212, RCW
120, Sh 2-217 \citep{2003A&A...408L..25D, 2008A&A...482..585D,2006A&A...446..171Z,2010A&A...518L..81Z,2011A&A...527A..62B}. 
Recently, \citet{2020ApJ...905...61P} studied a blistered H\,{\sc ii} region Sh 2-112 and discussed the 
feedback effect of a massive star in triggering star formation. We have also explored the possibility of triggered star formation in the S301 region.    
Figure \ref{multi} and  \ref{ionis} shows the morphology of the S301 star forming complex. The most massive star ALS 207 located in the region is responsible for creating the H\,{\sc ii} region (traced by radio and H$\alpha$ emission) and  arc-like MIR shell/PDRs towards the south and eastern directions of ALS 207. 
The H\,{\sc i} (21cm) map confirms the presence of low density neutral hydrogen material at the western border that is devoid of gas and dust.
 The extinction map  shows the presence of molecular clouds in the north-east direction between the NE-cluster  
and the ionization layer of gas. We can also see the presence of YSOs still embedded in the molecular cloud. 
These facts point to a star formation process triggered by the massive central star ALS 207 in the north-east direction of the S301 region.
Earlier in Section 3.8, we have found that the pressure due to the H\,{\sc ii} region  `$P_{HII}$' is relatively higher than the other components, i.e,  $P_{rad}$ and $P_{wind}$.
Hence, we can argue that the photoionized gas associated with the H\,{\sc ii} region (against the radiation and wind) can be a major 
contributor in the feedback process in the S301 region.  The total pressure due to a massive star ALS 207  at the center of the NE-cluster 
was found to be higher than that of a typical cool molecular cloud ($P_{MC}$$\sim$10$^{-11}$--10$^{-12}$ dynes cm$^{-2}$ for a 
temperature $\sim$20 K and particle density $\sim$10$^{3}$--10$^{4}$ cm$^{-3}$) \citep[see Table 7.3 of][]{1980pim..book.....D}
and hence can initiate the collapse of the molecular cloud.

Many authors \citep{2020ApJ...891...81P, 2020ApJ...896...29K} have given age gradient as an argument to support the triggered star formation scenario. 
To check this, we have done a comparison between the age 
of the YSOs, age of the NE-cluster, dynamical age of the H\,{\sc ii} region and age of the massive star. 
The peaks in the distributions of the ages of the YSOs and NE-cluster members were found to be at $\sim$2.5 - 1.5 Myr (see Section 3.2.3 and 3.3).
The MF slopes for the cluster region was also found to be shallower ($\Gamma=-0.85\pm0.07$)
than the \citet{1955ApJ...121..161S} value i.e., $\Gamma=-1.35$.
Usually, the higher mass stars mostly follow the Salpeter MF \citep{1955ApJ...121..161S}. At lower
masses, the MF is less well constrained, but appears to flatten
below 1 M$_\odot$  and exhibits fewer stars of the lowest masses
\citep{2002Sci...295...82K, 2003PASP..115..763C, 2015arXiv151101118L, 2016ApJ...827...52L}.
The MF distribution of the NE-cluster indicates that there is a larger population of relatively massive stars suggesting the youth of this cluster.
The dynamical age of the H\,{\sc ii} region S301 is found to be  $\sim$4.8 Myr (see Section 3.7). 

Since ALS 207 has the spectral type of O6.5 and is still in the MS, hence the age of the ALS 207 
should be $\sim$5 Myr \citep{1994A&AS..103...97M} (see also Section 3.2.3). The age of ALS 207 is more or less similar to the  dynamical age of the H\,{\sc ii} region S301.
Thus, H\,{\sc ii} region S301 is old enough to initiate the formation of a second generation of stars in its surroundings, which in fact are younger in age. 
We have also calculated the time frame in which the ionization front from the central massive star (ALS 207) with a speed of 9 kms$^{-1}$ \citep{1976RMxAA...1..373P} will reach the NE-cluster (4.24 pc), as 1.4 Myr. Hence, we can safely assume that the stars/YSOs in the NE-cluster formed after the ionization front from the massive star
reached there.   

The distribution of MIR swept up shell ahead of the layer of ionized gas along with the NE-cluster being on the opposite side embedded in the molecular gas (Figure \ref{multi}(b))
hint towards  a `collect-and-collapse' scenario of star formation in this region \citep{2020ApJ...898..172D, 2009ARep...53..611K, 1977ApJ...214..725E}.
This scenario predicts the shell of MIR emission around the H\,{\sc ii} region and existence of young stellar clusters behind that.

\begin{figure*}
\centering
\includegraphics[width=0.49\textwidth]{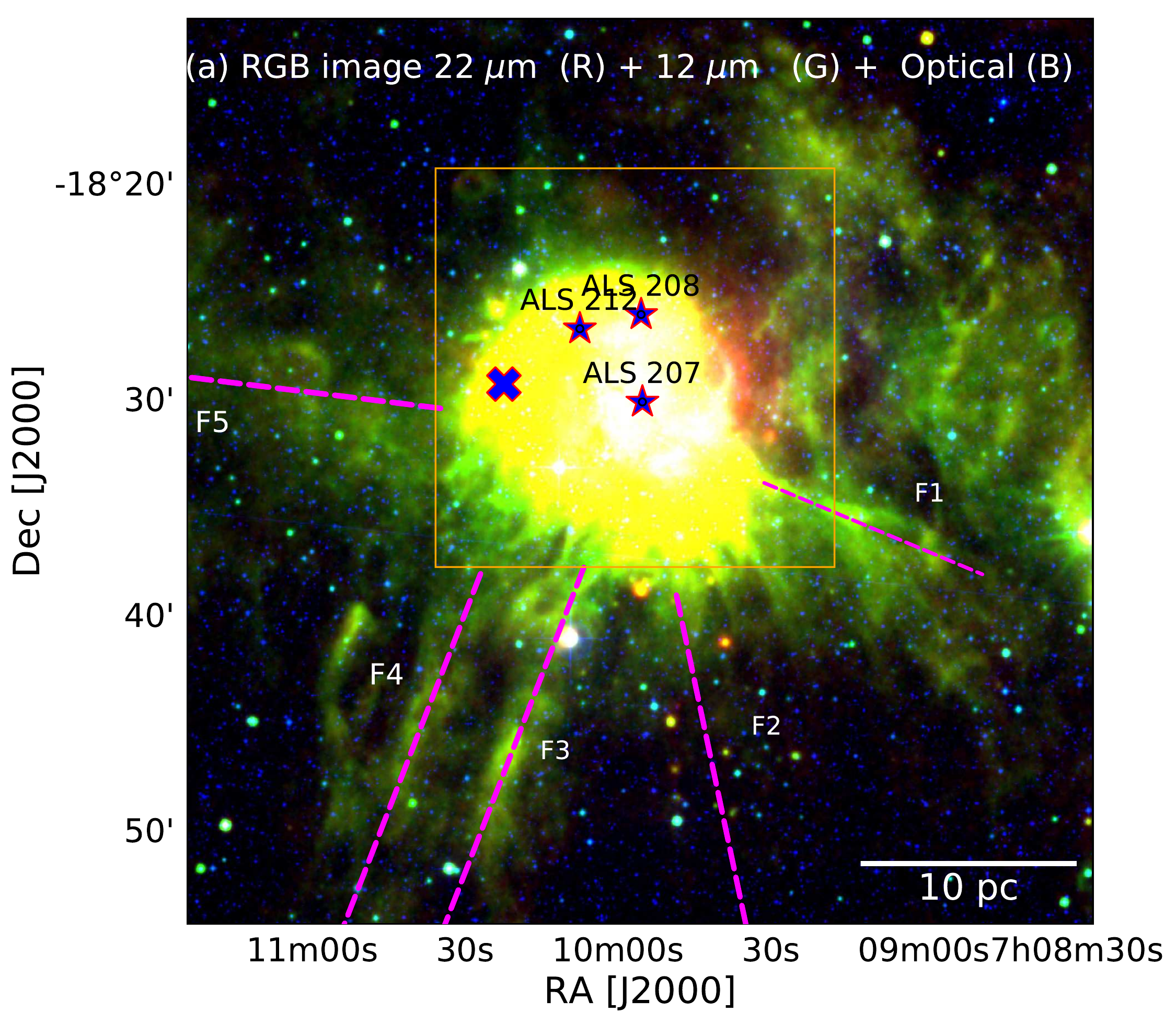}
\includegraphics[width=0.49\textwidth]{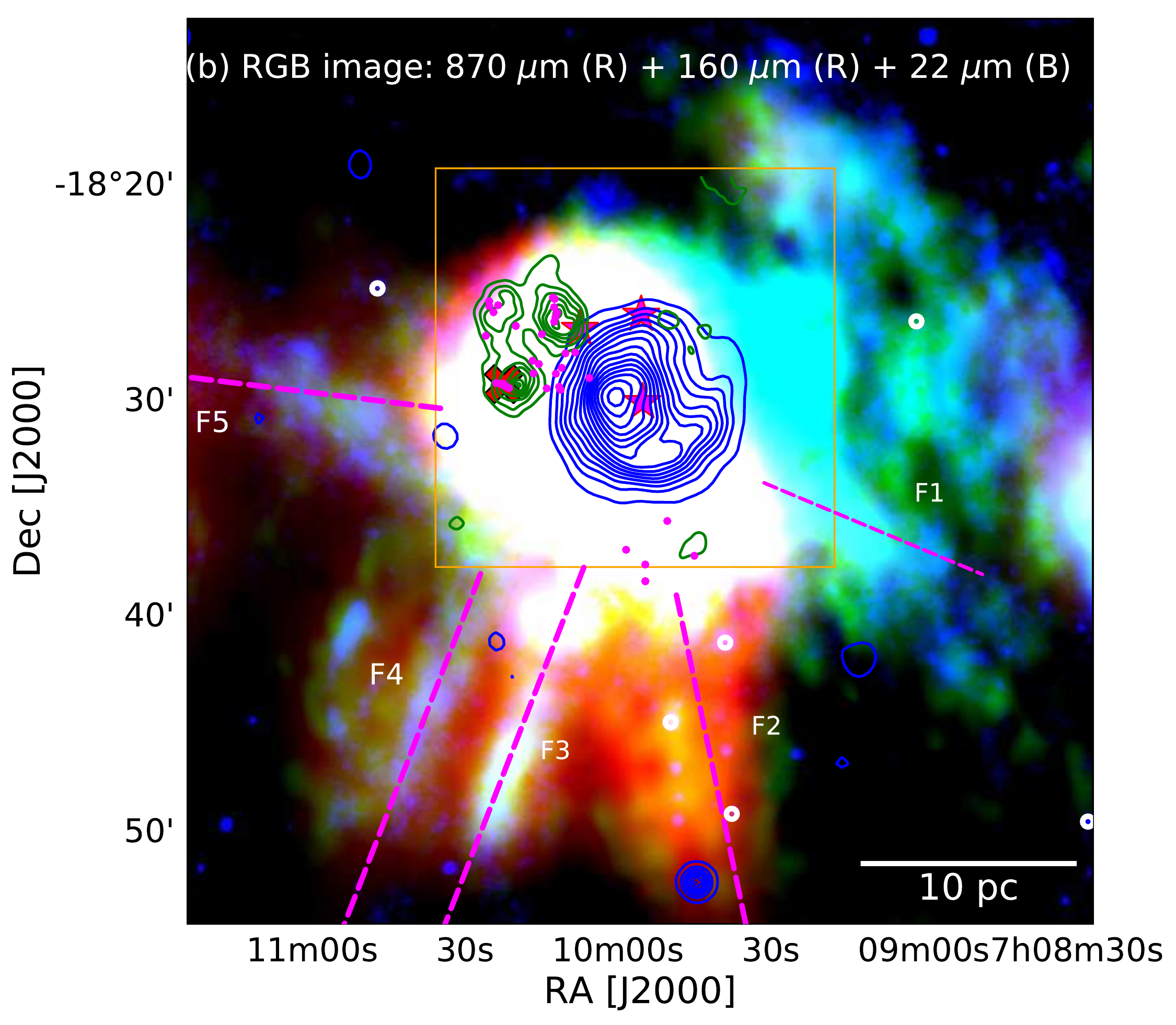}
\caption{\label{hub} (a) Large-scale view ($0.7^{\circ}\times 0.7^{\circ}$) of the region S301 shown with the color-composite image made
by  using the WISE 22 $\mu$m (red), WISE 12 $\mu$m (green) and optical V band (blue) images. (b) Color-composite image of the S301 region  made by 
using the Planck 353 MHz (red), AKARI (160 $\mu$m, green) and WISE 22 $\mu$m (blue) images.
The NVSS 1.4 GHz contours and stellar surface density contours are also over plotted over the image in blue and green colors, respectively. 
Locations of the massive stars and the YSOs are also shown by stars symbol and filled circles. The region inside the box is the probable central hub part of the system. Possible filamentary structures are shown by the magenta dashed lines. The cross symbol in both the panels is showing reflection nebula Bran 7D (see section 1)}
\end{figure*}

\subsection{Formation of massive stars: Possibility of a Hub-Filament System}

MIR/FIR surveys like \textit{Herschel}, AKARI etc. had revealed that the cold ISM, known as the birthplace of stars, 
is organized in filamentary structures \citep{2015A&A...580A..49I,2010A&A...518L.103M}.
There have been many studies in which hub-filament system (HFS) is attributed as the possible 
progenitors of high mass star formation \citep{2012A&A...540L..11S, 2017ApJ...851..140D}.
Recently, \citet{2020A&A...642A..87K} have given a detailed explanation regarding the various stages of massive star formation in the HFS. 
In the HFS, gas flows to the hub via filaments which are the objects of high aspect ratio and low density. 
Hub is a place where filaments converge and are of low aspect ratio and high density. 
Initially, flow-driven filaments approach each other due to intra-molecular cloud velocity dispersion, which join to form a hub system. 
Afterwards, a dense core can form at the junction followed by the formation of massive star/s.
These massive stars can further create H\,{\sc ii} region/s which can trigger formation of the second generation of  stars. 
In literature, there are multiple examples of H\,{\sc ii} regions 
found in the HFSs \citep{2009ApJ...700.1609M,2015A&A...582A...1D,2017ApJ...844...15D,2020ApJ...903...13D}.
 
To explore the possible formation scenario of the massive stars in the S301 region, in Figure \ref{hub}, 
we show the large scale  ($0.7^{\circ} \times 0.7^{\circ}$) multi-wavelength  view of the S301 region.
In Figure \ref{hub}a, we show the color-composite image made by using  the 
WISE 22 $\mu$m (red), WISE 12 $\mu$m (green) and optical $V$ band (blue) images. 
The Figure \ref{hub}b shows the color-composite image made by using the Planck 325 MHz (red), 
AKARI 160 $\mu$m (green), and WISE 22 $\mu$m images. It is also superimposed with the NVSS radio and stellar surface density contours.  
In both the figures, there is a hint of a central hub of gas and dust along with small filamentary structures (shown with magenta lines).
The location of the main ionizing star ALS 207 at the center of this hub suggests that it may be formed via the HFS related physical processes. To further confirm the hypothesis of HFS at S301 getting fed up by the flow driven filaments (traced in MIR/FIR wavelengths), 
we need a high resolution molecular data. The P-V (position-velocity) diagrams along the filaments showing velocity gradient would confirm this hypothesis. 
Unfortunately, there is no high resolution data available currently. This will be explored in further studies.

\section{Conclusion}

We have presented a multiwavelength study of the H\,{\sc ii} region S301 using the deep optical/NIR data, 
radio continuum data, spectroscopic data along with MIR/FIR archival data taken from different telescopes.
From the analyses performed in the present work, we made the following conclusions.

\begin{itemize}

\item
Spectral classifications of the massive stars ALS 207 (O6.5V), ALS 208 (B1III) and ALS 212 (B1V) have been done using the optical spectroscopic data taken from the HCT. 

\item
We have identified  a new NE-cluster in the north-east direction of the S301 region by doing stellar density distribution analysis. 
Gaia DR3 data have been used to identify 194 members  of this cluster. We constrained the distance and reddening of this NE-cluster as 3.54 kpc and $E(B-V)$=0.5 mag, respectively.
The distance of the NE-cluster is similar to that of the H\,{\sc ii} region S301.
Most of the members of the NE-cluster have age and mass around 1.5 Myr and 1.3 M$_\odot$.
The slope of the MF `$\Gamma$' in the mass range $0.4 <$M/M$_\odot < 7$ for the member stars of the NE-cluster is found to be $-0.85\pm0.07$, which is shallower than
the \citet{1955ApJ...121..161S} value -1.35.

\item
We have identified 37 YSOs in the $\sim18^\prime.5\times18^\prime.5$ FOV around the massive star ALS 207. 
Majority of YSOs were found to be spatially associated  with the NE-cluster.
Out of them, four are Class\,{\sc i} and the rest are Class\,{\sc ii} sources. 
The mean value of age and mass of the YSOs is found to be 2.5$\pm$1.6 Myr and 2.3$\pm$0.8 M$_\odot$, respectively.

\item
The morphology of the S301 H\,{\sc ii} region has been analyzed using the H$\alpha$, MIR to FIR and radio continuum images. The S301 region is showing distribution of hot ionised gas near the massive star ALS 207 which is bounded by an arc-like structure of gas and dust
from the south-eastern direction. 
We have found the dense molecular material between the massive star  ALS 207 and the NE-cluster using the extinction map.
The north-western region seems to be devoid of gas and dust material.
The distributions of MIR dust, ionized gas and neutral hydrogen suggest a blister morphology
for the S301 with the massive star ALS 207 being near the edge of the cloud and the NE-cluster is embedded in the cold molecular cloud.

\item
The pressure calculation, distribution of PDRs/YSOs and the age difference between the massive star and the NE cluster provide supporting
arguments for the positive feedback of a massive star ALS 207 in the S301 H\,{\sc ii} region.
Our analyses also suggest that the massive stars in the S301 region might have formed in a HFS.


\end{itemize}

\section*{Acknowledgments}
We thank the anonymous referee for several constructive, useful comments and suggestions, which greatly enhanced the quality of the paper. 
The observations reported in this paper were obtained by using the 1.3m telescope, Nainital, India and  the 2 m HCT at IAO, Hanle, the High Altitude Station of Indian Institute of Astrophysics, Bangalore, India. We also acknowledge the TIFR Near Infrared Spectrometer and Imager mounted on 2 m HCT  using which we have made NIR observations. This publication makes use of data from the Two Micron All Sky Survey, which is a joint project of the University of Massachusetts and the Infrared Processing and Analysis Center/California Institute of Technology, funded by the National Aeronautics and Space Administration and the National Science Foundation. This work is based on observations made with the $Spitzer$ Space Telescope, which is operated by the Jet Propulsion Laboratory, California Institute of Technology under a contract with National Aeronautics and Space Administration. This publication makes use of data products from the Wide-field Infrared Survey Explorer, which is a joint project of the University of California, Los Angeles, and the Jet Propulsion Laboratory/California Institute of Technology, funded by the National Aeronautics and Space Administration. DKO acknowledges the support of the Department of Atomic Energy, Government of India, under project identification No. RTI 4002.

\bibliography{301-finalapj-referee-asph.bib}{}

\begin{thebibliography}{}
\expandafter\ifx\csname natexlab\endcsname\relax\def\natexlab#1{#1}\fi
\providecommand{\url}[1]{\href{#1}{#1}}
\providecommand{\dodoi}[1]{doi:~\href{http://doi.org/#1}{\nolinkurl{#1}}}
\providecommand{\doeprint}[1]{\href{http://ascl.net/#1}{\nolinkurl{http://ascl.net/#1}}}
\providecommand{\doarXiv}[1]{\href{https://arxiv.org/abs/#1}{\nolinkurl{https://arxiv.org/abs/#1}}}

\bibitem[{{Allen} {et~al.}(2007){Allen}, {Megeath}, {Gutermuth}, {Myers},
  {Wolk}, {Adams}, {Muzerolle}, {Young}, \& {Pipher}}]{2007prpl.conf..361A}
{Allen}, L., {Megeath}, S.~T., {Gutermuth}, R., {et~al.} 2007, Protostars and
  Planets V, 361

\bibitem[{{Anderson} {et~al.}(2018){Anderson}, {Hogg}, {Leistedt},
  {Price-Whelan}, \& {Bovy}}]{2018AJ....156..145A}
{Anderson}, L., {Hogg}, D.~W., {Leistedt}, B., {Price-Whelan}, A.~M., \&
  {Bovy}, J. 2018, \aj, 156, 145, \dodoi{10.3847/1538-3881/aad7bf}

\bibitem[{{Arenou} {et~al.}(2017){Arenou}, {Luri}, {Babusiaux}, {Fabricius},
  {Helmi}, {Robin}, {Vallenari}, {Blanco-Cuaresma}, {Cantat-Gaudin},
  {Findeisen}, {Reyl{\'e}}, {Ruiz-Dern}, {Sordo}, {Turon}, {Walton}, {Shih},
  {Antiche}, {Barache}, {Barros}, {Breddels}, {Carrasco}, {Costigan},
  {Diakit{\'e}}, {Eyer}, {Figueras}, {Galluccio}, {Heu}, {Jordi},
  {Krone-Martins}, {Lallement}, {Lambert}, {Leclerc}, {Marrese}, {Moitinho},
  {Mor}, {Romero-G{\'o}mez}, {Sartoretti}, {Soria}, {Soubiran}, {Souchay},
  {Veljanoski}, {Ziaeepour}, {Giuffrida}, {Pancino}, \&
  {Bragaglia}}]{2017A&A...599A..50A}
{Arenou}, F., {Luri}, X., {Babusiaux}, C., {et~al.} 2017, \aap, 599, A50,
  \dodoi{10.1051/0004-6361/201629895}

\bibitem[{{Avedisova} \& {Palous}(1989)}]{1989BAICz..40...42A}
{Avedisova}, V.~S., \& {Palous}, J. 1989, Bulletin of the Astronomical
  Institutes of Czechoslovakia, 40, 42

\bibitem[{{Bailer-Jones} {et~al.}(2021){Bailer-Jones}, {Rybizki}, {Fouesneau},
  {Demleitner}, \& {Andrae}}]{2021AJ....161..147B}
{Bailer-Jones}, C.~A.~L., {Rybizki}, J., {Fouesneau}, M., {Demleitner}, M., \&
  {Andrae}, R. 2021, \aj, 161, 147, \dodoi{10.3847/1538-3881/abd806}

\bibitem[{{Bailer-Jones} {et~al.}(2018){Bailer-Jones}, {Rybizki}, {Fouesneau},
  {Mantelet}, \& {Andrae}}]{2018AJ....156...58B}
{Bailer-Jones}, C.~A.~L., {Rybizki}, J., {Fouesneau}, M., {Mantelet}, G., \&
  {Andrae}, R. 2018, \aj, 156, 58, \dodoi{10.3847/1538-3881/aacb21}

\bibitem[{{Balaguer-N{\'u}nez} {et~al.}(1998){Balaguer-N{\'u}nez}, {Tian}, \&
  {Zhao}}]{1998A&AS..133..387B}
{Balaguer-N{\'u}nez}, L., {Tian}, K.~P., \& {Zhao}, J.~L. 1998, \aaps, 133,
  387, \dodoi{10.1051/aas:1998324}

\bibitem[{{Beltr{\'a}n} {et~al.}(2006){Beltr{\'a}n}, {Brand}, {Cesaroni},
  {Fontani}, {Pezzuto}, {Testi}, \& {Molinari}}]{2006A&A...447..221B}
{Beltr{\'a}n}, M.~T., {Brand}, J., {Cesaroni}, R., {et~al.} 2006, \aap, 447,
  221, \dodoi{10.1051/0004-6361:20053999}

\bibitem[{{Bessell} \& {Brett}(1988)}]{1988PASP..100.1134B}
{Bessell}, M.~S., \& {Brett}, J.~M. 1988, \pasp, 100, 1134,
  \dodoi{10.1086/132281}

\bibitem[{{Binder} \& {Povich}(2018)}]{2018ApJ...864..136B}
{Binder}, B.~A., \& {Povich}, M.~S. 2018, \apj, 864, 136,
  \dodoi{10.3847/1538-4357/aad7b2}

\bibitem[{{Bisbas} {et~al.}(2009){Bisbas}, {W{\"u}nsch}, {Whitworth}, \&
  {Hubber}}]{2009A&A...497..649B}
{Bisbas}, T.~G., {W{\"u}nsch}, R., {Whitworth}, A.~P., \& {Hubber}, D.~A. 2009,
  \aap, 497, 649, \dodoi{10.1051/0004-6361/200811522}

\bibitem[{{Brand} {et~al.}(2011){Brand}, {Massi}, {Zavagno}, {Deharveng}, \&
  {Lefloch}}]{2011A&A...527A..62B}
{Brand}, J., {Massi}, F., {Zavagno}, A., {Deharveng}, L., \& {Lefloch}, B.
  2011, \aap, 527, A62, \dodoi{10.1051/0004-6361/201015389}

\bibitem[{{Bressert} {et~al.}(2012){Bressert}, {Ginsburg}, {Bally},
  {Battersby}, {Longmore}, \& {Testi}}]{2012ApJ...758L..28B}
{Bressert}, E., {Ginsburg}, A., {Bally}, J., {et~al.} 2012, \apjl, 758, L28,
  \dodoi{10.1088/2041-8205/758/2/L28}

\bibitem[{{Cantat-Gaudin} {et~al.}(2018){Cantat-Gaudin}, {Jordi}, {Vallenari},
  {Bragaglia}, {Balaguer-N{\'u}{\~n}ez}, {Soubiran}, {Bossini}, {Moitinho},
  {Castro-Ginard}, {Krone-Martins}, {Casamiquela}, {Sordo}, \&
  {Carrera}}]{2018A&A...618A..93C}
{Cantat-Gaudin}, T., {Jordi}, C., {Vallenari}, A., {et~al.} 2018, \aap, 618,
  A93, \dodoi{10.1051/0004-6361/201833476}

\bibitem[{{Chabrier}(2003)}]{2003PASP..115..763C}
{Chabrier}, G. 2003, \pasp, 115, 763, \dodoi{10.1086/376392}

\bibitem[{{Chauhan} {et~al.}(2009){Chauhan}, {Pandey}, {Ogura}, {Ojha},
  {Bhatt}, {Ghosh}, \& {Rawat}}]{2009MNRAS.396..964C}
{Chauhan}, N., {Pandey}, A.~K., {Ogura}, K., {et~al.} 2009, \mnras, 396, 964,
  \dodoi{10.1111/j.1365-2966.2009.14756.x}

\bibitem[{{Cutri} {et~al.}(2003){Cutri}, {Skrutskie}, {van Dyk}, {Beichman},
  {Carpenter}, {Chester}, {Cambresy}, {Evans}, {Fowler}, {Gizis}, {Howard},
  {Huchra}, {Jarrett}, {Kopan}, {Kirkpatrick}, {Light}, {Marsh}, {McCallon},
  {Schneider}, {Stiening}, {Sykes}, {Weinberg}, {Wheaton}, {Wheelock}, \&
  {Zacarias}}]{2003yCat.2246....0C}
{Cutri}, R.~M., {Skrutskie}, M.~F., {van Dyk}, S., {et~al.} 2003, VizieR Online
  Data Catalog, 2246, 0

\bibitem[{{Dale} {et~al.}(2015){Dale}, {Haworth}, \&
  {Bressert}}]{2015MNRAS.450.1199D}
{Dale}, J.~E., {Haworth}, T.~J., \& {Bressert}, E. 2015, \mnras, 450, 1199,
  \dodoi{10.1093/mnras/stv396}

\bibitem[{{Deharveng} {et~al.}(2008){Deharveng}, {Lefloch}, {Kurtz}, {Nadeau},
  {Pomar{\`e}s}, {Caplan}, \& {Zavagno}}]{2008A&A...482..585D}
{Deharveng}, L., {Lefloch}, B., {Kurtz}, S., {et~al.} 2008, \aap, 482, 585,
  \dodoi{10.1051/0004-6361:20079233}

\bibitem[{{Deharveng} {et~al.}(2003){Deharveng}, {Lefloch}, {Zavagno},
  {Caplan}, {Whitworth}, {Nadeau}, \& {Mart{\'\i}n}}]{2003A&A...408L..25D}
{Deharveng}, L., {Lefloch}, B., {Zavagno}, A., {et~al.} 2003, \aap, 408, L25,
  \dodoi{10.1051/0004-6361:20031157}

\bibitem[{{Deharveng} {et~al.}(2010){Deharveng}, {Schuller}, {Anderson},
  {Zavagno}, {Wyrowski}, {Menten}, {Bronfman}, {Testi}, {Walmsley}, \&
  {Wienen}}]{2010A&A...523A...6D}
{Deharveng}, L., {Schuller}, F., {Anderson}, L.~D., {et~al.} 2010, \aap, 523,
  A6, \dodoi{10.1051/0004-6361/201014422}

\bibitem[{{Deharveng} {et~al.}(2015){Deharveng}, {Zavagno}, {Samal},
  {Anderson}, {LeLeu}, {Brevot}, {Duarte-Cabral}, {Molinari}, {Pestalozzi},
  {Foster}, {Rathborne}, \& {Jackson}}]{2015A&A...582A...1D}
{Deharveng}, L., {Zavagno}, A., {Samal}, M.~R., {et~al.} 2015, \aap, 582, A1,
  \dodoi{10.1051/0004-6361/201423835}

\bibitem[{{Dewangan} {et~al.}(2017{\natexlab{a}}){Dewangan}, {Ojha}, \&
  {Baug}}]{2017ApJ...844...15D}
{Dewangan}, L.~K., {Ojha}, D.~K., \& {Baug}, T. 2017{\natexlab{a}}, \apj, 844,
  15, \dodoi{10.3847/1538-4357/aa79a5}

\bibitem[{{Dewangan} {et~al.}(2020{\natexlab{a}}){Dewangan}, {Ojha}, {Sharma},
  {Palacio}, {Bhadari}, \& {Das}}]{2020ApJ...903...13D}
{Dewangan}, L.~K., {Ojha}, D.~K., {Sharma}, S., {et~al.} 2020{\natexlab{a}},
  \apj, 903, 13, \dodoi{10.3847/1538-4357/abb827}

\bibitem[{{Dewangan} {et~al.}(2017{\natexlab{b}}){Dewangan}, {Ojha}, \&
  {Zinchenko}}]{2017ApJ...851..140D}
{Dewangan}, L.~K., {Ojha}, D.~K., \& {Zinchenko}, I. 2017{\natexlab{b}}, \apj,
  851, 140, \dodoi{10.3847/1538-4357/aa9be2}

\bibitem[{{Dewangan} {et~al.}(2020{\natexlab{b}}){Dewangan}, {Sharma},
  {Pandey}, {Palacio}, {Ojha}, {Benaglia}, {Baug}, \&
  {Das}}]{2020ApJ...898..172D}
{Dewangan}, L.~K., {Sharma}, S., {Pandey}, R., {et~al.} 2020{\natexlab{b}},
  \apj, 898, 172, \dodoi{10.3847/1538-4357/ab9c27}

\bibitem[{{Dias} {et~al.}(2021){Dias}, {Monteiro}, {Moitinho}, {L{\'e}pine},
  {Carraro}, {Paunzen}, {Alessi}, \& {Villela}}]{2021MNRAS.504..356D}
{Dias}, W.~S., {Monteiro}, H., {Moitinho}, A., {et~al.} 2021, \mnras, 504, 356,
  \dodoi{10.1093/mnras/stab770}

\bibitem[{{Duronea} {et~al.}(2012){Duronea}, {Vasquez}, {Cappa}, {Corti}, \&
  {Arnal}}]{2012A&A...537A.149D}
{Duronea}, N.~U., {Vasquez}, J., {Cappa}, C.~E., {Corti}, M., \& {Arnal}, E.~M.
  2012, \aap, 537, A149, \dodoi{10.1051/0004-6361/201117958}

\bibitem[{{Dyson} \& {Williams}(1980)}]{1980pim..book.....D}
{Dyson}, J.~E., \& {Williams}, D.~A. 1980, {Physics of the interstellar medium}

\bibitem[{{Elmegreen} \& {Lada}(1977)}]{1977ApJ...214..725E}
{Elmegreen}, B.~G., \& {Lada}, C.~J. 1977, \apj, 214, 725,
  \dodoi{10.1086/155302}

\bibitem[{{Fich} {et~al.}(1990){Fich}, {Treffers}, \&
  {Dahl}}]{1990AJ.....99..622F}
{Fich}, M., {Treffers}, R.~R., \& {Dahl}, G.~P. 1990, \aj, 99, 622,
  \dodoi{10.1086/115356}

\bibitem[{{Gaia Collaboration} {et~al.}(2016){Gaia Collaboration}, {Prusti},
  {de Bruijne}, {Brown}, {Vallenari}, {Babusiaux}, {Bailer-Jones}, {Bastian},
  {Biermann}, {Evans}, \& et~al.}]{2016A&A...595A...1G}
{Gaia Collaboration}, {Prusti}, T., {de Bruijne}, J.~H.~J., {et~al.} 2016,
  \aap, 595, A1, \dodoi{10.1051/0004-6361/201629272}

\bibitem[{{Gaia Collaboration} {et~al.}(2018){Gaia Collaboration}, {Brown},
  {Vallenari}, {Prusti}, {de Bruijne}, {Babusiaux}, {Bailer-Jones}, {Biermann},
  {Evans}, {Eyer}, \& et~al.}]{2018A&A...616A...1G}
{Gaia Collaboration}, {Brown}, A.~G.~A., {Vallenari}, A., {et~al.} 2018, \aap,
  616, A1, \dodoi{10.1051/0004-6361/201833051}

\bibitem[{{Garmany} {et~al.}(2015){Garmany}, {Glaspey}, {Bragan{\c{c}}a},
  {Daflon}, {Borges Fernandes}, {Oey}, {Bensby}, \&
  {Cunha}}]{2015AJ....150...41G}
{Garmany}, C.~D., {Glaspey}, J.~W., {Bragan{\c{c}}a}, G.~A., {et~al.} 2015,
  \aj, 150, 41, \dodoi{10.1088/0004-6256/150/2/41}

\bibitem[{{Grasha} {et~al.}(2017){Grasha}, {Elmegreen}, {Calzetti}, {Adamo},
  {Aloisi}, {Bright}, {Cook}, {Dale}, {Fumagalli}, {Gallagher}, {Gouliermis},
  {Grebel}, {Kahre}, {Kim}, {Krumholz}, {Lee}, {Messa}, {Ryon}, \&
  {Ubeda}}]{2017ApJ...842...25G}
{Grasha}, K., {Elmegreen}, B.~G., {Calzetti}, D., {et~al.} 2017, \apj, 842, 25,
  \dodoi{10.3847/1538-4357/aa740b}

\bibitem[{{Grasha} {et~al.}(2018){Grasha}, {Calzetti}, {Bittle}, {Johnson},
  {Donovan Meyer}, {Kennicutt}, {Elmegreen}, {Adamo}, {Krumholz}, {Fumagalli},
  {Grebel}, {Gouliermis}, {Cook}, {Gallagher}, {Aloisi}, {Dale}, {Linden},
  {Sacchi}, {Thilker}, {Walterbos}, {Messa}, {Wofford}, \&
  {Smith}}]{2018MNRAS.481.1016G}
{Grasha}, K., {Calzetti}, D., {Bittle}, L., {et~al.} 2018, \mnras, 481, 1016,
  \dodoi{10.1093/mnras/sty2154}

\bibitem[{{Guetter} \& {Vrba}(1989)}]{1989AJ.....98..611G}
{Guetter}, H.~H., \& {Vrba}, F.~J. 1989, \aj, 98, 611, \dodoi{10.1086/115161}

\bibitem[{{Gutermuth} {et~al.}(2009){Gutermuth}, {Megeath}, {Myers}, {Allen},
  {Pipher}, \& {Fazio}}]{2009ApJS..184...18G}
{Gutermuth}, R.~A., {Megeath}, S.~T., {Myers}, P.~C., {et~al.} 2009, \apjs,
  184, 18, \dodoi{10.1088/0067-0049/184/1/18}

\bibitem[{{Gutermuth} {et~al.}(2005){Gutermuth}, {Megeath}, {Pipher},
  {Williams}, {Allen}, {Myers}, \& {Raines}}]{2005ApJ...632..397G}
{Gutermuth}, R.~A., {Megeath}, S.~T., {Pipher}, J.~L., {et~al.} 2005, \apj,
  632, 397, \dodoi{10.1086/432460}

\bibitem[{{Gutermuth} {et~al.}(2011){Gutermuth}, {Pipher}, {Megeath}, {Myers},
  {Allen}, \& {Allen}}]{2011ApJ...739...84G}
{Gutermuth}, R.~A., {Pipher}, J.~L., {Megeath}, S.~T., {et~al.} 2011, \apj,
  739, 84, \dodoi{10.1088/0004-637X/739/2/84}

\bibitem[{{Henney} {et~al.}(2005){Henney}, {Arthur}, \&
  {Garc{\'\i}a-D{\'\i}az}}]{2005ApJ...627..813H}
{Henney}, W.~J., {Arthur}, S.~J., \& {Garc{\'\i}a-D{\'\i}az}, M.~T. 2005, \apj,
  627, 813, \dodoi{10.1086/430593}

\bibitem[{{Hur} {et~al.}(2012){Hur}, {Sung}, \&
  {Bessell}}]{2012AJ....143...41H}
{Hur}, H., {Sung}, H., \& {Bessell}, M.~S. 2012, \aj, 143, 41,
  \dodoi{10.1088/0004-6256/143/2/41}

\bibitem[{{Inoue}(2001)}]{2001AJ....122.1788I}
{Inoue}, A.~K. 2001, \aj, 122, 1788, \dodoi{10.1086/323095}

\bibitem[{{Inutsuka} {et~al.}(2015){Inutsuka}, {Inoue}, {Iwasaki}, \&
  {Hosokawa}}]{2015A&A...580A..49I}
{Inutsuka}, S.-i., {Inoue}, T., {Iwasaki}, K., \& {Hosokawa}, T. 2015, \aap,
  580, A49, \dodoi{10.1051/0004-6361/201425584}

\bibitem[{{Jacoby} {et~al.}(1984){Jacoby}, {Hunter}, \&
  {Christian}}]{1984lss..book.....J}
{Jacoby}, G.~H., {Hunter}, D.~H., \& {Christian}, C.~A. 1984, {A library of
  stellar spectra}

\bibitem[{{Jose} {et~al.}(2012){Jose}, {Pandey}, {Ogura}, {Samal}, {Ojha},
  {Bhatt}, {Chauhan}, {Eswaraiah}, {Mito}, {Kobayashi}, \&
  {Yadav}}]{2012MNRAS.424.2486J}
{Jose}, J., {Pandey}, A.~K., {Ogura}, K., {et~al.} 2012, \mnras, 424, 2486,
  \dodoi{10.1111/j.1365-2966.2012.21175.x}

\bibitem[{{Kaur} {et~al.}(2020){Kaur}, {Sharma}, {Dewangan}, {Ojha},
  {Durgapal}, \& {Panwar}}]{2020ApJ...896...29K}
{Kaur}, H., {Sharma}, S., {Dewangan}, L.~K., {et~al.} 2020, \apj, 896, 29,
  \dodoi{10.3847/1538-4357/ab9122}

\bibitem[{{Kendrew} {et~al.}(2012){Kendrew}, {Simpson}, {Bressert}, {Povich},
  {Sherman}, {Lintott}, {Robitaille}, {Schawinski}, \&
  {Wolf-Chase}}]{2012ApJ...755...71K}
{Kendrew}, S., {Simpson}, R., {Bressert}, E., {et~al.} 2012, \apj, 755, 71,
  \dodoi{10.1088/0004-637X/755/1/71}

\bibitem[{{Kirsanova} {et~al.}(2009){Kirsanova}, {Wiebe}, \&
  {Sobolev}}]{2009ARep...53..611K}
{Kirsanova}, M.~S., {Wiebe}, D.~S., \& {Sobolev}, A.~M. 2009, Astronomy
  Reports, 53, 611, \dodoi{10.1134/S106377290907004X}

\bibitem[{{Koenig} \& {Leisawitz}(2014)}]{2014ApJ...791..131K}
{Koenig}, X.~P., \& {Leisawitz}, D.~T. 2014, \apj, 791, 131,
  \dodoi{10.1088/0004-637X/791/2/131}

\bibitem[{{Kroupa}(2002)}]{2002Sci...295...82K}
{Kroupa}, P. 2002, Science, 295, 82, \dodoi{10.1126/science.1067524}

\bibitem[{{Krumholz} \& {Matzner}(2009)}]{2009ApJ...703.1352K}
{Krumholz}, M.~R., \& {Matzner}, C.~D. 2009, \apj, 703, 1352,
  \dodoi{10.1088/0004-637X/703/2/1352}

\bibitem[{{Kumar} {et~al.}(2014){Kumar}, {Sharma}, {Manfroid}, {Gosset},
  {Rauw}, {Naz{\'e}}, \& {Kesh Yadav}}]{2014A&A...567A.109K}
{Kumar}, B., {Sharma}, S., {Manfroid}, J., {et~al.} 2014, \aap, 567, A109,
  \dodoi{10.1051/0004-6361/201323027}

\bibitem[{{Kumar} {et~al.}(2020){Kumar}, {Palmeirim}, {Arzoumanian}, \&
  {Inutsuka}}]{2020A&A...642A..87K}
{Kumar}, M.~S.~N., {Palmeirim}, P., {Arzoumanian}, D., \& {Inutsuka}, S.~I.
  2020, \aap, 642, A87, \dodoi{10.1051/0004-6361/202038232}

\bibitem[{{Kwan}(1997)}]{1997ApJ...489..284K}
{Kwan}, J. 1997, \apj, 489, 284, \dodoi{10.1086/304773}

\bibitem[{{Lada} \& {Lada}(2003)}]{2003ARAA..41...57L}
{Lada}, C.~J., \& {Lada}, E.~A. 2003, \araa, 41, 57,
  \dodoi{10.1146/annurev.astro.41.011802.094844}

\bibitem[{{Landolt}(1992)}]{1992AJ....104..340L}
{Landolt}, A.~U. 1992, \aj, 104, 340, \dodoi{10.1086/116242}

\bibitem[{{Lim} {et~al.}(2015){Lim}, {Sung}, {Hur}, \&
  {Park}}]{2015arXiv151101118L}
{Lim}, B., {Sung}, H., {Hur}, H., \& {Park}, B.-G. 2015, arXiv e-prints,
  arXiv:1511.01118.
\newblock \doarXiv{1511.01118}

\bibitem[{{Lim} {et~al.}(2011){Lim}, {Sung}, {Karimov}, \&
  {Ibrahimov}}]{2011JKAS...44...39L}
{Lim}, B., {Sung}, H.~S., {Karimov}, R., \& {Ibrahimov}, M. 2011, Journal of
  Korean Astronomical Society, 44, 39.
\newblock \doarXiv{1103.4927}

\bibitem[{{Lopez} {et~al.}(2014){Lopez}, {Krumholz}, {Bolatto}, {Prochaska},
  {Ramirez-Ruiz}, \& {Castro}}]{2014ApJ...795..121L}
{Lopez}, L.~A., {Krumholz}, M.~R., {Bolatto}, A.~D., {et~al.} 2014, \apj, 795,
  121, \dodoi{10.1088/0004-637X/795/2/121}

\bibitem[{{Luhman} {et~al.}(2016){Luhman}, {Esplin}, \&
  {Loutrel}}]{2016ApJ...827...52L}
{Luhman}, K.~L., {Esplin}, T.~L., \& {Loutrel}, N.~P. 2016, \apj, 827, 52,
  \dodoi{10.3847/0004-637X/827/1/52}

\bibitem[{{Ma{\'\i}z Apell{\'a}niz} {et~al.}(2016){Ma{\'\i}z Apell{\'a}niz},
  {Sota}, {Arias}, {Barb{\'a}}, {Walborn}, {Sim{\'o}n-D{\'\i}az}, {Negueruela},
  {Marco}, {Le{\~a}o}, {Herrero}, {Gamen}, \& {Alfaro}}]{2016ApJS..224....4M}
{Ma{\'\i}z Apell{\'a}niz}, J., {Sota}, A., {Arias}, J.~I., {et~al.} 2016,
  \apjs, 224, 4, \dodoi{10.3847/0067-0049/224/1/4}

\bibitem[{{Marcolino} {et~al.}(2009){Marcolino}, {Bouret}, {Martins},
  {Hillier}, {Lanz}, \& {Escolano}}]{2009A&A...498..837M}
{Marcolino}, W.~L.~F., {Bouret}, J.-C., {Martins}, F., {et~al.} 2009, \aap,
  498, 837, \dodoi{10.1051/0004-6361/200811289}

\bibitem[{{Martins} \& {Palacios}(2017)}]{2017A&A...598A..56M}
{Martins}, F., \& {Palacios}, A. 2017, \aap, 598, A56,
  \dodoi{10.1051/0004-6361/201629538}

\bibitem[{{Men'shchikov} {et~al.}(2010){Men'shchikov}, {Andr{\'e}}, {Didelon},
  {K{\"o}nyves}, {Schneider}, {Motte}, {Bontemps}, {Arzoumanian}, {Attard},
  {Abergel}, {Baluteau}, {Bernard}, {Cambr{\'e}sy}, {Cox}, {di Francesco}, {di
  Giorgio}, {Griffin}, {Hargrave}, {Huang}, {Kirk}, {Li}, {Martin}, {Minier},
  {Miville-Desch{\^e}nes}, {Molinari}, {Olofsson}, {Pezzuto}, {Roussel},
  {Russeil}, {Saraceno}, {Sauvage}, {Sibthorpe}, {Spinoglio}, {Testi},
  {Ward-Thompson}, {White}, {Wilson}, {Woodcraft}, \&
  {Zavagno}}]{2010A&A...518L.103M}
{Men'shchikov}, A., {Andr{\'e}}, P., {Didelon}, P., {et~al.} 2010, \aap, 518,
  L103, \dodoi{10.1051/0004-6361/201014668}

\bibitem[{{Meyer} {et~al.}(1997){Meyer}, {Calvet}, \&
  {Hillenbrand}}]{1997AJ....114..288M}
{Meyer}, M.~R., {Calvet}, N., \& {Hillenbrand}, L.~A. 1997, \aj, 114, 288,
  \dodoi{10.1086/118474}

\bibitem[{{Meynet} {et~al.}(1994){Meynet}, {Maeder}, {Schaller}, {Schaerer}, \&
  {Charbonnel}}]{1994A&AS..103...97M}
{Meynet}, G., {Maeder}, A., {Schaller}, G., {Schaerer}, D., \& {Charbonnel}, C.
  1994, \aaps, 103, 97

\bibitem[{{Moffat} {et~al.}(1979){Moffat}, {Fitzgerald}, \&
  {Jackson}}]{1979A&AS...38..197M}
{Moffat}, A.~F.~J., {Fitzgerald}, M.~P., \& {Jackson}, P.~D. 1979, \aaps, 38,
  197

\bibitem[{{Myers}(2009)}]{2009ApJ...700.1609M}
{Myers}, P.~C. 2009, \apj, 700, 1609, \dodoi{10.1088/0004-637X/700/2/1609}

\bibitem[{{Ninan} {et~al.}(2014){Ninan}, {Ojha}, {Ghosh}, {D'Costa}, {Naik},
  {Poojary}, {Sand imani}, {Meshram}, {Jadhav}, {Bhagat}, {Gharat}, {Bakalkar},
  {Prabhu}, {Anupama}, \& {Toomey}}]{2014JAI.....350006N}
{Ninan}, J.~P., {Ojha}, D.~K., {Ghosh}, S.~K., {et~al.} 2014, Journal of
  Astronomical Instrumentation, 3, 1450006, \dodoi{10.1142/S2251171714500068}

\bibitem[{{Ojha} {et~al.}(2004){Ojha}, {Tamura}, {Nakajima}, {Fukagawa},
  {Sugitani}, {Nagashima}, {Nagayama}, {Nagata}, {Sato}, {Pickles}, \&
  {Ogura}}]{2004ApJ...608..797O}
{Ojha}, D.~K., {Tamura}, M., {Nakajima}, Y., {et~al.} 2004, \apj, 608, 797,
  \dodoi{10.1086/420876}

\bibitem[{{Paladini} {et~al.}(2012){Paladini}, {Umana}, {Veneziani},
  {Noriega-Crespo}, {Anderson}, {Piacentini}, {Pinheiro Gon{\c{c}}alves},
  {Paradis}, {Tibbs}, {Bernard}, \& {Natoli}}]{2012ApJ...760..149P}
{Paladini}, R., {Umana}, G., {Veneziani}, M., {et~al.} 2012, \apj, 760, 149,
  \dodoi{10.1088/0004-637X/760/2/149}

\bibitem[{{Panagia}(1973)}]{1973AJ.....78..929P}
{Panagia}, N. 1973, \aj, 78, 929, \dodoi{10.1086/111498}

\bibitem[{{Pandey} {et~al.}(2000){Pandey}, {Ogura}, \&
  {Sekiguchi}}]{2000PASJ...52..847P}
{Pandey}, A.~K., {Ogura}, K., \& {Sekiguchi}, K. 2000, \pasj, 52, 847

\bibitem[{{Pandey} {et~al.}(2020{\natexlab{a}}){Pandey}, {Sharma}, {Kobayashi},
  {Sarugaku}, \& {Ogura}}]{2020MNRAS.492.2446P}
{Pandey}, A.~K., {Sharma}, S., {Kobayashi}, N., {Sarugaku}, Y., \& {Ogura}, K.
  2020{\natexlab{a}}, \mnras, 492, 2446, \dodoi{10.1093/mnras/stz3596}

\bibitem[{{Pandey} {et~al.}(2008){Pandey}, {Sharma}, {Ogura}, {Ojha}, {Chen},
  {Bhatt}, \& {Ghosh}}]{2008MNRAS.383.1241P}
{Pandey}, A.~K., {Sharma}, S., {Ogura}, K., {et~al.} 2008, \mnras, 383, 1241,
  \dodoi{10.1111/j.1365-2966.2007.12641.x}

\bibitem[{{Pandey} {et~al.}(2003){Pandey}, {Upadhyay}, {Nakada}, \&
  {Ogura}}]{2003AA...397..191P}
{Pandey}, A.~K., {Upadhyay}, K., {Nakada}, Y., \& {Ogura}, K. 2003, \aap, 397,
  191, \dodoi{10.1051/0004-6361:20021509}

\bibitem[{{Pandey} {et~al.}(2013){Pandey}, {Eswaraiah}, {Sharma}, {Samal},
  {Chauhan}, {Chen}, {Jose}, {Ojha}, {Kesh Yadav}, \&
  {Chandola}}]{2013ApJ...764..172P}
{Pandey}, A.~K., {Eswaraiah}, C., {Sharma}, S., {et~al.} 2013, \apj, 764, 172,
  \dodoi{10.1088/0004-637X/764/2/172}

\bibitem[{{Pandey} {et~al.}(2020{\natexlab{b}}){Pandey}, {Sharma}, {Panwar},
  {Dewangan}, {Ojha}, {Bisen}, {Sinha}, {Ghosh}, \&
  {Pandey}}]{2020ApJ...891...81P}
{Pandey}, R., {Sharma}, S., {Panwar}, N., {et~al.} 2020{\natexlab{b}}, \apj,
  891, 81, \dodoi{10.3847/1538-4357/ab6dc7}

\bibitem[{{Panwar} {et~al.}(2020){Panwar}, {Sharma}, {Ojha}, {Baug},
  {Dewangan}, {Bhatt}, \& {Pandey}}]{2020ApJ...905...61P}
{Panwar}, N., {Sharma}, S., {Ojha}, D.~K., {et~al.} 2020, \apj, 905, 61,
  \dodoi{10.3847/1538-4357/abc42e}

\bibitem[{{Pastorelli} {et~al.}(2019){Pastorelli}, {Marigo}, {Girardi}, {Chen},
  {Rubele}, {Trabucchi}, {Aringer}, {Bladh}, {Bressan}, {Montalb{\'a}n},
  {Boyer}, {Dalcanton}, {Eriksson}, {Groenewegen}, {H{\"o}fner}, {Lebzelter},
  {Nanni}, {Rosenfield}, {Wood}, \& {Cioni}}]{2019MNRAS.485.5666P}
{Pastorelli}, G., {Marigo}, P., {Girardi}, L., {et~al.} 2019, \mnras, 485,
  5666, \dodoi{10.1093/mnras/stz725}

\bibitem[{{Phelps} \& {Janes}(1994)}]{1994ApJS...90...31P}
{Phelps}, R.~L., \& {Janes}, K.~A. 1994, \apjs, 90, 31, \dodoi{10.1086/191857}

\bibitem[{{Pismis} \& {Moreno}(1976)}]{1976RMxAA...1..373P}
{Pismis}, P., \& {Moreno}, M.~A. 1976, \rmxaa, 1, 373

\bibitem[{{Pomar{\`e}s} {et~al.}(2009){Pomar{\`e}s}, {Zavagno}, {Deharveng},
  {Cunningham}, {Jones}, {Kurtz}, {Russeil}, {Caplan}, \&
  {Comer{\'o}n}}]{2009A&A...494..987P}
{Pomar{\`e}s}, M., {Zavagno}, A., {Deharveng}, L., {et~al.} 2009, \aap, 494,
  987, \dodoi{10.1051/0004-6361:200811050}

\bibitem[{{Quireza} {et~al.}(2006){Quireza}, {Rood}, {Balser}, \&
  {Bania}}]{2006ApJS..165..338Q}
{Quireza}, C., {Rood}, R.~T., {Balser}, D.~S., \& {Bania}, T.~M. 2006, \apjs,
  165, 338, \dodoi{10.1086/503901}

\bibitem[{{Robitaille} {et~al.}(2007){Robitaille}, {Whitney}, {Indebetouw}, \&
  {Wood}}]{2007ApJS..169..328R}
{Robitaille}, T.~P., {Whitney}, B.~A., {Indebetouw}, R., \& {Wood}, K. 2007,
  \apjs, 169, 328, \dodoi{10.1086/512039}

\bibitem[{{Robitaille} {et~al.}(2006){Robitaille}, {Whitney}, {Indebetouw},
  {Wood}, \& {Denzmore}}]{2006ApJS..167..256R}
{Robitaille}, T.~P., {Whitney}, B.~A., {Indebetouw}, R., {Wood}, K., \&
  {Denzmore}, P. 2006, \apjs, 167, 256, \dodoi{10.1086/508424}

\bibitem[{{Russeil} {et~al.}(1995){Russeil}, {Georgelin}, {Georgelin}, {Le
  Coarer}, \& {Marcelin}}]{1995A&AS..114..557R}
{Russeil}, D., {Georgelin}, Y.~M., {Georgelin}, Y.~P., {Le Coarer}, E., \&
  {Marcelin}, M. 1995, \aaps, 114, 557

\bibitem[{{Salpeter}(1955)}]{1955ApJ...121..161S}
{Salpeter}, E.~E. 1955, \apj, 121, 161, \dodoi{10.1086/145971}

\bibitem[{{Schmidt-Kaler}(1982)}]{Schmidt-Kaler1982}
{Schmidt-Kaler}, T. 1982, {in Landolt-B{\"o}rnstein: Numerical Data and
  Functional Relationship in Science and Technology, Vol. 2b. eds. Schaifers
  K., Voigt H. H., Landolt H. (Springer-Verlag), Berlin, p. 19}, ed. V.~H.~H.
  L.~H. {Schaifers}, K.

\bibitem[{{Schmiedeke} {et~al.}(2016){Schmiedeke}, {Schilke}, {M{\"o}ller},
  {S{\'a}nchez-Monge}, {Bergin}, {Comito}, {Csengeri}, {Lis}, {Molinari},
  {Qin}, \& {Rolffs}}]{2016A&A...588A.143S}
{Schmiedeke}, A., {Schilke}, P., {M{\"o}ller}, T., {et~al.} 2016, \aap, 588,
  A143, \dodoi{10.1051/0004-6361/201527311}

\bibitem[{{Schneider} {et~al.}(2012){Schneider}, {Csengeri}, {Hennemann},
  {Motte}, {Didelon}, {Federrath}, {Bontemps}, {Di Francesco}, {Arzoumanian},
  {Minier}, {Andr{\'e}}, {Hill}, {Zavagno}, {Nguyen-Luong}, {Attard},
  {Bernard}, {Elia}, {Fallscheer}, {Griffin}, {Kirk}, {Klessen}, {K{\"o}nyves},
  {Martin}, {Men'shchikov}, {Palmeirim}, {Peretto}, {Pestalozzi}, {Russeil},
  {Sadavoy}, {Sousbie}, {Testi}, {Tremblin}, {Ward-Thompson}, \&
  {White}}]{2012A&A...540L..11S}
{Schneider}, N., {Csengeri}, T., {Hennemann}, M., {et~al.} 2012, \aap, 540,
  L11, \dodoi{10.1051/0004-6361/201118566}

\bibitem[{{Sharma} {et~al.}(2008){Sharma}, {Pandey}, {Ogura}, {Aoki}, {Pandey},
  {Sandhu}, \& {Sagar}}]{2008AJ....135.1934S}
{Sharma}, S., {Pandey}, A.~K., {Ogura}, K., {et~al.} 2008, \aj, 135, 1934,
  \dodoi{10.1088/0004-6256/135/5/1934}

\bibitem[{{Sharma} {et~al.}(2006){Sharma}, {Pandey}, {Ogura}, {Mito},
  {Tarusawa}, \& {Sagar}}]{2006AJ....132.1669S}
---. 2006, \aj, 132, 1669, \dodoi{10.1086/507094}

\bibitem[{{Sharma} {et~al.}(2017){Sharma}, {Pandey}, {Ojha}, {Bhatt}, {Ogura},
  {Kobayashi}, {Yadav}, \& {Pandey}}]{2017MNRAS.467.2943S}
{Sharma}, S., {Pandey}, A.~K., {Ojha}, D.~K., {et~al.} 2017, \mnras, 467, 2943,
  \dodoi{10.1093/mnras/stx014}

\bibitem[{{Sharma} {et~al.}(2012){Sharma}, {Pandey}, {Pandey}, {Chauhan},
  {Ogura}, {Ojha}, {Borrissova}, {Mito}, {Verdugo}, \&
  {Bhatt}}]{2012PASJ...64..107S}
{Sharma}, S., {Pandey}, A.~K., {Pandey}, J.~C., {et~al.} 2012, \pasj, 64, 107.
\newblock \doarXiv{1204.2897}

\bibitem[{{Sharma} {et~al.}(2020){Sharma}, {Ghosh}, {Ojha}, {Pandey}, {Sinha},
  {Pandey}, {Ghosh}, {Panwar}, \& {Pandey}}]{2020MNRAS.498.2309S}
{Sharma}, S., {Ghosh}, A., {Ojha}, D.~K., {et~al.} 2020, \mnras, 498, 2309,
  \dodoi{10.1093/mnras/staa2412}

\bibitem[{{Shima} {et~al.}(2017){Shima}, {Tasker}, \&
  {Habe}}]{2017MNRAS.467..512S}
{Shima}, K., {Tasker}, E.~J., \& {Habe}, A. 2017, \mnras, 467, 512,
  \dodoi{10.1093/mnras/stw3279}

\bibitem[{{Sinha} {et~al.}(2020){Sinha}, {Sharma}, {Pandey}, {Yadav}, {Ogura},
  {Matsunaga}, {Kobayashi}, {Bisht}, {Pandey}, \&
  {Ghosh}}]{2020MNRAS.493..267S}
{Sinha}, T., {Sharma}, S., {Pandey}, A.~K., {et~al.} 2020, \mnras, 493, 267,
  \dodoi{10.1093/mnras/staa206}

\bibitem[{{Stahler} \& {Palla}(2005)}]{2005fost.book.....S}
{Stahler}, S.~W., \& {Palla}, F. 2005, {The Formation of Stars}, 865

\bibitem[{{Stetson}(1992)}]{1992ASPC...25..297S}
{Stetson}, P.~B. 1992, in Astronomical Society of the Pacific Conference
  Series, Vol.~25, Astronomical Data Analysis Software and Systems I, ed. D.~M.
  {Worrall}, C.~{Biemesderfer}, \& J.~{Barnes}, 297

\bibitem[{{Tenorio-Tagle}(1979)}]{1979A&A....71...59T}
{Tenorio-Tagle}, G. 1979, \aap, 71, 59

\bibitem[{{Thompson} {et~al.}(2012){Thompson}, {Urquhart}, {Moore}, \&
  {Morgan}}]{2012MNRAS.421..408T}
{Thompson}, M.~A., {Urquhart}, J.~S., {Moore}, T.~J.~T., \& {Morgan}, L.~K.
  2012, \mnras, 421, 408, \dodoi{10.1111/j.1365-2966.2011.20315.x}

\bibitem[{{Walborn} \& {Fitzpatrick}(1990)}]{1990PASP..102..379W}
{Walborn}, N.~R., \& {Fitzpatrick}, E.~L. 1990, \pasp, 102, 379,
  \dodoi{10.1086/132646}

\bibitem[{{Whitney} {et~al.}(2004){Whitney}, {Indebetouw}, {Bjorkman}, \&
  {Wood}}]{2004ApJ...617.1177W}
{Whitney}, B.~A., {Indebetouw}, R., {Bjorkman}, J.~E., \& {Wood}, K. 2004,
  \apj, 617, 1177, \dodoi{10.1086/425608}

\bibitem[{{Whitney} {et~al.}(2003{\natexlab{a}}){Whitney}, {Wood}, {Bjorkman},
  \& {Cohen}}]{2003ApJ...598.1079W}
{Whitney}, B.~A., {Wood}, K., {Bjorkman}, J.~E., \& {Cohen}, M.
  2003{\natexlab{a}}, \apj, 598, 1079, \dodoi{10.1086/379068}

\bibitem[{{Whitney} {et~al.}(2003{\natexlab{b}}){Whitney}, {Wood}, {Bjorkman},
  \& {Wolff}}]{2003ApJ...591.1049W}
{Whitney}, B.~A., {Wood}, K., {Bjorkman}, J.~E., \& {Wolff}, M.~J.
  2003{\natexlab{b}}, \apj, 591, 1049, \dodoi{10.1086/375415}

\bibitem[{{Whittet}(2003)}]{2003dge..conf.....W}
{Whittet}, D.~C.~B., ed. 2003, {Dust in the galactic environment, 2nd ed. by
  D.C.B. Whittet. Bristol: Institute of Physics (IOP) Publishing, 2003 Series
  in Astronomy and Astrophysics, ISBN 0750306246.}

\bibitem[{{Wouterloot} {et~al.}(1988){Wouterloot}, {Brand}, \&
  {Henkel}}]{1988A&A...191..323W}
{Wouterloot}, J.~G.~A., {Brand}, J., \& {Henkel}, C. 1988, \aap, 191, 323

\bibitem[{{Wright} {et~al.}(2010){Wright}, {Eisenhardt}, {Mainzer}, {Ressler},
  {Cutri}, {Jarrett}, {Kirkpatrick}, {Padgett}, {McMillan}, {Skrutskie},
  {Stanford}, {Cohen}, {Walker}, {Mather}, {Leisawitz}, {Gautier}, {McLean},
  {Benford}, {Lonsdale}, {Blain}, {Mendez}, {Irace}, {Duval}, {Liu}, {Royer},
  {Heinrichsen}, {Howard}, {Shannon}, {Kendall}, {Walsh}, {Larsen}, {Cardon},
  {Schick}, {Schwalm}, {Abid}, {Fabinsky}, {Naes}, \&
  {Tsai}}]{2010AJ....140.1868W}
{Wright}, E.~L., {Eisenhardt}, P.~R.~M., {Mainzer}, A.~K., {et~al.} 2010, \aj,
  140, 1868, \dodoi{10.1088/0004-6256/140/6/1868}

\bibitem[{{Yadav} {et~al.}(2016){Yadav}, {Pandey}, {Sharma}, {Ojha}, {Samal},
  {Mallick}, {Jose}, {Ogura}, {Richichi}, {Irawati}, {Kobayashi}, \&
  {Eswaraiah}}]{2016MNRAS.461.2502Y}
{Yadav}, R.~K., {Pandey}, A.~K., {Sharma}, S., {et~al.} 2016, \mnras, 461,
  2502, \dodoi{10.1093/mnras/stw1356}

\bibitem[{{Zavagno} {et~al.}(2006){Zavagno}, {Deharveng}, {Comer{\'o}n},
  {Brand}, {Massi}, {Caplan}, \& {Russeil}}]{2006A&A...446..171Z}
{Zavagno}, A., {Deharveng}, L., {Comer{\'o}n}, F., {et~al.} 2006, \aap, 446,
  171, \dodoi{10.1051/0004-6361:20053952}

\bibitem[{{Zavagno} {et~al.}(2010){Zavagno}, {Russeil}, {Motte}, {Anderson},
  {Deharveng}, {Rod{\'o}n}, {Bontemps}, {Abergel}, {Baluteau}, {Sauvage},
  {Andr{\'e}}, {Hill}, \& {White}}]{2010A&A...518L..81Z}
{Zavagno}, A., {Russeil}, D., {Motte}, F., {et~al.} 2010, \aap, 518, L81,
  \dodoi{10.1051/0004-6361/201014623}

\end{thebibliography}
\bibliographystyle{aasjournal}


\begin{table}
\centering
\caption{\label{log}  Log of  observations.}
\begin{tabular}{@{}rrr@{}}
\hline
Telescope/Instrument & Comments & Exp. (sec)$\times$ No. of frames\\
$[$Date of observations$]$& $[$Filter$]$   &\\
\hline
 1.3 m DFOT/2K CCD& Optical imaging of S301&\\
$[$ 2019 Dec 25$]$&$U$   &  $300\times3$\\
$"$&$B$   &  $300\times3$\\
$"$&$V$   &  $180\times3,60\times3$\\
$"$&$R_c$ &  $180\times3,60\times3,10\times3$\\
$"$&$I_c$ &  $180\times3,60\times3,10\times3$ \\
$[$2018 Jan 16 $]$&$I_c$    & $1200\times5$\\
$[$2018 Jan 16$]$&$V$ &    $1800\times5$\\
& Optical imaging of Standard field (SA98)&\\
$[$2019 Dec 25$]$&$U$   &  $300\times4$\\
$"$&$B$   &  $300\times1,120\times4,60\times1$\\
$"$&$V$   &  $60\times6,30\times6$\\
$"$&$R_c$ &  $30\times6,20\times6$\\
$"$&$I_c$ &  $30\times3,20\times4$\\
2m HCT/TIRSPEC&NIR imaging of S301 (Nine pointings)&\\
$[$2018 Dec 25  \& 2018 Dec 26$]$&$J$   &  $20\times35$\\
$"$&$H$   &  $20\times35$\\
$"$&$K$   &  $20\times35$\\
2m HCT & Optical spectroscopy of ALS 207,ALS 208 ans ALS 212 &\\
$[$2019 Aug 20$]$ & GRISM 7 &  $600\times1, 900\times1, 900\times1$ \\
\hline
\end{tabular}
\end{table}

\begin{table}
\centering
\caption{\label{cftt} Completeness of the photometric data}
\begin{tabular}{@{}rcccc@{}}
\hline
Band& Number of & Detection &   \multicolumn{2}{c}{Completeness limit$^d$ (upto 80 \%)}  \\
    &   sources$^a$ & Limit (mag)&(mag)&  Mass ($M_\odot$)\\
\hline
$U$    &596                 &19.5&$-$&$-$\\
$B$    &1156                &20.2&$-$&$-$\\
$V$    &5236                &22.7&21.0&0.4 \\
$R_c$  &2965                &20.7&$-$&$-$ \\
$I_c$  &5624                &21.6&20.0& 0.3\\
$J$    &2277$^b$+ 1846$^c$  &18.4&15.3& 0.4  \\
$H$    &2811$^b$+ 1534$^c$  &18.1&15.0&  0.4 \\
$K$    &2809$^b$+ 970$^c$   &17.8&15.0& 0.3 \\
\hline
\end{tabular}

$^a$: for $18^\prime.5\times18^\prime.5$ FOV;\\
$^b$: data from $TIRSPEC$;\\
$^c$: data from  2MASS for the bright stars;\\
$^d$: for distance = 3.54 kpc and $E(B-V)$ = 0.5 mag.
\end{table}

\begin{table*}
\centering
\caption{\label{PMT} Sample of 137 stars identified as a member of the NE-cluster (having optical counterparts). Magnitudes in different bands along with age 
and mass derived using the CMD analysis are also provided in the table.
The complete table is available in the electronic form only.}
\begin{tabular}{ccrcccccccc}
\hline
ID& $\alpha_{(2000)}$&$\delta_{(2000)}$& Parallax$\pm\sigma~~~~~$&$\mu_\alpha\pm\sigma$&$\mu_\delta\pm\sigma$ & $G$ & $G_{BP}-G_{RP}$ & Probability &  Age$\pm \sigma$ \\
& {\rm $(degrees)$} & {\rm $(degrees) $} & (mas)&  (mas/yr)& (mas/yr) & (mag) & (mag) & (Percentage) & (Myr)  \\
\hline
1 & 107.583885&-18.488205& $0.103\pm0.130$ &$-1.691\pm0.103$ &$2.728\pm0.108$&$18.124$&  2.058  &  100  &$1.2\pm0.32$  \\
2 & 107.591286&-18.478752& $0.265\pm0.523$ &$-2.020\pm0.330$ &$3.343\pm0.496$&$19.305$&  2.514  &  93   &$0.66\pm0.08$  \\ 
3 & 107.585960&-18.478432& $0.709\pm0.374$ &$-2.003\pm0.278$ &$2.309\pm0.302$&$19.388$&  2.356  &  94   &$1.89\pm0.47$ \\
4 & 107.588402&-18.483150& $0.337\pm0.085$ &$-1.758\pm0.060$ &$2.652\pm0.073$&$16.939$&  1.963  &  100  &$0.76\pm0.26$ \\
\hline
\end{tabular}
\end{table*}

\begin{table*}
\centering
\begin{tabular}{@{}r@{ }c@{ }r@{ }l@{ }l@{ }l@{ }l@{ }l@{ }l@{}}
\hline
Mass$\pm\sigma$ &$U\pm\sigma$ & $B\pm\sigma$& $V\pm\sigma$& $R\pm\sigma$& $I\pm\sigma$&   $J\pm\sigma$&     $H\pm\sigma$&  $K\pm\sigma$   \\
($M_{\odot}$)  & (mag)       &  (mag)         & (mag)             &    (mag)         & (mag)           &(mag)            &(mag)     &(mag) \\
\hline
 $0.99\pm0.1$ & $-$ &$-$              &$18.922\pm0.090$&    $17.771\pm0.020$      &$16.940\pm0.006$      &$15.519\pm0.011$  &$14.735\pm0.008$  &$14.466\pm0.009$ \\
 $0.53\pm0.04$ & $-$ &$-$              &$19.591\pm0.031$    &$18.248\pm0.028$    &$17.181\pm0.017$     &$15.487\pm0.009$  &$14.585\pm0.006$  &$14.275\pm0.008$  \\
 $0.81\pm0.08$& $-$ &$-$              &$19.769\pm0.021$    &$18.641\pm0.080$     &$17.652\pm0.022$     &$16.073\pm0.033$  &$15.258\pm0.014$  &$14.928\pm0.013$  \\
 $4.2\pm0.72$ & $-$ &$13.993\pm0.003$ &$13.186\pm0.004$    &$12.638\pm0.008$    &$12.109\pm0.008$     &$11.214\pm0.003$  &$10.738\pm0.003$  &$10.427\pm0.003$  \\
\hline
\end{tabular}
\end{table*}

\begin{table*}
\centering
\caption{\label{data1_yso}  A sample table containing information for 37 YSOs
identified in the S301 region ($18^\prime.5\times18^\prime.5$ FOV).  Magnitudes in different bands along with age 
and mass derived using the SED analysis are also provided in the table.
The complete table is available in an electronic form only.}
\begin{tabular}{@{}r@{ }c@{ }r@{ }l@{ }l@{ }l@{ }l@{ }l@{ }l@{ }l@{ }l@{ }l@{ }l@{ }c@{ }c@{ }c@{ }}
\hline
ID& $\alpha_{(2000)}$ & $\delta_{(2000)}$ & $U\pm\sigma$ & $B\pm\sigma$ &   $V\pm\sigma$ & $R\pm\sigma$& $I\pm\sigma$ & $J\pm\sigma$ & $H\pm\sigma$  \\
      &$(degrees)$    & $(degrees)$       & (mag)  &(mag)&  (mag)       &  (mag)         & (mag)  & (mag)  &(mag)  \\
\hline
1& 107.493520 &  -18.616960  &$-$  &$-$  &$21.842\pm0.065$ &$-$                &$19.055\pm0.014$              &$16.983\pm0.045$  &$15.990\pm0.026 $  \\
2& 107.546871 &  -18.493279  &$-$  &$-$  &$21.725\pm0.078$ &$20.249\pm0.096$  &$18.703\pm0.012$              &$15.978\pm0.013$  &$14.666\pm0.007$   \\
3& 107.596327 &  -18.488375  &$-$  &$-$  &$21.441\pm0.050$  &$-$               &$18.772\pm0.035$              &$16.859\pm0.031$  &$15.612\pm0.014$  \\
4& 107.546036 &  -18.476207  &$-$  &$-$  &$22.420\pm0.101$  &$-$               &$18.425\pm0.009$              &$15.330\pm0.012$  &$13.800\pm0.009 $   \\
\hline
\end{tabular}
\end{table*}

\begin{table*}
\centering
\begin{tabular}{@{}r@{ }c@{ }r@{ }l@{ }l@{ }l@{ }l@{ }l@{ }l@{ }l@{ }l@{ }l@{ }l@{ }c@{ }c@{ }c@{ }}
\hline
$K\pm\sigma$ & $[3.4]\pm\sigma$& $[4.6]\pm\sigma$& $[12]\pm\sigma$& $[22]\pm\sigma$& Class &   $N_{data}$ & $\chi^2_{min}$    &  Mass$\pm \sigma$ &  Age$\pm \sigma$ \\
(mag)       &  (mag)         & (mag)             &    (mag)         & (mag)           &            &     &            &($M_{\odot}$) & (Myr) \\
\hline
 $15.330\pm0.026 $ &$14.5\pm0.04$    &$14.116\pm0.048$ &$9.33\pm0.102$    &$7.653\pm0.168$  & II &  9  & 11.3 &$2.01\pm0.89 $&$4.16\pm3.35$ \\
 $13.773\pm0.008$ &$-$              &$-$ &$-$  &$-$& II &  6  & 0.1  &$2.54\pm1.68 $&$1.33\pm2.21$ \\
 $14.853\pm0.013$ &$-$              &$-$ &$-$  &$-$& II &  5  & 0.1  &$1.51\pm1.26 $&$2.26\pm2.22$ \\   
 $12.822\pm0.009$ &$-$              &$-$ &$-$  &$-$& II &  $-$  & $-$ & $-$  & $-$\\
\hline
\end{tabular}
\end{table*}


\clearpage

\appendix

\section{Reddening Law}

\begin{figure*}[h]
\centering
\includegraphics[width=0.45\textwidth]{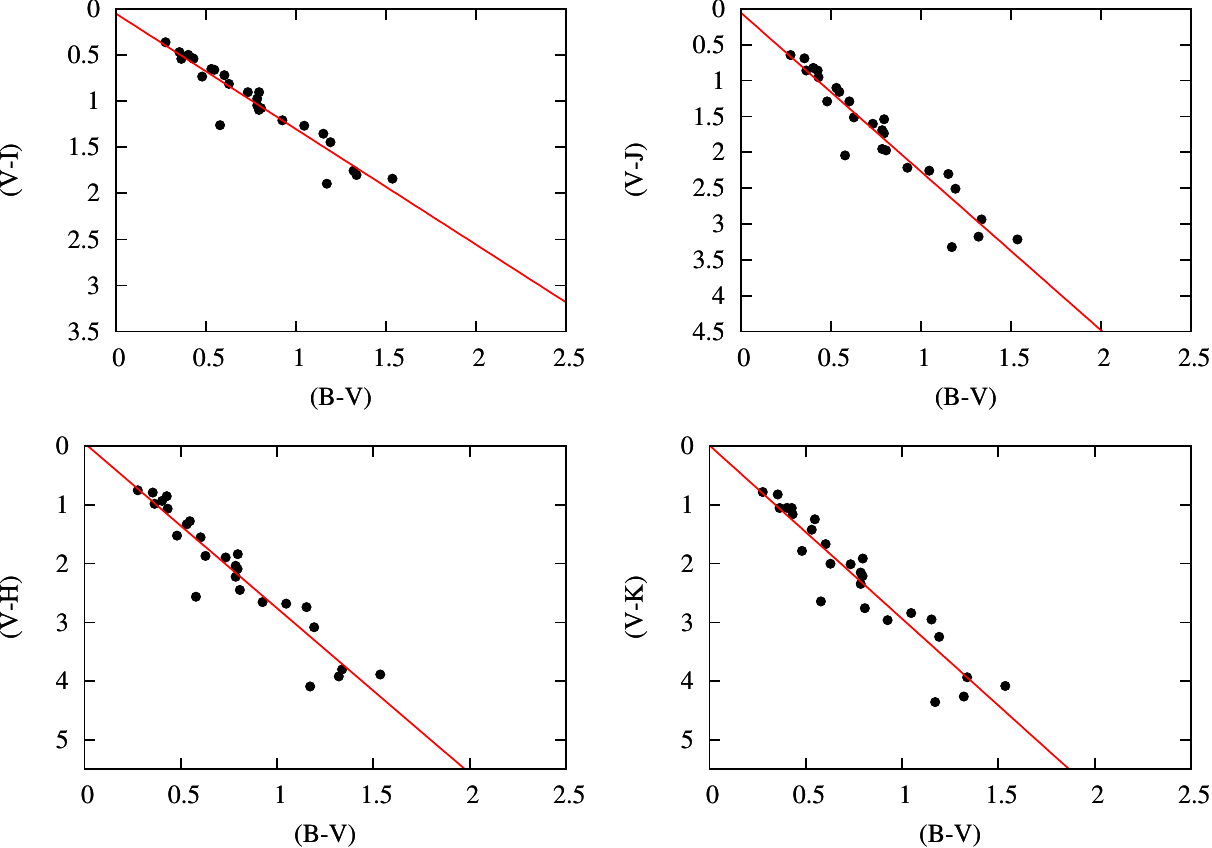}
\caption{\label{2color} $(V-I_c),(V - J), (V -H), (V - K)$ vs. $(B - V)$ TCDs
for the member stars associated with the S301 region. 
Straight lines show the least-square fit to the distribution of  stars.}
\end{figure*}

To study the nature of the diffuse interstellar medium (ISM) associated with the S301 region we have used TCDs to derive the ratio of total-to-selective extinction 
$R_V$ = $A_V$/$E(B-V)$ \citep{2003AA...397..191P}. Although we have a normal reddening law $R_V$ = 3.1$\pm$ 0.2 \citep{2003dge..conf.....W, 1989AJ.....98..611G, 2011JKAS...44...39L} in solar neighborhood but some anomaly has been found in case of few SFRs \citep[see e.g.,][]{2000PASJ...52..847P, 2008MNRAS.383.1241P,2012AJ....143...41H,2013ApJ...764..172P,2014A&A...567A.109K}. 
The $(V - \lambda)$ versus $(B - V)$ TCDs where $\lambda$ represents the wavelengths from one of the broad-band filters ($R, I, J, H, K, L$), are shown in Figure \ref{2color}. We used all member stars of S301 detected in optical and near-infrared bands, except YSOs (source of contamination due to excess in infra-red) to generate the TCDs. The method has been discussed in the appendix of \citet{2020ApJ...891...81P}.
The slopes of the the $(V-I_c),(V-J),(V-H)$ and $(V-K)$ versus $(B-V)$ TCDs fitted using the least squares fit, are found to be $1.27\pm0.15, 2.19\pm0.17, 2.78\pm0.19$ and $2.92\pm0.21$, respectively.
These value are higher than those found for the general ISM \citep[1.10, 1.96, 2.42 and 2.60; cf.,][]{2003AA...397..191P} by a factor of $\sim$1.1, but considering the errors in the fitting values,
we have assumed normal reddening law ($R_V$ =3.1) for the S301 region.

\section{YSOs identification and classification}
\begin{figure*}
\includegraphics[width=0.5\textwidth]{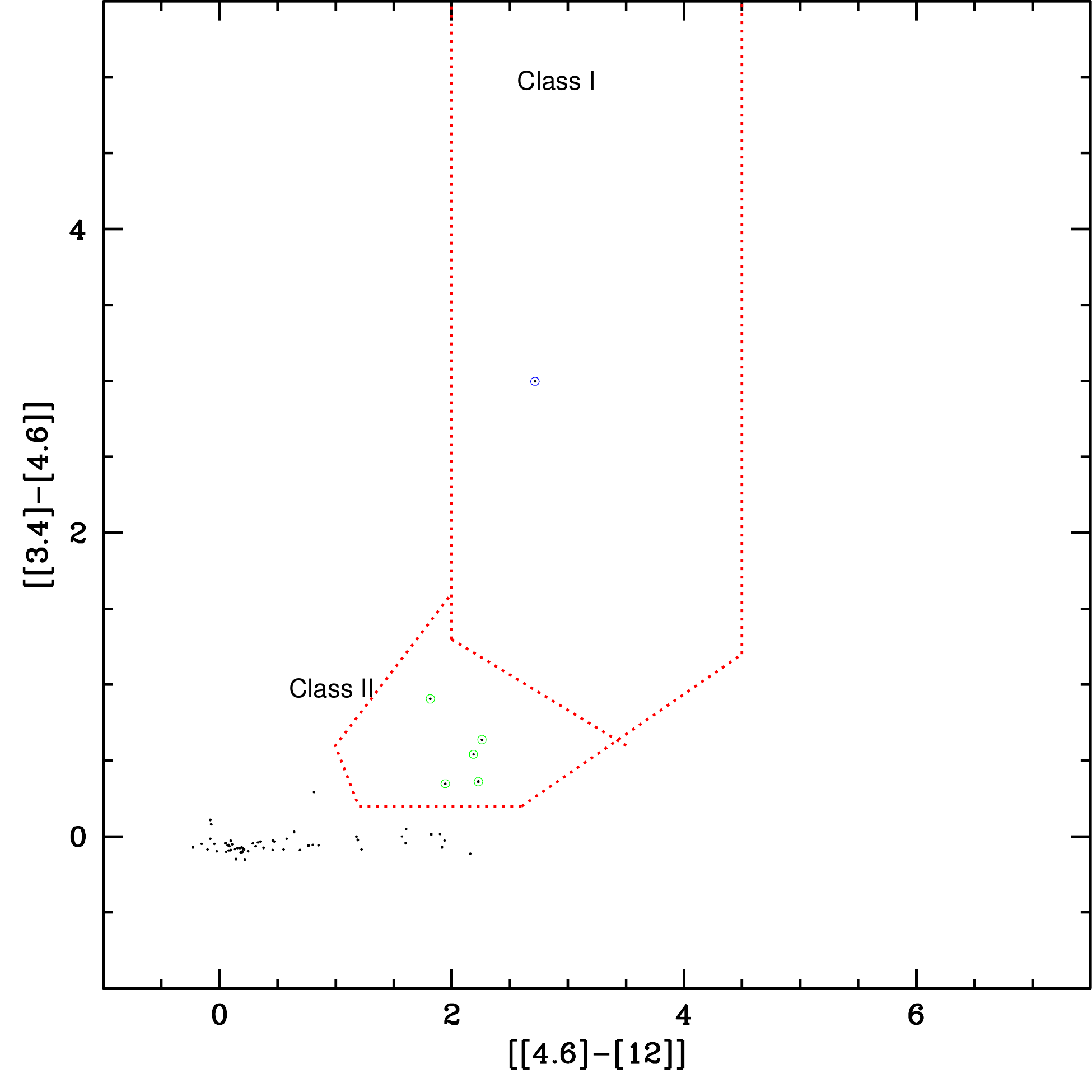}
\includegraphics[width=0.5\textwidth]{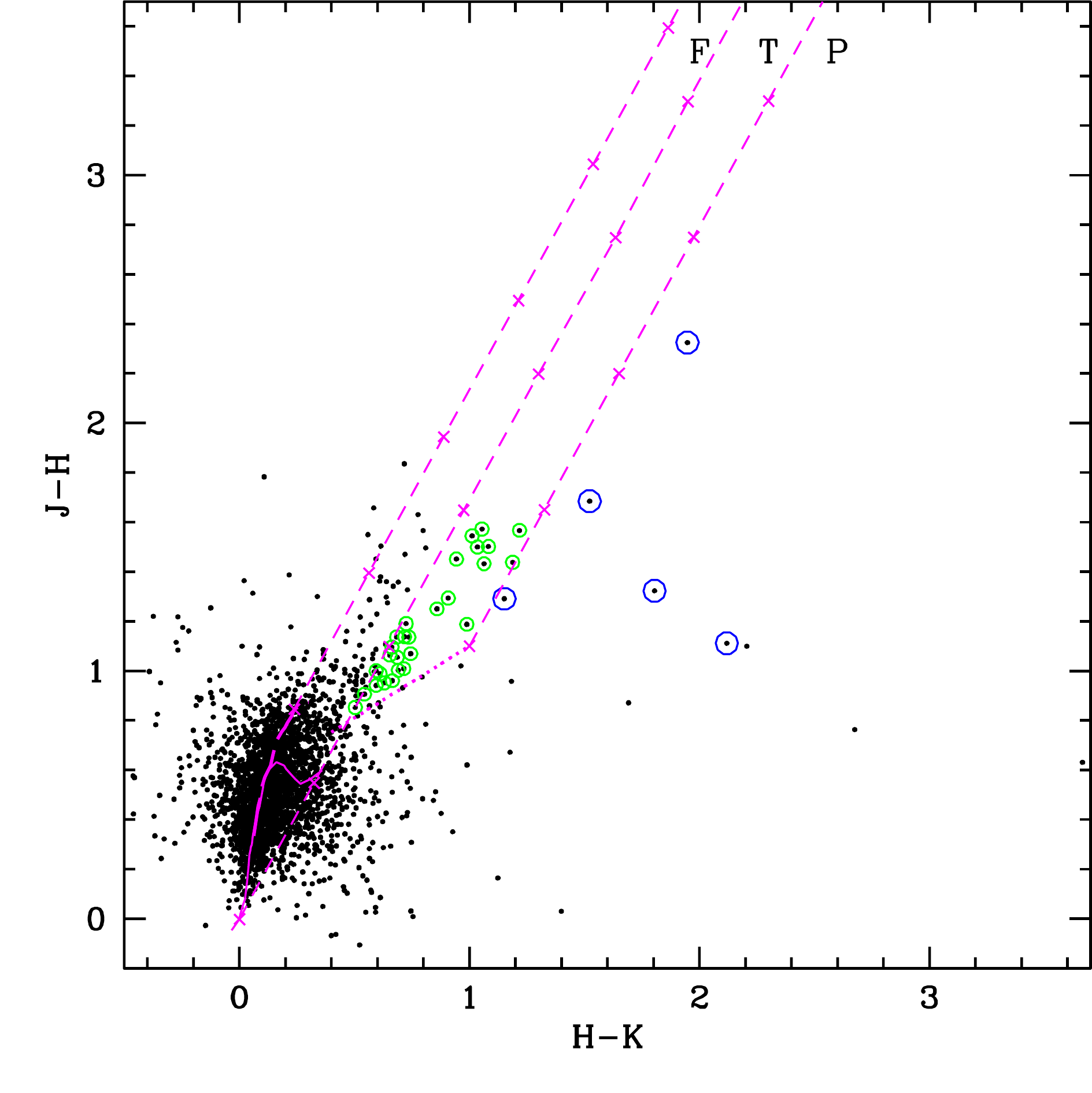}
\caption{\label{wise}  Left panel: ${[4.6] - [12]]}$ vs. ${[[3.4] - [4.6]]}$ TCD for the sources in the $\sim18^\prime.5\times18^\prime.5$ FOV of S301.The YSOs classified as Class~I and Class~II, are based on the color criteria by \citet{2014ApJ...791..131K}.
Right panel:  ${[H-K]}$ vs. ${[J - H]}$ TCD \citep{2004ApJ...608..797O} for the sources in the same FOV. 
The continuous and thick magenta dashed curves represent the reddened MS and giant branches \citep{1988PASP..100.1134B}, 
respectively. The dotted magenta line indicates the locus of dereddened CTTSs \citep{1997AJ....114..288M}. 
The parallel magenta dashed lines are the reddening lines drawn from the tip (spectral type M4) of the 
giant branch (left reddening line), from the base (spectral type A0) of the MS branch (middle reddening
line) and from the tip of the intrinsic CTTS line (right reddening line). 
The crosses on the reddening lines show an increment of $A_V$ = 5 mag. 
The YSOs classified as Class\,{\sc i} and Class\,{\sc ii} are shown with blue  and  green circles, respectively.  
 }
\end{figure*}

To make a census of YSOs in the present study,  we used the observed NIR and archival MIR data of S301 in $\sim$ $18^\prime.5\times18^\prime.5$ field of view 
around the central massive star ALS 207. We used the following schemes to identify and classify YSOs in the present study :
   We took the MIR data of S301 from the ALLWISE catalog of WISE and used the procedure outlined in \citet{2014ApJ...791..131K} to identify YSOs. The procedure contains several photometric quality criteria for different WISE bands as well as other selection criteria to isolate candidates like extra galactic contaminants like AGN, AGB stars and star-forming galaxies from the sample. In the Figure \ref{wise} (left panel), we have shown $([3.4] - [4.6])$ versus  $([4.6] - [12])$ TCD for all the sources belonging to this region. Using this procedure, we have identified a single Class\,{\sc i} and five Class\,{\sc ii} sources shown with blue and green colors respectively in  Figure \ref{wise} (left panel).
We also used the observed NIR data from TIRSPEC along with the 2MASS data to identify YSOs in the region. For that we made a combined catalog using the observed NIR data and 2MASS data as the brighter sources are saturated in the TIRSPEC observation and were replaced with the 2MASS data. We also removed those sources from the catalog which have their counterparts in the ALLWISE catalog to cancel out any possible redundancy in YSOs identification. The NIR TCD has been used to identify YSOS using the final NIR catalog and the scheme is well described in \citet{2004ApJ...608..797O}. The NIR TCD is plotted in Figure \ref{wise} (right panel) with all the stars in our final NIR catalog. In Figure one can see the thick broken curves representing the MS and giant branches \citep{1988PASP..100.1134B}, the locus of unreddened Classical T Tauri Stars (CTTS) \citep{1997AJ....114..288M} is shown  by a dotted line. The parallel dashed lines drawn from the tip of the giant branch, base of the MS branch and the tip of the intrinsic CTTS line, are the reddening vectors. The sources belonging to the `F' region in the figure are recognized as either field stars or Class\,{\sc iii} sources while the sources in the `T' region (between the middle and lower reddening lines) are categorized as the  CTTS or Class\,{\sc ii} sources. The sources falling in the `P' region are classified as Class\,{\sc i} sources. Using the NIR data, we finally identified 5 Class\,{\sc i} and 26 Class\,{\sc ii} sources marked with blue and green colors, respectively in Figure \ref{wise} (right panel). In total, 37 YSOs were identified in the $18^\prime.5\times18^\prime.5$ FOV
around the central massive star ALS 207.

\end{document}